\pdfoutput=1
\documentclass[12pt,a4paper]{article}
{\def\usepackage{ws-procs9x6}}

\ifnum\pdfoutput=1\else
\PassOptionsToPackage{hypertex}{hyperref}
\PassOptionsToPackage{draft}{graphicx}
\usepackage{showkeys}
\fi

\usepackage{amsmath,amssymb}
\usepackage[bookmarks=true,hyperfigures=true]{hyperref}
\usepackage{graphicx}
\usepackage[nosort]{cite}
\usepackage[bulletsep]{collref}
\usepackage{verbatim}

\usepackage[a4paper,text={450pt,650pt},centering]{geometry}

\let\oldbfseries=\bfseries
\let\oldmdseries=\mdseries
\let\oldnormalfont=\normalfont
\renewcommand{\bfseries}{\oldbfseries\boldmath}
\renewcommand{\mdseries}{\oldmdseries\unboldmath}
\renewcommand{\normalfont}{\oldnormalfont\unboldmath}

\allowdisplaybreaks[3]

\numberwithin{equation}{section}

\usepackage[font=small,labelfont=bf,width=0.85\textwidth]{caption}

\providecommand{\hypersetup}[1]{}
\providecommand{\texorpdfstring}[2]{#1}

\hypersetup{plainpages=false}
\hypersetup{pdfpagemode=UseNone}
\hypersetup{bookmarksnumbered=true}
\hypersetup{pdfstartview=FitH}
\hypersetup{colorlinks=false}
\hypersetup{citebordercolor={.5 1 .5}}
\hypersetup{urlbordercolor={.5 1 1}}
\hypersetup{linkbordercolor={1 .7 .7}}

\DeclareMathSymbol{\Gamma}{\mathalpha}{letters}{"00}
\DeclareMathSymbol{\Delta}{\mathalpha}{letters}{"01}
\DeclareMathSymbol{\Theta}{\mathalpha}{letters}{"02}
\DeclareMathSymbol{\Lambda}{\mathalpha}{letters}{"03}
\DeclareMathSymbol{\Xi}{\mathalpha}{letters}{"04}
\DeclareMathSymbol{\Pi}{\mathalpha}{letters}{"05}
\DeclareMathSymbol{\Sigma}{\mathalpha}{letters}{"06}
\DeclareMathSymbol{\Upsilon}{\mathalpha}{letters}{"07}
\DeclareMathSymbol{\Phi}{\mathalpha}{letters}{"08}
\DeclareMathSymbol{\Psi}{\mathalpha}{letters}{"09}
\DeclareMathSymbol{\Omega}{\mathalpha}{letters}{"0A}

\newcommand{\gen}[1]{\mathrm{#1}}
\newcommand{\superN}{\mathcal{N}}

\newcommand{\Tr}{\mathop{\mathrm{Tr}}}
\renewcommand{\Re}{\mathop{\mathrm{Re}}}

\newcommand{\Li}{\mathop{\mathrm{Li}}\nolimits}
\newcommand{\disc}{\mathop{\mathrm{disc}}\nolimits}

\newcommand{\order}[1]{\mathcal{O}(#1)}

\newcommand{\Complex}{\mathbb{C}}

\newcommand{\sign}{\mathop{\mathrm{sign}}}

\ifx\genfrac\sdflkaj\else\fi
\newcommand{\sfrac}[2]{{\textstyle\frac{#1}{#2}}}
\newcommand{\half}{\sfrac{1}{2}}

\newcommand{\quarter}{\sfrac{1}{4}}

\newcommand{\indup}[1]{_{\mathrm{#1}}}

\newcommand{\lrbrk}[1]{\left(#1\right)}
\newcommand{\bigbrk}[1]{\bigl(#1\bigr)}

\newcommand{\tvev}[1]{\langle#1\rangle}
\newcommand{\vev}[1]{\langle#1\rangle}

\newcommand{\comm}[2]{[#1,#2]}

\newcommand{\state}[1]{\mathopen{|}#1\mathclose{\rangle}}
\newcommand{\costate}[1]{\mathopen{\langle}#1\mathclose{|}}

\newcommand{\sprod}[2]{\langle#1,#2\rangle}

\newcommand{\cprod}[2]{[#1,#2]}

\newcommand{\tprod}[2]{\langle #1,#2]}
\newcommand{\tprods}[2]{\langle #1#2]}

\newcommand{\alg}[1]{\mathfrak{#1}}
\newcommand{\grp}[1]{\mathrm{#1}}
\newcommand{\genY}[1]{\widehat{\mathrm{#1}}}

\newcommand{\nln}{\nonumber\\}

\def\[{\begin{equation}}
\def\]{\end{equation}}
\def\<{\begin{eqnarray}}
\def\>{\end{eqnarray}}
\def\nln{\nonumber\\}

\makeatletter
\let\@keywords\@empty
\let\@subject\@empty
\providecommand{\keywords}[1]{\gdef\@keywords{#1}}
\providecommand{\subject}[1]{\gdef\@subject{#1}}
\def\thetitle{\@title}
\def\theauthor{\@author}
\def\thesubject{\@subject}
\def\thedate{\@date}
\def\thekeywords{\@keywords}
\makeatother
\AtBeginDocument{
\hypersetup{pdftitle={\thetitle}}%
\hypersetup{pdfauthor={\theauthor}}%
\hypersetup{pdfsubject={\thesubject}}%
\hypersetup{pdfkeywords={\thekeywords}}%
}

\makeatletter
\def\mr@ignsp#1 {\ifx\:#1\@empty\else #1\expandafter\mr@ignsp\fi}%
\newcommand{\multiref}[1]{\begingroup
\xdef\mr@no@sparg{\expandafter\mr@ignsp#1 \: }%
\def\mr@comma{}%
\@for\mr@refs:=\mr@no@sparg\do{\mr@comma\def\mr@comma{,}\ref{\mr@refs}}%
\endgroup}
\makeatother

\renewcommand{\eqref}[1]{(\multiref{#1})}

\newcommand{\namedref}[2]{\hyperref[#2]{#1~\ref*{#2}}}
\newcommand{\Secref}[1]{\namedref{Section}{#1}}
\newcommand{\secref}[1]{\namedref{Sec.}{#1}}

\newcommand{\appref}[1]{\namedref{App.}{#1}}

\newcommand{\figref}[1]{\namedref{Fig.}{#1}}

\makeatletter
\newlength{\apb@width}
\newcommand{\autoparbox}[2][c]{\settowidth{\apb@width}{#2}\parbox[#1]{\apb@width}{#2}}
\newcommand{\includegraphicsbox}[2][]{\autoparbox{\includegraphics[#1]{#2}}}
\makeatother

\ifx\href\asklfhas\newcommand{\href}[2]{#2}\fi
\newcommand{\arxivlink}[1]{\href{http://arxiv.org/abs/#1}{arxiv:#1}}

\title{Null Polygonal Wilson Loops\texorpdfstring{\\}{} in Full \texorpdfstring{$\mathcal{N}=4$}{N=4} Superspace}
\author{Niklas Beisert\texorpdfstring{$^{a,b}$}{},
Song He\texorpdfstring{$^{b}$}{},
 Burkhard U.W.\ Schwab\texorpdfstring{$^{a,b}$}{},
Cristian Vergu\texorpdfstring{$^{a,c}$}{}}
\begin{document}

\pdfbookmark[1]{Title Page}{title}

\thispagestyle{empty}
\begin{flushright}\footnotesize
\texttt{\arxivlink{1203.1443}}\\
\texttt{AEI-2012-007}\\%
\texttt{NSF-KITP-12-011}\\%
\end{flushright}
\vspace{1cm}

\begin{center}%
\begingroup\Large\bfseries\thetitle\par\endgroup
\vspace{1cm}%

\begingroup\scshape\theauthor\par\endgroup
\vspace{5mm}%

\begingroup\itshape
$^a$
Institut f\"ur Theoretische Physik,\\
Eidgen\"ossische Technische Hochschule Z\"urich\\
Wolfgang-Pauli-Strasse 27, 8093 Z\"urich, Switzerland
\vspace{3mm}

$^b$
Max-Planck-Institut f\"ur Gravitationsphysik\\
Albert-Einstein-Institut\\
Am M\"uhlenberg 1, 14476 Potsdam, Germany
\vspace{3mm}

$^c$
Department of Physics, Brown University\\
Box 1843, Providence, RI 02912, USA
\vspace{3mm}

\par\endgroup
\vspace{5mm}

\begingroup\ttfamily
\verb+{nbeisert,schwabbu,verguc}@itp.phys.ethz.ch, song.he@aei.mpg.de+
\par\endgroup

\vspace{1cm}

\textbf{Abstract}\vspace{7mm}

\begin{minipage}{12.7cm}
We compute the one-loop expectation value of light-like polygonal Wilson loops in $\mathcal{N}=4$ super-Yang--Mills
theory in full superspace.
When projecting to chiral superspace we recover the known results for tree-level
next-to-maximally-helicity-violating (NMHV) scattering amplitude.
The one-loop MHV amplitude is also included in our result
but there are additional terms which do not
immediately correspond to scattering amplitudes.
We finally discuss different regularizations and their Yangian anomalies.
\end{minipage}

\end{center}

\newpage

\section{Introduction}
\label{sec:intro}

The $\mathcal{N}=4$ super-Yang--Mills theory exhibits integrable features
in the planar limit~\cite{Beisert:2010jr}.  This integrability has been used very successfully
for finding the spectrum of anomalous dimensions of single-trace local
operators.

One would like to go beyond this, and compute other physical
quantities.  The $\mathcal{N}=4$ super-Yang--Mills theory being conformal,
the correlation functions of local, gauge invariant operators
are natural quantities to consider.
However, while some partial results
have been obtained concerning the correlation functions, we are still
very far from having an all-order understanding.

Part of the problem is that, even after using the superconformal symmetry, the correlation functions depend on a large number of invariants. One
can consider special limits in which the kinematics simplify.  For example, one can take the operators in the correlation functions to be
pairwise light-like separated.  In that limit, the correlation functions are essentially squares of the Wilson loop in the fundamental
representation of the $SU(N)$ gauge group (or the Wilson loop in the adjoint representation, which is the same in the large $N$ limit), defined
on a polygonal light-like contour~\cite{arXiv:1007.3246, arXiv:1007.3243, arXiv:1103.3714, arXiv:1103.4353}.

One can also consider scattering amplitudes in $\mathcal{N}=4$ super-Yang--Mills.  In the vacuum where all the scalars have zero expectation
values, these amplitudes are infrared (IR) divergent and need to be regularized.  There are two favored options for performing the regularization.  One
is a supersymmetry preserving variant of dimensional regularization, the other is the so-called ``mass regularization'', which consists in
giving vacuum expectation values to some of the scalars~\cite{arXiv:0705.0303, Alday:2009zm}.

The correlation functions in the light-like limit and the polygonal light-like Wilson loops have ultraviolet (UV) divergences so they also need
to be regularized.  How to perform this regularization is not entirely obvious, and some difficulties have been reported in the literature (see
ref.~\cite{arXiv:1103.3008}), concerning the use of dimensional regularization.

Even though the scattering amplitudes, the light-like polygonal Wilson loops and the correlation functions in the light-like limit seem to be
very different, it has been shown that, in fact, they contain essentially the same information.\footnote{To be more precise, the scattering
amplitudes and the correlation functions contain parts which are odd under parity transformations~\cite{Bern:2006vw, arXiv:0803.1465}.  As shown
in ref.~\cite{arXiv:0803.1465}, the scattering amplitudes also contain so-called ``$\mu$ terms'' which are curious integrals such that the
integrand vanishes when the dimensional regularization parameter $\epsilon = (4-D)/2$ goes to zero, but the \emph{integral} diverges. However,
it turns out that when taking the logarithm, all of these complicated contributions disappear and the result matches the Wilson loop result. The
$\mu$ terms also cancel for the two-loop NMHV amplitudes, as shown in~\cite{Kosower:2010yk}.}  There are several arguments that strongly support
this.  At strong coupling this can be understood from supersymmetric T-duality (see~\cite{arXiv:0705.0303,arXiv:0807.3196}), which maps Wilson
loops to scattering amplitudes and at the same time exchanges the UV and IR regimes.  At weak coupling this is supported by explicit
perturbative
computations~\cite{Anastasiou:2003kj,Bern:2005iz,Bern:2006vw,arXiv:0803.1465,arXiv:0707.1153,arXiv:0709.2368,arXiv:0712.1223,arXiv:0712.4138,Anastasiou:2009kna}.
Note that, when relating Wilson loops to correlation functions there is no need to exchange UV and IR\@.  This is only needed when relating them
to the scattering amplitudes.

The interchange of UV and IR makes it more challenging to match the answers for Wilson loops (or correlation functions) and scattering
amplitudes. For example, in dimensional regularization one has to match $\epsilon_{\text{UV}}$, which is used for regularizing the UV
divergences of the Wilson loop with $\epsilon_{\text{IR}}$, which is used for regularizing the IR divergences of the scattering amplitudes.

In the planar limit one can unambiguously define a notion of integrand~\cite{ArkaniHamed:2010kv,ArkaniHamed:2010gh} for scattering amplitudes/Wilson loops.  The integrand is a rational differential form which is well defined even in the absence of a regulator.  Because of this, it has been more fruitful to compare the integrands of scattering amplitudes and Wilson loops and in refs.~\cite{Mason:2010yk,arXiv:1010.1167} it was shown that the integrands coincide.

So far, all of these quantities have been mostly studied in chiral
superspace (ref.~\cite{CaronHuot:2011ky} by Caron-Huot is an exception).  The motivation
was that, for describing the on-shell states used in scattering
amplitudes one uses an on-shell superspace which is very naturally
described chirally.  However, the chiral superspace has a big downside:
it obscures some of the symmetries of the answers.  This goes beyond
just the obvious breaking of manifest parity symmetry since for chiral Wilson
loops the $\bar{\gen{Q}}$ operator is broken as well.  However, as
has been shown recently (see~\cite{Bullimore:2011kg,CaronHuot:2011kk}), one can repair the non-invariance under the
$\bar{\gen{Q}}$ operator and use it to build higher-loop answers from
lower-loop ones.

It has been shown in ref.~\cite{arXiv:0807.1095} that the tree level
scattering amplitudes in $\mathcal{N}=4$ super-Yang--Mills are invariant
under a hidden dual superconformal symmetry.
In ref.~\cite{arXiv:0902.2987} the superconformal symmetry
and the dual superconformal symmetry were shown to generate
an infinite-dimensional Yangian symmetry.
In the case of $\mathcal{N}=4$ super-Yang--Mills this is the
Yangian $\gen{Y}[\alg{psu}(2,2\vert 4)]$.
At loop level, this Yangian symmetry is broken by IR divergences
for scattering amplitudes or by UV divergences for Wilson loops.
The integrands are Yangian invariant
up to total derivatives~\cite{arXiv:1001.3348,ArkaniHamed:2010kv}.

In the chiral formulation, the momentum twistors~\cite{Hodges:2009hk,Mason:2009qx}
$W_i = (w_i \vert \chi_i) = (\lambda_i, \mu_i \vert \chi_i)$, which are points in
$\mathbb{CP}^{3 \vert 4}$ play an important role.  They provide
unconstrained variables for the kinematics and the superconformal group
acts linearly on their homogeneous coordinates.  The results are
expressed in terms of two kinds of basic objects: four-brackets
$\langle i j k l\rangle = \varepsilon_{abcd} w_i^a w_j^b w_k^c w_l^d$, and
$R$-invariants~\cite{arXiv:0807.1095,Mason:2009qx}
\begin{equation}
  [i j k l m] = \frac {\delta^{0 \vert 4} (\langle i j k l\rangle \chi_m
+ \langle j k l m\rangle \chi_i + \langle k l m i\rangle \chi_j +
\langle l m i j\rangle \chi_k + \langle m i j k\rangle \chi_l)}{\langle
i j k l\rangle \langle j k l m\rangle \langle k l m i\rangle \langle l m
i j\rangle \langle m i j k\rangle}\,.
\end{equation}

The $R$-invariants are superconformal invariant and in fact Yangian
invariant, but the four-brackets are only conformally invariant.  In the
answer for scattering amplitudes the $R$-invariants enter somewhat
trivially, as global multiplicative factors, but the four-brackets enter
in a much more non-trivial way, as arguments of transcendental
functions.  Therefore, the superconformal symmetry is much less obvious
in this presentation.

Motivated by these shortcomings of the chiral formalism, in this paper
we study the Wilson loops in full superspace, where both $\gen{Q}$ and
$\bar{\gen{Q}}$ symmetry operators play the same role.

In the non-chiral approach, which we will present in more detail below,
we have two sets of momentum twistors, $W_i$ and their conjugate $\bar{W}_i$.  
Using them one can easily form superconformal invariants $W_i \cdot \bar{W}_j$ 
(see \appref{app:invariants} for a discussion of superconformal invariants).  
In the chiral formulation the answers are written in terms of twistor four-brackets.  
These four-brackets are conformal but not superconformal invariant.  
If we want to make superconformal symmetry manifest, 
we need to use quantities like $W_i \cdot \bar{W}_j$ instead.  
When performing the Grassmann expansion 
of the superconformal invariants $W_i \cdot \bar{W}_j$ 
we recover the usual four-brackets at the first order.

The momentum twistors and their conjugates are not unconstrained, but they
satisfy some relations $W_i \cdot \bar{W}_i = W_i \cdot \bar{W}_{i+1} =
W_i \cdot \bar{W}_{i-1} = 0$.

We define and compute a Wilson loop in full superspace to one-loop order.  At this order the answer contains a rational part which is the same
as the tree-level NHMV scattering amplitude,\footnote{The tree-level NMHV amplitudes can be written in several different forms. The form we
obtained is the same as the CSW-like form of Mason and Skinner~\cite{Mason:2010yk}.}
and a transcendental piece which is similar to the one-loop
MHV scattering amplitude.  The transcendental part of the answer is of transcendentality two and it contains dilogarithms and products of
logarithms of superconformal invariants $W_i \cdot \bar{W}_j$.\footnote{As we will show, its divergent parts in a certain regularization contain
terms like $\langle i j\rangle$, or  $[i j]$, which break superconformal symmetry.}  We believe that this form of the answer is more
satisfactory than the chiral presentation, since the superconformal symmetry is manifest,
except for some ``boundary'' cases which appear when the propagator
approaches a null edge.
So the breaking of the symmetry is localized to the regions where the UV divergences arise.

We should note that the transcendentality two part of the answer, when
expanded out in powers of Grassmann variables, yields the one-loop
answer at zeroth order in the expansion.  In ref.~\cite{CaronHuot:2011ky}, Caron-Huot also considered the next order in the $\bar{\theta}$ expansion.

The answer we obtain is not in the form
that is usually presented in the literature, but it is related to it via
dilogarithm identities.  Another noteworthy feature is that the rational
and transcendental parts are computed by two kinds of propagators, which
are related by a Grassmann Fourier transform.

We have also studied the superconformal and
Yangian anomalies of the answer.
In order to avoid dealing with divergent quantities, we have used a framing
regularization, conjectured a super-Poincar\'e invariant expression, and defined a finite quantity from the Wilson loop which is similar to one
defined in ref.~\cite{Alday:2010ku} for studying the near collinear limit of Wilson loops.
Then we defined and computed the action of the Yangian
on this quantity.

It is important to stress that our computation applies only to
non-chiral Wilson loops $W(x_i,\theta_i,\bar{\theta}_i)$ but not to scattering amplitudes.  One can obtain the scattering
amplitudes by setting $\bar{\theta}_i = 0$ but there is no obvious way
to define non-chiral scattering amplitudes such that they are dual to
the non-chiral Wilson loops.

The organization of the paper is as follows.  In \secref{sec:superspace} we review $\mathcal{N}=4$ super-Yang--Mills theory in $\mathcal{N}=4$
superspace.  In the next \Secref{sec:gaugefieldprop} we introduce some prepotentials for the gauge connection and compute their two-point
functions in light-cone gauge.
This puts us in the position to carry out simple computations in this quantum field theory.
In \secref{sec:vertexcor} we perform the one-loop computations in momentum space and in \secref{sec:twistor}
we perform the same computations in momentum twistor space.  In \secref{sec:regularization} we present the regularizations we use. In \secref{sec:yangian} we compute the Yangian anomalies.  We end in \secref{sec:conclusions} with some conclusions.  Our
conventions and some computational details can be found in the appendices.

\section{\texorpdfstring{$\superN=4$}{N=4} SYM in Superspace}
\label{sec:superspace}

We would like to compute the Wilson loop expectation value
with as much manifest supersymmetry as possible.
The obvious choice is to use the $\mathcal{N}=4$ superspace.
We therefore review a formulation of classical on-shell $\mathcal{N}=4$
super-Yang--Mills theory (SYM) in this full (non-chiral) superspace \cite{Sohnius:1978wk}.

\subsection{\texorpdfstring{$\superN=4$}{N=4} Superspace}

Superspace has coordinates $z^A = (x^\mu,
\theta^{a \alpha}, \bar{\theta}^{\dot{\alpha}}_a)$.  Here $\alpha,
\dot{\alpha}$ are Lorentz indices, and $a$ are flavor symmetry indices
ranging from $1$ to $4$ and transforming in the $\mathbf{4}$ or
$\bar{\mathbf{4}}$ representations of $\grp{SU(4)}$.

The supersymmetry transformations are
\begin{equation}
  \label{eq:susy-generators}
  \gen{Q}_{a \alpha} = \frac {\partial}{\partial \theta^{a \alpha}}
- i \bar{\theta}_a^{\dot{\alpha}} \sigma^\mu_{\alpha \dot{\alpha}} \frac {\partial}{\partial x^\mu}\,, \quad
  \bar{\gen{Q}}_{\dot{\alpha}}^a = -\frac {\partial}{\partial \bar{\theta}^{\dot{\alpha}}_a}
+ i \theta^{a \alpha} \sigma^\mu_{\alpha \dot{\alpha}} \frac {\partial}{\partial x^\mu}\,,
\end{equation}
Under a supersymmetry transformation $\zeta^{a \alpha} \gen{Q}_{a \alpha} + \gen{\bar{Q}}^a_{\dot{\alpha}} \bar{\zeta}_a^{\dot{\alpha}}$, the superspace
coordinates transform like
\begin{gather}
  \delta x^\mu = i \left(\theta^{a \alpha} \sigma_{\alpha \dot{\alpha}}^\mu \bar{\zeta}_a^{\dot{\alpha}} - \zeta^{a \alpha} \sigma_{\alpha \dot{\alpha}}^\mu \bar{\theta}_a^{\dot{\alpha}} \right),\\
  \delta \theta^{a \alpha} = \zeta^{a \alpha}, \quad \delta \bar{\theta}_a^{\dot{\alpha}} = \bar{\zeta}_a^{\dot{\alpha}}.
\end{gather}
The supersymmetry covariant derivatives are
\begin{equation}
  \label{eq:susy-covariant-derivatives}
  D_{a \alpha} = \frac {\partial}{\partial \theta^{a \alpha}} + i \bar{\theta}_a^{\dot{\alpha}} \sigma^\mu_{\alpha \dot{\alpha}} \frac {\partial}{\partial x^\mu}\,, \quad
  \bar{D}_{\dot{\alpha}}^a = -\frac {\partial}{\partial \bar{\theta}^{\dot{\alpha}}_a} - i \theta^{a \alpha} \sigma^\mu_{\alpha \dot{\alpha}} \frac {\partial}{\partial x^\mu}\,, \quad
  \partial_{\alpha \dot{\alpha}} = \sigma^\mu_{\alpha \dot{\alpha}} \frac {\partial}{\partial x^\mu}\,.
\end{equation}
These supersymmetry covariant derivatives form the following algebra
\begin{equation}
  \lbrace D_{a \alpha}, D_{b \beta}\rbrace = 0, \quad
  \lbrace \bar{D}_{\dot{\alpha}}^a, \bar{D}_{\dot{\beta}}^b\rbrace = 0, \quad
  \lbrace D_{a \alpha}, \bar{D}_{\dot{\alpha}}^b\rbrace = -2 i \delta_a^b \partial_{\alpha \dot{\alpha}}.
\end{equation}
These derivatives have the following behavior under hermitian conjugation
\begin{equation}
  \partial_{\alpha \dot{\alpha}}^\dagger = -\partial_{\alpha \dot{\alpha}}, \quad
  D_{a \alpha}^\dagger = \bar{D}^a_{\dot{\alpha}}.
\end{equation}

A naive interval $x_1 - x_2$ is invariant under translations, but not
under superspace translations.  A quantity which is
invariant under superspace translations is
\begin{equation}
  \label{eq:susy-interval}
  x_{j,k}^\mu \equiv x_k^\mu - x_j^\mu - i \theta_k^{a \alpha} \sigma_{\alpha \dot{\alpha}}^\mu \bar{\theta}_{j a}^{\dot{\alpha}} + i \theta_j^{a \alpha} \sigma_{\alpha \dot{\alpha}}^\mu \bar{\theta}_{k a}^{\dot{\alpha}}.
\end{equation}
We emphasize here that our notation $x_{j,k}$ does \emph{not} stand
for $x_k-x_j$.

It is usual to define chiral and antichiral combinations as $x^{\pm \mu}
= x^\mu \pm i \theta^{a \alpha} \sigma_{\alpha \dot{\alpha}}^\mu
\bar{\theta}^{\dot{\alpha}}_a$.  The chiral/antichiral combinations
satisfy $D_{a \alpha} x^{-\mu} = 0$, $\bar{D}_{\dot{\alpha}}^a x^{+
\mu} = 0$.
There are chiral and antichiral versions of the above superspace interval
defined simply by $x_{j,k}^{\pm} = x_k^\pm - x_j^\pm$.
We can also define a mixed-chiral interval
$x_{j,k}^{+- \mu} \equiv x_k^{- \mu} - x_j^{+ \mu} + 2 i \theta_j^{a
\alpha} \sigma_{\alpha \dot{\alpha}}^\mu \bar{\theta}_{k
a}^{\dot{\alpha}}$, which has the property that $\bar{D}_j
x_{j,k}^{+-} = 0$, $D_k x_{j,k}^{+-} = 0$.  Here we have schematically
denoted by $\bar{D}_j$ the antichiral derivative with respect to the
superspace coordinates $(x_j, \theta_j, \bar{\theta}_j)$ and by
$D_k$ the chiral derivative with respect to the superspace
coordinates $(x_k, \theta_k, \bar{\theta}_k)$.
The chiral-antichiral interval can also be written as
\begin{equation}\label{eq:mixedinterval}
  x_{j,k}^{+- \mu} = x_{j,k}^\mu + i \theta_{jk}^{a \alpha} \sigma_{\alpha \dot{\alpha}}^\mu \bar{\theta}_{jk a}^{\dot{\alpha}},
\end{equation}
where $\theta_{jk} = \theta_k - \theta_j$, $\bar{\theta}_{jk} = \bar{\theta}_k - \bar{\theta}_j$.
This writing makes it clear that the chiral-antichiral interval is invariant under superspace translations.

\subsection{Superspace Vielbein}

The supersymmetry covariant derivatives can be written more compactly as
\begin{equation}
  D_M = E_M^{\hphantom{M} A} \frac \partial {\partial z^A},
\end{equation} where $E_M^{\hphantom{M} A}$ is called the inverse supervielbein
\begin{equation}
  E_M^{\hphantom{M} A} = \bordermatrix{
  {}_M \backslash {}^A & \boxed{{}_\nu} & \boxed{{}_{b \beta}} & \boxed{{}^b_{\dot{\beta}}}\cr
  \boxed{{}_\mu} & \delta_\mu^\nu & 0 & 0\cr
  \boxed{{}_{a \alpha}} & i \bar{\theta}_a^{\dot{\alpha}} \sigma^\nu_{\alpha \dot{\alpha}} & \delta_a^b \delta_\alpha^\beta & 0\cr
  \boxed{{}^a_{\dot{\alpha}}} & -i \theta^{a \alpha} \sigma^\nu_{\alpha \dot{\alpha}} & 0 & -\delta_b^a \delta_{\dot{\alpha}}^{\dot{\beta}}}.
\end{equation}
The  supervielbein is
\begin{equation}
  E_A^{\hphantom{A} N} = \bordermatrix{
  {}_A \backslash {}^N & \boxed{{}_\rho} & \boxed{{}_{c \gamma}} & \boxed{{}^c_{\dot{\gamma}}}\cr
  \boxed{{}_\nu} & \delta_\nu^\rho & 0 & 0\cr
  \boxed{{}_{b \beta}} & -i \bar{\theta}_b^{\dot{\beta}} \sigma^\rho_{\beta \dot{\beta}} & \delta_b^c \delta_\beta^\gamma & 0\cr
  \boxed{{}^b_{\dot{\beta}}} & -i \theta^{b \beta} \sigma^\rho_{\beta \dot{\beta}} & 0 & -\delta_c^b \delta_{\dot{\beta}}^{\dot{\gamma}}}.
\end{equation}
Now we can define the supervielbein as a differential form by $E^M = d z^A E_A^{\hphantom{A} M}$.
In components, this reads
\begin{equation}
  E^\rho = d x^\rho - i d \theta^{a \alpha} \sigma^\rho_{\alpha \dot{\alpha}} \bar{\theta}_a^{\dot{\alpha}} - i d \bar{\theta}_a^{\dot{\alpha}} \sigma^\rho_{\alpha \dot{\alpha}} \theta^{a \alpha}, \quad
  E^{a \alpha} = d \theta^{a \alpha}, \quad
  E_a^{\dot{\alpha}} = - d \bar{\theta}_a^{\dot{\alpha}}.
\end{equation}
Our conventions for differential calculus with Grassmann numbers
are such that $d z^M \wedge d z^N = - (-)^{M N} d z^N \wedge d z^M$,
where $(-)^{M N}$ is the product of gradings of $z^M$ and $z^N$.
Therefore, $d \theta \wedge d \bar{\theta} = d \bar{\theta} \wedge d \theta$
and $d \theta \wedge d x = -d x \wedge d \theta$.

Putting together the covariant derivatives and the supervielbein,
there are two alternative forms for the exterior derivative
\begin{align}\label{eq:exterior}
  d &= E^\rho \partial_\rho + E^{a \alpha} D_{a \alpha} + E^{\dot{\alpha}}_a \bar{D}_{\dot{\alpha}}^a\\
    &= d x^\rho \frac{\partial}{\partial x^\rho} + d \theta^{a \alpha} \frac \partial {\partial \theta^{a \alpha}} + d \bar{\theta}_a^{\dot{\alpha}} \frac \partial {\partial \bar{\theta}_a^{\dot{\alpha}}}.
\end{align}

Finally, note that the supervielbein has the following torsion components
\begin{equation}\label{eq:torsion}
   d E^\rho = 2 i \sigma^\rho_{\alpha \dot{\alpha}} E^{a \alpha} \wedge E^{\dot{\alpha}}_a, \quad
  d E^{a \alpha} = 0, \quad
  d E^{\dot{\alpha}}_a = 0.
\end{equation}

\subsection{Superspace Connection}

We introduce a gauge connection one-form $A$ on superspace.
It is conveniently expanded in a basis of the supervielbein
\begin{equation}
  A = E^\rho A_\rho + E^{a \alpha} A_{a \alpha} + E_a^{\dot{\alpha}} \bar{A}_{\dot{\alpha}}^a.
\end{equation}
The components $A_{\alpha \dot{\alpha}}(x, \theta,
\bar{\theta})$, $A_{a \alpha}(x, \theta, \bar{\theta})$,
$\bar{A}_{\dot{\alpha}}^a(x, \theta, \bar{\theta})$
are used to define gauge and supersymmetry covariant derivatives,
as follows
\begin{equation}
  \nabla_{a \alpha} \bullet = D_{a \alpha} \bullet + [A_{a \alpha}, \bullet\rbrace, \quad
  \bar{\nabla}_{\dot{\alpha}}^a \bullet = \bar{D}_{\dot{\alpha}}^a \bullet +
[\bar{A}_{\dot{\alpha}}^a, \bullet\rbrace, \quad
  \nabla_{\alpha \dot{\alpha}} \bullet = \partial_{\alpha \dot{\alpha}} \bullet + \left[A_{\alpha \dot{\alpha}}, \bullet\right].
\end{equation}
We take the gauge connection to be antihermitian, $A=-A^\dagger$, and the components satisfy the following
reality conditions
\[
(A_{\alpha\dot\beta})^\dagger=-A_{\beta\dot\alpha},
\quad
(A_{a\alpha})^\dagger=\bar A^a_{\dot\alpha}.
\]

The gauge potentials have infinitesimal gauge transformations given by
\begin{equation}
  \delta A_{a \alpha} = \nabla_{a \alpha} \Lambda, \quad
  \delta \bar{A}_{\dot{\alpha}}^a = \bar{\nabla}_{\dot{\alpha}}^a \Lambda, \quad
  \delta A_{\alpha \dot{\alpha}} = \nabla_{\alpha \dot{\alpha}} \Lambda,
\end{equation} where $\Lambda(x, \theta, \bar{\theta})$ is some antihermitian superfield ($\Lambda^\dagger = -\Lambda$).

Starting with the gauge connections, one can define gauge covariant field strengths as the components of $F = d A + A \wedge A$ in the expansion in terms of the vielbeins $E^{\dot{\alpha} \alpha}$, $E^{a \alpha}$, $E_a^{\dot{\alpha}}$.
 We find
\begin{align}
  F =&
  \mathrel{\phantom{+}}E^{a \alpha} \wedge E_b^{\dot{\alpha}} \bigbrk{2 i \delta_a^b A_{\alpha \dot{\alpha}} + \bar{D}_{\dot{\alpha}}^b A_{a \alpha} + D_{a \alpha} \bar{A}_{\dot{\alpha}}^b + \lbrace A_{a \alpha},
  \bar{A}_{\dot{\alpha}}^b\rbrace}\nln
  &+\tfrac 1 2 E^{a \alpha} \wedge E^{b \beta} \bigbrk{D_{b \beta} A_{a \alpha} + D_{a \alpha} A_{b \beta} + \lbrace A_{a \alpha}, A_{b
  \beta}\rbrace}\nln
  &+\tfrac 1 2 E^{\dot{\alpha}}_a \wedge E^{\dot{\beta}}_b \bigbrk{\bar{D}_{\dot{\beta}}^b \bar{A}_{\dot{\alpha}}^a + \bar{D}_{\dot{\alpha}}^a \bar{A}_{\dot{\beta}}^b + \lbrace\bar{A}_{\dot{\alpha}}^a,
  \bar{A}_{\dot{\beta}}^b\rbrace}\nln
  &+E^{\dot{\alpha} \alpha} \wedge E^{b \beta} \bigbrk{D_{b \beta} A_{\alpha \dot{\alpha}} - \partial_{\alpha \dot{\alpha}} A_{b \beta} + [  A_{b \beta}, A_{\alpha
  \dot{\alpha}}]}\nln
  &+E^{\dot{\alpha} \alpha} \wedge E_b^{\dot{\beta}} \bigbrk{\bar{D}^b_{\dot{\beta}} A_{\alpha \dot{\alpha}} - \partial_{\alpha \dot{\alpha}} \bar{A}_{\dot{\beta}}^b + [\bar{A}_{\dot{\beta}}^b, A_{\alpha
  \dot{\alpha}}]}\nln
  &+\tfrac 1 2 E^{\dot{\alpha} \alpha} \wedge E^{\dot{\beta} \beta} \bigbrk{\partial_{\beta \dot{\beta}} A_{\alpha \dot{\alpha}} - \partial_{\alpha \dot{\alpha}} A_{\beta \dot{\beta}} + [A_{\beta \dot{\beta}}, A_{\alpha \dot{\alpha}}]}.
\end{align}

When expanded in components, the gauge connections defined above contain too many fields to match the degrees of freedom in $\mathcal{N}=4$
super Yang--Mills.  Said differently, these superfields form reducible representations of the supersymmetry algebra and we will have to impose
constraints on them in order to obtain irreducible representations. The constraints are imposed by demanding that certain components of the
field strength $F$ vanish
(see \cite{Sohnius:1978wk} as well as ref.~\cite[Chap.~12]{Galperin:2001uw} for
a textbook treatment of the $\mathcal{N}=3$ extended supersymmetry)
\begin{subequations}\label{eq:neq4-constraints}
\begin{align}
  \label{eq:neq4-constraints-a}
  D_{a \alpha} A_{b \beta} + D_{b \beta} A_{a \alpha} + \lbrace  A_{a \alpha}, A_{b \beta}\rbrace &= \varepsilon_{\alpha \beta} \bar{W}_{a b},\\
  \label{eq:neq4-constraints-b}
  \bar{D}_{\dot{\alpha}}^a \bar{A}_{\dot{\beta}}^b + \bar{D}_{\dot{\beta}}^b \bar{A}_{\dot{\alpha}}^a + \lbrace \bar{A}_{\dot{\alpha}}^a, \bar{A}_{\dot{\beta}}^b\rbrace &= \varepsilon_{\dot{\alpha} \dot{\beta}} W^{a b},\\
  \label{eq:neq4-constraints-c}
  2 i \delta_a^b A_{\alpha \dot{\alpha}}+D_{a \alpha} \bar{A}_{\dot{\alpha}}^b + \bar{D}_{\dot{\alpha}}^b A_{a \alpha} + \lbrace A_{a \alpha}, \bar{A}_{\dot{\alpha}}^b \rbrace &= 0.
\end{align}
\end{subequations}
These are at the same time definitions for the scalar superfields $W^{a b}$ and $\bar{W}_{a b}$
and constraints for the gauge connections.  For example,
the first constraint in eq.~\eqref{eq:neq4-constraints-a}
means that the left-hand side transforms as a singlet under
Lorentz transformations and as a $\mathbf{6}$ under $SU(4)$
flavor transformations.
In other words, $\bar{W}_{a b}$ is a rank two antisymmetric tensor.
It obeys the hermiticity condition $\bar{W}_{ab}=(W^{a b})^\dagger$.

Let us note here a crucial difference to $\mathcal{N}=1$ superfields.
In that case, the first two constraints in eq.~\eqref{eq:neq4-constraints-a,eq:neq4-constraints-b}
have a trivial right-hand side.
This allows to solve the constraints in this case.

The superfields $W^{a b}$ and $\bar{W}_{a b}$ are very natural superfields.  They have mass dimension one and their flavor symmetry transformations
are such that their bottom component in the $\theta$, $\bar{\theta}$ expansion
are the scalars fields $\phi_{a b}$ in the $\mathcal{N}=4$ supermultiplet.
More precisely, the scalars $\phi_{a b}$ are the bottom component in the $\bar{W}_{a b}$ multiplet
while the conjugate scalars $\phi^{a b} = \left(\phi_{a b}\right)^\dagger$
are part of the $W^{a b}$ multiplet.
The higher components contain the fermions $\psi_\alpha^a$,
$\bar{\psi}_{a \dot{\alpha}}=(\psi_\alpha^a)^\dagger$ and the field strength
$F_{\alpha \beta}$ and $\bar{F}_{\dot{\alpha} \dot{\beta}}=(F_{\alpha \beta})^\dagger$.

The scalar fields in $\mathcal{N}=4$ super Yang--Mills satisfy a reality condition
$\phi^{a b} = \left(\phi_{a b}\right)^\dagger = \sfrac 1 2 \varepsilon^{a b c d} \phi_{c d}$.
 The superfields themselves are related by a similar relation%
\footnote{This relation permits the insertion of a complex phase
which has no impact on physical quantities.}
\begin{equation}
  \label{eq:scalarconjugate}
  W^{a b} = \sfrac 1 2 \varepsilon^{a b c d} \bar{W}_{c d}.
\end{equation}  The constraint on the superfield imposes proper reality constraints
on the members $\psi$, $\bar{\psi}$, $F$ and $\bar{F}$ of the multiplet.

\section{Gauge Field Propagator}
\label{sec:gaugefieldprop}

In this section we derive a two-point function for the gauge fields
of $\superN=4$ SYM in superspace.
This is the relevant object for the one-loop contribution to a
Wilson loop expectation value.
Quantization of gauge fields in extended superspace is troublesome
due to the constraints, and we start by sketching our procedure and results
in terms of a simple example.
Subsequently we will lift the results to $\superN=4$ SYM.

\subsection{Sketch for a Scalar Field}
\label{sec:vevs}

The first problem we have to face is that the constraints for the gauge field in superspace force it on shell. A standard Feynman propagator
takes the form $1/(p^2-i\epsilon)$, which clearly is ill-defined when $p^2=0$. Nevertheless, there exists a well-defined on-shell propagator
which we can use for the calculation of the Wilson loop expectation value. This is the vacuum expectation value (VEV) of two fields in canonical
QFT,
\[
\label{eq:vevphiphi}
\Delta(x-y)=\costate{0}\phi(x)\phi(y)\state{0}.
\]
Here, we explicitly mean the VEV \emph{without time-ordering}.
This is not the same as the expectation value in a path integral
which equals the \emph{time-ordered} VEV
\[
\label{eq:feynmanphiphi}
i \Delta\indup{F}(x-y)=\costate{0}\mathrm{T}[\phi(x)\phi(y)]\state{0}
=\vev{\phi(x)\phi(y)}.
\]
There is no obvious formulation for the VEV without time-ordering
in the path integral formalism,
and thus we have to use the language of quantized fields.

Consider a real scalar field $\phi(x)$
and the Klein--Gordon equation $\partial^2\phi+m^2\phi=0$
with mass $m$.
The standard mode expansion for the field equation reads
\[
\phi(\vec{x},t)=\int \frac{d^3\vec p}{(2 \pi)^3 2 E(\vec p)}\bigbrk{
\exp(i\vec p\cdot\vec x+iE(\vec{p})t)\,a^\dagger(\vec p)
+\exp(-i\vec p\cdot\vec x-iE(\vec{p})t)\,a(\vec p)
},
\]
with the energy $E(\vec{p})=+\sqrt{\vec{p}^2+m^2}$.
The canonical commutator of two modes equals their VEV (without time-ordering)
and reads
\[
\comm{a(\vec p)}{a^\dagger(\vec q)}=
(2 \pi)^3 \; 2E(\vec p)\,\delta^3(\vec p-\vec {q})=
\costate{0}a(\vec p)a^\dagger(\vec q)\state{0}.
\]
The resulting VEV of two fields in position space reads
\[
\Delta(\vec{x},t)=
\int \frac{d^3\vec p}{(2 \pi)^3 2E(\vec p)}
\exp\bigbrk{-i\vec p\cdot\vec x-iE(\vec{p})t}.
\]

All of the above relations are on-shell.
In the massless case there is a convenient and covariant
formulation in terms of unconstrained spinor variables $\lambda,\bar\lambda$.
The corresponding mode expansion now reads\footnote{The factor of $\tfrac 12$ in the exponent has its origin in the identity
$x \cdot y = x_\mu y^\mu = \tfrac 1 2 x_{\alpha \dot{\alpha}}
y^{\dot{\alpha} \alpha}$.  Also, we are using a shorthand notation for spinor index contraction, as detailed in \appref{app:conventions}.}
\[
\phi(x)=\frac 1 {8 \pi^2} \int d^2\lambda\, d^2\bar\lambda\,
\exp\bigbrk{-\sfrac{i}{2}\langle \lambda|x|\bar\lambda]}\,A(\lambda,\bar\lambda).
\]
The field $A(\lambda,\bar\lambda)$ contains both the positive
and negative energy modes $a(\vec{p})$ and $a^\dagger(\vec{p})$
for $\bar\lambda=\pm\lambda^\dagger$,
and the integral is also over positive and negative energies.
Furthermore, the field obeys the scaling
$A(z\lambda,z^{-1}\lambda)=A(\lambda,\bar\lambda)$.
The corresponding VEV reads
\[
\label{eq:vevAA}
\costate{0}A(\lambda,\bar\lambda)A(\lambda',\bar\lambda')\state{0}
=
\theta(E(\lambda,\bar\lambda))
\int \frac{dz}{2\pi i z}\,
\delta^2(\lambda'+z^{-1}\lambda)\,
\delta^2(\bar\lambda'-z\bar\lambda).
\]
Here $z=e^{i\alpha}$ is a pure complex phase.
Furthermore, $E(\lambda,\bar\lambda)$ refers to the energy described
by the pair of spinors $\lambda,\bar\lambda$.
It appears only as an argument to the step function $\theta$.
Consequently, only the sign of
$\bar\lambda=\pm \lambda^\dagger$ is relevant,
and thus the VEV remains manifestly Lorentz covariant.
The resulting VEV in position space reads
\[
\Delta(x)=
\frac 1 {64 \pi^4} \int_+ d^2\lambda\, d^2\bar\lambda\,
\exp\bigbrk{-\sfrac{i}{2}\langle \lambda|x|\bar\lambda]}.
\]
The $+$ subscript of the integral means we restrict the integration to positive energy
by means of a factor $\theta(E(\lambda,\bar\lambda))$.
It equals the above position-space two-point function for $m=0$.

The above Gaussian integral can be performed easily,
but proper attention should be paid to singular contributions in
the imaginary part
\[
\label{eq:2pt-function}
\Delta(x)=
-\frac 1 {4 \pi^2} \frac{1}{x^2 - i\sign(x^0)\epsilon}=
-\frac 1 {4 \pi^2} \left(\frac{1}{x^2} + i\pi \sign(x^0)\,\delta(x^2)\right).
\]
Note that this expression is not symmetric under $x\to-x$;
due to the non-commutativity of quantum fields this is not necessary.
The Feynman propagator is the time-ordering of the same expression
\[
\label{eq:feynmanposition}
\Delta\indup{F}(x)=
-i \left(\theta(x^0)\Delta(x)+
\theta(-x^0)\Delta(-x)\right)
=
\frac i {4 \pi^2} \frac{1}{x^2 - i\epsilon}=
\frac i {4 \pi^2} \left(\frac{1}{x^2} + i\pi \delta(x^2)\right).
\]
Curiously, the VEV differs
from the Feynman propagator only by a distributional amount
in position space.
This fact will become important for the Wilson loop calculation.
The situation in momentum space is quite different:
\begin{align}
\Delta(p)&=
-\frac 1 {4 \pi^2} \int d^4 x\,
\frac{\exp(ix\cdot p)}{x^2 - i\sign(x^0)\epsilon}
= 2 \pi \theta(p_0)\delta(p^2),
\nln
\Delta\indup{F}(p)&=
\frac i {4 \pi^2} \int d^4 x\,
\frac{\exp(ix\cdot p)}{x^2 - i\epsilon}
=\frac{1}{p^2 + i\epsilon}\,.
\end{align}
Here the VEV is defined on-shell while the Feynman propagator is
clearly off-shell.

Our strategy for $\superN=4$ SYM is to derive the VEV's
of gauge fields in the spinor formalism.
This can be done on shell while fully respecting the
superspace constraints.
The VEV's can be converted to position space,
from which Feynman propagators follow.
This will give us all the information needed to compute
a Wilson loop expectation value at one loop.

\subsection{Gauge Prepotentials}
\label{sec:gauge-prepotentials}

In the following we will compute the supersymmetric Wilson loop in full superspace to one loop order.
To this order, apart from a global color factor, there is no difference between the abelian and non-abelian theory.
Therefore it is sufficient to consider the linearized theory.

To solve the linearized version of the constraints
\eqref{eq:neq4-constraints}
we make an ansatz for the fermionic
components of the gauge field
(see also \cite{Siegel:2011sy})
\begin{equation}\label{eq:prepotentials}
  A_{a \alpha} = D_{a \beta} B^\beta_{\hphantom{\beta} \alpha}+D_{a\alpha}\Lambda, \quad
  \bar{A}_{\dot{\alpha}}^a = -\bar{D}_{\dot{\beta}}^a \bar{B}^{\dot{\beta}}_{\hphantom{\dot{\beta}} \dot{\alpha}}
+\bar{D}_{\dot\alpha}^a\Lambda,
\end{equation}
in terms of
a pair of chiral and antichiral prepotentials $B^{\alpha \beta}(x^+,\theta)$
and $\bar{B}^{\dot{\alpha} \dot{\beta}}(x^-, \bar{\theta})$
with symmetric indices
as well as an explicit gauge transformation $\Lambda(x,\theta,\bar\theta)$.
The prepotentials $B$ and $\bar{B}$ are hermitian conjugates, $B^\dagger=\bar B$,
while $\Lambda$ is antihermitian.

The constraint \eqref{eq:neq4-constraints-c} defines
the bosonic components of the gauge field
\[
  \label{eq:a-bosonic}
A_{\alpha \dot{\alpha}}=
\partial_{\beta\dot\alpha} B^\beta_{\hphantom{\beta} \alpha}
-\partial_{\alpha\dot\beta} \bar{B}^{\dot{\beta}}_{\hphantom{\dot{\beta}} \dot{\alpha}}
+\partial_{\alpha\dot\alpha}\Lambda.
\]
The constraints
\eqref{eq:neq4-constraints-a,eq:neq4-constraints-b}
imply that the prepotentials are chiral harmonic functions
\[
  \label{eq:constraintharm}
  D_{a}{}^\alpha D_{b \alpha} B_{\beta \gamma} = 0, \quad
  \bar{D}^{a\dot{\alpha}} \bar{D}_{\dot{\alpha}}^b \bar{B}_{\dot{\beta} \dot{\gamma}} = 0.
\]
Applying further fermionic derivatives to these equations
shows that $B$ and $\bar B$ also obey the massless wave equation.
Finally, together with \eqref{eq:scalarconjugate} the constraints
\eqref{eq:neq4-constraints-a,eq:neq4-constraints-b} imply
a relationship between the two prepotentials
\[\label{eq:constraintpre}
  -\bar{D}_{\dot{\alpha}}^a \bar{D}_{\dot{\beta}}^b \bar{B}^{\dot{\alpha} \dot{\beta}}
= \half \varepsilon^{a b c d} D_{c \alpha} D_{d \beta} B^{\alpha \beta}.
\]

It is important to note that there is a redundancy in the definition of the prepotentials
\[\label{eq:redundant}
\delta B_{\alpha\beta} =
\partial_{\alpha\dot\beta} \Gamma^{\dot\beta}_{\beta}
+\partial_{\beta\dot\beta} \Gamma^{\dot\beta}_{\alpha},
\quad
\delta \bar B_{\dot\alpha\dot\beta} =
\partial_{\beta\dot\alpha} \bar \Gamma^{\beta}_{\dot\beta}
+\partial_{\beta\dot\beta} \bar \Gamma^{\beta}_{\dot\alpha},
\quad
\delta \Lambda=
-\partial_{\alpha\dot\alpha}
(\Gamma^{\alpha\dot\alpha}-\bar \Gamma^{\alpha\dot\alpha}),
\]
where $\Gamma_{\alpha\dot\alpha}$ is a chiral harmonic function and
$\bar \Gamma_{\alpha\dot\alpha}=(\Gamma_{\alpha\dot\alpha})^\dagger$ its hermitian conjugate.
This transformation leaves the gauge potentials $A$ and $\bar A$ invariant.

The prepotentials $B$ or $\bar{B}$ have an interesting analog in the case of bosonic Yang--Mills theory.
We refer to \appref{app:bos_reduction} for more details.

\subsection{On-Shell Momentum Space}

The prepotentials are harmonic functions on chiral superspace
and thus obey the massless wave equation.
They can be written as the Fourier transformation
\begin{align}
\label{eq:b-fields} B^{\alpha\beta}(x^+,\theta)&= \frac 1 {8 \pi^2}\int d^2 \lambda \,d^2\bar\lambda\,d^{0|4}\bar\eta\, \exp\left(\tfrac i
2\langle\lambda| x^+ |\bar\lambda]+\langle\lambda| \theta |\bar\eta]\right)\, C^{\alpha\beta}(\lambda,\bar\lambda,\bar\eta), \nln \bar
B^{\dot\alpha\dot\beta}(x^-,\bar\theta)&= \frac 1 {8 \pi^2} \int d^2 \lambda\,d^2\bar\lambda\,d^{0|4} \eta\, \exp\left(\tfrac i 2
\langle\lambda|x^- |\bar\lambda]-\langle\eta| \bar\theta |\bar\lambda]\right)\, \bar{C}^{\dot\alpha\dot\beta}(\lambda,\bar\lambda,\eta),
\end{align}
in terms of on-shell momentum space fields
$C^{\alpha \beta}(\lambda, \bar{\lambda}, \bar{\eta})$ and $\bar{C}^{\dot{\alpha} \dot{\beta}}(\lambda, \bar{\lambda}, \eta)$.
 We have used the shorthand notation $\langle \lambda \vert x
\vert \bar{\lambda}] = \lambda^\alpha x_{\alpha \dot{\alpha}}
\bar{\lambda}^{\dot{\alpha}}$, $\langle \lambda \vert \theta \vert
\bar{\eta}] = \lambda^\alpha \theta_\alpha^a \bar{\eta}_a$ and
$\langle \eta \vert \bar{\theta} \vert \bar{\lambda}] = \eta^a
\bar{\theta}_{a \dot{\alpha}} \bar{\lambda}^{\dot{\alpha}}$.
The on-shell Fourier transformation in eqs.\ \eqref{eq:b-fields}
includes states with both positive and negative energies
for $\bar\lambda=\pm\lambda^\dagger$.
Reality conditions imply the following conjugation property of the modes
\[
C^{\alpha\beta}(\lambda,\bar\lambda,\bar\eta)^\dagger
=\bar{C}^{\dot\alpha\dot\beta}(\lambda,-\bar\lambda,\eta).
\]

The harmonic constraints in \eqref{eq:constraintharm} are satisfied
because the two derivatives each pull a $\lambda$ which are
subsequently contracted to $\langle \lambda \lambda \rangle=0$.
The constraint \eqref{eq:constraintpre}
relates the two mode expansions
\[\label{eq:constraintpre2}
\bar\lambda_{\dot\alpha}\bar\lambda_{\dot\beta}\bar{C}^{\dot\alpha\dot\beta}(\lambda,\bar\lambda,\eta) = 4 \int d^{0|4}\bar\eta
\exp \left(\tfrac 1 2\eta\bar\eta\right)
\lambda_{\alpha}\lambda_{\beta}C^{\alpha\beta}(\lambda,\bar\lambda,\bar\eta).
\]

The above spinor integrals have the following scaling symmetry
\[
(\lambda, \bar{\lambda}, \eta, \bar{\eta}) \to
(z \lambda, z^{-1} \bar{\lambda}, z \eta, z^{-1} \bar{\eta}).
\]
Consequently, the fields $C$ and $\bar{C}$ have to obey the scaling property
\[\label{eq:scalingc}
C^{\alpha\beta}(z\lambda,z^{-1}\bar\lambda,z^{-1}\bar\eta)=z^{-4}
C^{\alpha\beta}(\lambda,\bar\lambda,\bar\eta), \quad
\bar{C}^{\dot\alpha\dot\beta}(z\lambda,z^{-1}\bar\lambda,z \eta)=
z^4 \bar{C}^{\dot\alpha\dot\beta}(\lambda,\bar\lambda,\eta).
\]
The reality conditions for spacetime with $(3,1)$ signature imply
that $z=e^{i\phi}$ is a pure complex phase.
Hence the compact scaling symmetry merely leads to a factor of $2\pi$
in the integral and does not need to be ``gauge fixed'' otherwise.

In terms of the fields $C^{\alpha \beta}$, $\bar{C}^{\dot{\alpha} \dot{\beta}}$, the redundancy of eq.~\eqref{eq:redundant} becomes
\begin{equation}\label{eq:redundancy}
  \delta C^{\alpha \beta} = i \left(\lambda^\alpha \bar{\lambda}^{\dot{\alpha}} \tilde{\Gamma}^\beta_{\dot{\alpha}} + \lambda^\beta \bar{\lambda}^{\dot{\alpha}} \tilde{\Gamma}^\alpha_{\dot{\alpha}}\right), \quad
  \delta \bar{C}^{\dot{\alpha} \dot{\beta}} = i \left(\bar{\lambda}^{\dot{\alpha}} \lambda^\alpha \tilde{\bar{\Gamma}}^{\dot{\beta}}_\alpha + \bar{\lambda}^{\dot{\beta}} \lambda^\alpha \tilde{\bar{\Gamma}}^{\dot{\alpha}}_\alpha\right),
\end{equation}
where $\tilde{\Gamma}$ and $\tilde{\bar{\Gamma}}$ are the Fourier transforms of $\Gamma$ and $\bar{\Gamma}$, respectively.
Note that the contractions of
$\tilde{\Gamma}$ and $\tilde{\bar{\Gamma}}$
with
$\bar\lambda$ and $\lambda$, respectively,
leave two redundant d.o.f.\ in $C$ and $\bar{C}$.
Effectively $C^{\alpha\beta}$ and $\bar{C}^{\dot\alpha\dot\beta}$ have only one physical component.

\subsection{Light-Cone Gauge}

In \eqref{eq:redundant} we have seen that the prepotential carries
some on-shell (chiral harmonic) redundant degrees of freedom.
To eliminate them we introduce a pair of reference spinors
$l^{\alpha},\bar{l}^{\dot\alpha}$
defining a null vector $l^\alpha \bar{l}^{\dot{\alpha}}$.
For a light-cone gauge we impose
that $l_\alpha B^{\alpha\beta}=0$,
$\bar{l}_{\dot{\alpha}} \bar{B}^{\dot{\alpha} \dot{\beta}} = 0$.
These conditions are solved by
\[
\label{eq:scalarC}
C^{\alpha\beta}(\lambda,\bar\lambda,\bar\eta)=
\frac{l^{\alpha}l^{\beta}}{\langle \lambda l \rangle^2}\,
C(\lambda,\bar\lambda,\bar{\eta}), \quad
\bar{C}^{\dot\alpha\dot\beta}(\lambda,\bar\lambda,\eta)= \frac{\bar
l^{\dot\alpha}\bar l^{\dot\beta}}{[ \bar l \bar\lambda ]^2}\,
\bar{C}(\lambda,\bar\lambda,\eta),
\]
where $C$ and $\bar{C}$ are on-shell physical modes.
The scaling property \eqref{eq:scalingc} translates to
\[
\label{eq:scalingC}
C(z\lambda,z^{-1}\bar\lambda,z^{-1}\bar\eta)=
z^{-2} C(\lambda,\bar\lambda,\bar\eta), \quad
\bar{C}(z\lambda,z^{-1}\bar\lambda,z \eta)=
z^2 \bar{C}(\lambda,\bar\lambda,\eta).
\]
Furthermore, they are related by the constraint \eqref{eq:constraintpre2}
\[\label{eq:fermfourier}
\bar{C}(\lambda,\bar\lambda,\eta) = 4 \int d^{0|4}\bar\eta
\exp \left(\tfrac 1 2\eta\bar\eta\right) C(\lambda,\bar\lambda,\bar\eta).\]
The fields $\bar{C}$ and $C$ are also related by complex conjugation.
As a consequence of~\eqref{eq:b-fields} and $B^\dagger = \bar{B}$ have
\begin{equation}
  C^\dagger(\lambda,\bar{\lambda},\bar{\eta}) =
  \bar{C}(-\lambda,\bar{\lambda},-\eta) =
  \bar{C}(\lambda,-\bar{\lambda},\eta),
\end{equation} where the last equality follows from the scaling symmetry in~\eqref{eq:scalingC}.

It is physically evident that this mode expansion is complete because for every light-like momentum given in terms of $\lambda,\bar\lambda$, the
expansion of $C$ in terms of holomorphic $\eta\in\Complex^{0|4}$ yields the desired $16$ on-shell states of $\superN=4$ SYM\@. The conjugate field
$\bar{C}$ is fully determined by $C$ and does not carry additional degrees of freedom.

Note that the $l$-dependence in the above expressions is merely a gauge artifact.
The variation of $C$ w.r.t.\ the spinor $l^\alpha$ reads
\begin{equation}
  \delta C^{\alpha \beta} = \frac {\langle l \delta l\rangle}{\langle
l \lambda\rangle^3}\, (\lambda^\alpha l^\beta + \lambda^\beta l^\alpha) C.
\end{equation}
where we decomposed $\delta l$ on the basis $l$, $\lambda$.
This corresponds to the redundancy of the gauge fields
specified by \eqref{eq:redundancy} with
\begin{equation}
  \tilde{\Gamma}^{\alpha \dot{\alpha}} =  \frac {i l^\alpha
\bar{l}^{\dot{\alpha}}\,\langle l \delta
l\rangle}{[\bar{\lambda} \bar{l}] \langle l \lambda\rangle^3}\,C.
\end{equation}
The answer for $\tilde{\bar{\Gamma}}$ can be obtained by complex conjugation.
Another way to see that the $l$-dependence is gauge is to compute the quantities $\bar{W}_{ab} = -D_{a \alpha} D_{b \beta} B^{\alpha \beta}$ and
similarly for $W^{ab}$ and notice that they are independent of $l$ and also invariant with respect to the linearized gauge transformations. For
example,
\begin{equation}
  \bar{W}_{a b} = -\frac 1 {8 \pi^2} \int d^2 \lambda
\,d^2\bar\lambda\,d^{0|4}\bar\eta\, (\bar{\eta}_a + 2 \bar{\lambda}_{\dot{\alpha}} \bar{\theta}^{\dot{\alpha}}_a) (\bar{\eta}_b + 2 \bar{\lambda}_{\dot{\beta}} \bar{\theta}^{\dot{\beta}}_b) \exp \left(\tfrac i 2 \langle \lambda \vert x^+ \vert \bar{\lambda}] + \langle \lambda \vert \theta \vert \bar{\eta}]\right) C(\lambda, \bar{\lambda}, \bar{\eta}).
\end{equation}

The description of the on-shell states in $\mathcal{N}=4$ in terms
of the superfield $C$ (or $\bar{C}$) should be related to the light-cone description
by Mandelstam~\cite{Mandelstam:1982cb} and by Brink et al.\ \cite{Brink:1982pd}.
If we set $B$ or $\bar{B}$ to zero (thus breaking the reality condition relating $B$ to $\bar{B}$), we obtain an (anti-)selfdual theory.  Actions for this theory with $\mathcal{N}=4$ supersymmetry have been found in refs.~\cite{Parkes:1992rz, Siegel:1992za}.

\subsection{Quantization}

Conventionally the quantization of a theory
starts with the derivation of the propagator from
the kinetic terms in the action.
Unfortunately, it is far from trivial to write down an
action for extended supersymmetric Yang--Mills theory,
at least if supersymmetry is to be manifest.
Nevertheless we can construct a supersymmetric propagator,
and show that it agrees with our expectations.

The major problem we have to face is that the linearized constraints
for the gauge field force it on shell.
Consequently, we have expressed the solution
to the constraints through momentum space superfields $C$ and $\bar{C}$
which are manifestly on shell.
We now lift the VEV's discussed in \secref{sec:vevs}
to the gauge prepotentials of $\superN=4$ SYM.

The Grassmann components of the fields $C$, $\bar{C}$ contain precisely the
physical fields of $\superN=4$ SYM in light-cone gauge.
Hence, we could use their VEV's to define the
VEV's for the superfields $C$ and $\bar{C}$.
This is tedious, and instead we use
a number of constraints that the VEV
$\costate{0} CC\state{0}$
must satisfy.
It has to satisfy momentum
conservation $\lambda \bar{\lambda} + \lambda' \bar{\lambda}' = 0$.
Moreover, it has to conserve the
supersymmetric analog of momentum,
$\lambda \bar{\eta} + \lambda'\bar{\eta}' = 0$.
It has to have the right transformation under
$(\lambda,\bar{\lambda}, \bar{\eta}) \to (z \lambda, z^{-1} \bar{\lambda}, z^{-1} \bar{\eta})$.
Finally, it has to have right mass dimension.
A suitable expression, analogous to eq.~\eqref{eq:vevAA},
which satisfies all the constraints is
\begin{equation}
  \label{eq:cc-propagator}
\costate{0} C(\lambda,\bar\lambda,\bar\eta)\,C(\lambda',\bar\lambda',\bar\eta')\state{0} = \theta(E(\lambda,\bar\lambda)) \int \frac{d z}{2 \pi i z^3}\,
\delta^2(\lambda'+z^{-1}\lambda)\,
\delta^2(\bar\lambda'-z\bar\lambda)\,
\delta^{0|4}(\bar\eta'-z\bar\eta).
\end{equation}
Here the integral is over a pure complex phase $z=e^{i\phi}$.

By \eqref{eq:fermfourier}, the VEV of a $C$ field and a $\bar{C}$ field is
\[\costate{0}  C(\lambda,\bar\lambda,\bar\eta) \bar{C}(\lambda',\bar\lambda',\eta')\state{0}
= 4 \theta(E(\lambda,\bar\lambda)) \int \frac{dz}{2 \pi i z^3}\,
\delta^2(\lambda'+z^{-1}\lambda)
\delta^2(\bar\lambda'-z\bar\lambda)
\exp(\tfrac 1 2 z\eta'\bar\eta).\]
By using \eqref{eq:fermfourier} again, we find that the VEV of two $\bar{C}$'s is,
as expected, similar to that of two $C$'s
\begin{equation}
 \costate{0}  \bar{C}(\lambda, \bar{\lambda}, \eta) \bar{C}(\lambda', \bar{\lambda}', \eta')\state{0}
= \theta(E(\lambda,\bar\lambda)) \int \frac{z\,d z}{2 \pi i} \,
\delta^2(\lambda'+z^{-1}\lambda )
\delta^2(\bar{\lambda}' - z \bar{\lambda})
\delta^{0 \vert 4}(\eta'+z^{-1}\eta),
\end{equation}
which is a consistency check for our
$\costate{0} C C\state{0}$.

Let us test that this choice for the $\langle C C\rangle$ propagator yields the results we expect, by computing the scalar two-point functions.
To do this computation, notice that since $\phi_{a b}(x) = \bar{W}_{a b}(x, \theta = 0, \bar{\theta} = 0)$, we get
\begin{multline}
  \langle 0\vert \phi_{a b}(x) \phi_{c d}(x')\vert 0 \rangle = \frac 1 {64 \pi^4} \int d^2 \lambda
\,d^2\bar\lambda\,d^{0|4}\bar\eta d^2 \lambda'
\,d^2\bar{\lambda}'\,d^{0|4}\bar{\eta}' \; \bar{\eta}_a \bar{\eta}_b \bar{\eta}_c' \bar{\eta}_d' \\
\exp \left(\tfrac i 2 \langle \lambda \vert x \vert \bar{\lambda}] + \tfrac i 2 \langle \lambda' \vert x' \vert \bar{\lambda}']\right)
\langle 0\vert C(\lambda, \bar{\lambda}, \bar{\eta}) C(\lambda', \bar{\lambda}', \bar{\eta}')\vert 0\rangle.
\end{multline}
Using usual manipulations for the delta functions, we can show that this two-point function is
\begin{equation}
  \langle 0 \vert\phi_{a b}(x) \phi_{c d}(x')\vert 0\rangle = \frac {-1}{4 \pi^2} \frac {\varepsilon_{abcd}} {(x-x')^2} + \text{singular support},
\end{equation} which is the expected result (the missing distributional terms are given in eq.~\eqref{eq:2pt-function}).  This computation also allows us to fix the normalization of the $CC$ two-point function.

\section{Wilson Loop Expectation Value}
\label{sec:vertexcor}

We now turn to the calculation
of the one-loop expectation value of a
null polygonal Wilson loop
$\mathcal{W}[C]$
in full superspace%
\[\label{eq:WLexpval}
\frac {g^{2} N}{64 \pi^{2}} M^{(1)}[C]=\frac{1}{N} \oint_C \oint'_{C} \half\tvev{\Tr A A'}.
\]
In the following we will not explicitly write down the factors of $g^{2} N$, since they can easily be restored when needed.
Here $C$ is the contour of a null polygon in full $\superN=4$ superspace (see~\cite{Beisert:2012}),
and $A$ and $A'$ denote one copy of the gauge connection
for each of the two integrals.
At this perturbative level one needs only a two-point correlation function
which we obtained in the previous section.
Interaction vertices are not needed.
Furthermore, the color algebra can be performed to
reduce the computation to the abelian case.

\subsection{Chiral Decomposition}
\label{sec:chiraldecomp}

As a first step we write the gauge connection as a differential form on superspace
and substitute the prepotential ansatz discussed in \secref{sec:gauge-prepotentials}
\begin{align}
  A =&\,\tfrac 1 2 E^{\dot{\alpha} \alpha} A_{\alpha \dot{\alpha}} + E^{a \alpha} A_{a \alpha} + E^{\dot{\alpha}}_a
  \bar{A}_{\dot{\alpha}}^a\nln
  =&\,\tfrac 1 2 \left(d x^{\dot{\alpha} \alpha}
           - 2 i d \theta^{a \alpha} \bar{\theta}_a^{\dot{\alpha}}
            - 2 i d \bar{\theta}_a^{\dot{\alpha}} \theta^{a \alpha}\right)
   \bigbrk{-\partial_{\alpha \dot{\beta}} \bar{B}^{\dot{\beta}}_{\hphantom{\dot{\beta}} \dot{\alpha}}
         + \partial_{\beta \dot{\alpha}} B^\beta_{\hphantom{\beta} \alpha}
         + \partial_{\alpha\dot\alpha}\Lambda}
\nln &
  + d \theta^{a \alpha} \bigbrk{D_{a \beta} B^\beta_{\hphantom{\beta} \alpha} + D_{a\alpha}\Lambda}
  - d \bar{\theta}^{\dot{\alpha}}_a
    \bigbrk{-\bar{D}_{\dot{\beta}}^a \bar{B}^{\dot{\beta}}_{\hphantom{\dot{\beta}} \dot{\alpha}} +\bar D_{\dot\alpha}^a \Lambda}.
\end{align}
We use the relations%
\footnote{We use the notation $\partial^{\pm}_{\alpha\dot\alpha}\equiv \sigma^\mu_{\alpha\dot\alpha}\partial/\partial (x^\pm)^\mu$.}
\begin{align}
  d \left(x^\pm\right)^{\dot{\alpha} \alpha} &=
     d x^{\dot{\alpha} \alpha} \pm 2 i d \theta^{a \alpha} \bar{\theta}^{\dot{\alpha}}_a \mp 2 i d \bar{\theta}_a^{\dot{\alpha}} \theta^{a \alpha},\\
  D_{a \alpha} B^\alpha_{\hphantom{\alpha} \beta}(x^+, \theta) &=
     \left(\frac \partial {\partial \theta^{a \alpha}} + 2 i \bar{\theta}_a^{\dot{\alpha}} \partial^+_{\alpha \dot{\alpha}} \right) B^\alpha_{\hphantom{\alpha} \beta}(x^+, \theta),\\
  \bar{D}_{\dot{\alpha}}^a \bar{B}^{\dot{\alpha}}_{\hphantom{\dot{\alpha}} \dot{\beta}}(x^-, \bar{\theta}) &= \left(-\frac \partial {\partial \bar{\theta}_a^{\dot{\alpha}}} - 2 i \theta^{a \alpha} \partial^-_{\alpha \dot{\alpha}}\right) \bar{B}^{\dot{\alpha}}_{\hphantom{\dot{\alpha}} \dot{\beta}}(x^-, \bar{\theta}),
\end{align}
to simplify the connection
\begin{align}
  \label{eq:chiralconn}
A&=A^+ + A^- + d\Lambda,
\nln
  A^+ &= \frac 1 2 d \left(x^+\right)^{\dot{\alpha} \alpha} \partial^+_{\beta\dot{\alpha} } \,B^\beta_{\hphantom{\beta} \alpha} + d \theta^{a \alpha} \frac \partial {\partial \theta^{a \beta}}\, B^\beta_{\hphantom{\beta}   \alpha},\\
  A^- &= -\frac 1 2 d \left(x^-\right)^{\dot{\alpha} \alpha} \partial^-_{\alpha\dot{\beta}} \,\bar{B}^{\dot{\beta}}_{\hphantom{\dot{\beta}} \dot{\alpha}} - d \bar{\theta}^{\dot{\alpha}}_a \frac \partial {\partial \bar{\theta}^{\dot{\beta}}_a} \,\bar{B}^{\dot{\beta}}_{\hphantom{\dot{\beta}} \dot{\alpha}}.
\end{align}
So we see that the connection nicely splits into a connection on the chiral and antichiral part of the full superspace
and a non-chiral gauge transformation which has no impact on closed Wilson loops.

\begin{figure}\centering
$\displaystyle
M^{(1)}[C] =
\half\oint\oint
\left( \includegraphicsbox[height=3.0cm]{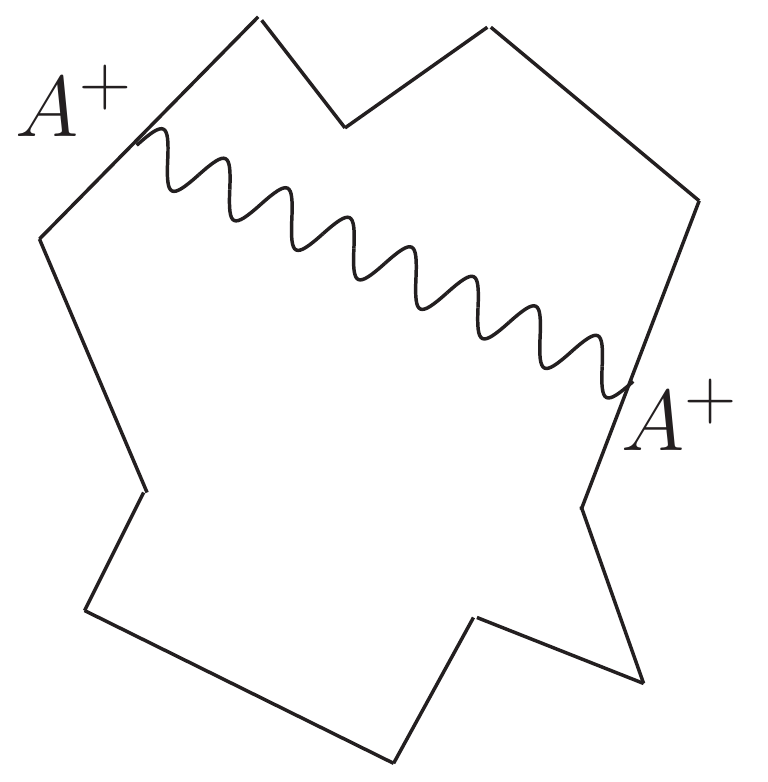}
      +\includegraphicsbox[height=3.0cm]{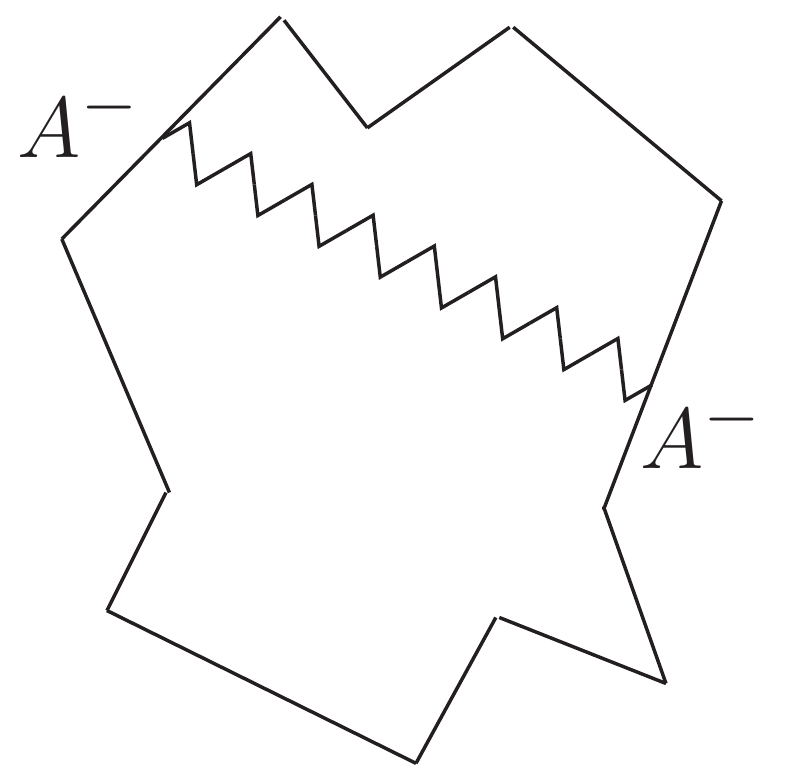}
     +2\includegraphicsbox[height=3.0cm]{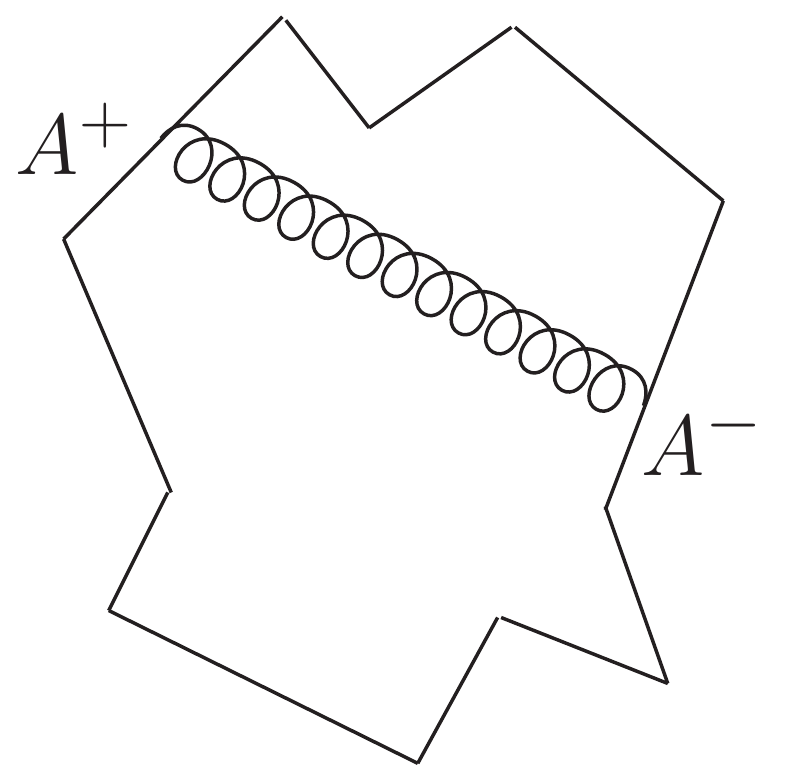}\right)
$ \caption{The one-loop expectation value of a Wilson loop, including chiral, antichiral and mixed-chirality contributions.} \label{fig:WLa}
\end{figure}

For the one-loop Wilson loop expectation value this implies three terms, as illustrated in \figref{fig:WLa},
\[
\label{eq:oneloop}
M^{(1)}[C]=
M^{(1)}_{++}[C]+M^{(1)}_{--}[C]+2M^{(1)}_{+-}[C],
\quad
\frac 1 {64 \pi^{2}} M^{(1)}_{\pm\pm'}[C]=
 \oint_C \oint'_{C} \half\tvev{A^\pm A'^{\pm'}}.
\]
The three types of contributions above have different forms which will not mix.
In particular, they can be distinguished
by a charge counting the number of $\theta$'s
minus the number of $\bar\theta$'s.

The former two terms in the above equation are fully chiral or
antichiral, respectively; they depend only on the projections of the Wilson loop
onto the chiral or antichiral subspaces of superspace. The chiral
part of the result, by construction, agrees with the expectation
value of the supersymmetric Wilson loop proposed
in~\cite{Mason:2010yk, arXiv:1010.1167}.
It is going to be a finite rational function. The antichiral part is
(almost\footnote{The imaginary part of the Feynman propagator causes
some subtle distributional discrepancy due to unitarity.}) the complex conjugate of
the chiral part. The latter term in the above equation is
mixed-chiral; it depends non-trivially on all superspace
coordinates. The bosonic truncation of this part, by construction,
agrees with the expectation value of a Wilson loop in ordinary
spacetime, see~\cite{arXiv:0707.1153, arXiv:0709.2368}.
It is going to be a divergent function of transcendentality two
($\Li_2$, $\log^2$).

\subsection{Use of Correlators}
\label{sec:onshellcorrelator}

Conventionally, Wilson loop expectation values $\tvev{W_C}$ are evaluated in
the path integral. In particular, the two-point correlation function
$\tvev{A A'}$ translates to a Feynman propagator $\Delta\indup{F}$.
The Feynman propagator almost obeys the equation of motion of the corresponding field.
Importantly however, the e.o.m.\ are violated at coincident points where
a delta distribution remains. A Feynman propagator is off-shell.
This is a mostly negligible effect in position space,
where Wilson loop expectation values are ordinarily computed.
For our supersymmetric Wilson loop it puts us in a slightly inconvenient position:
On the one hand, the gauge connection has to be constrained in such a way that
the equations of motion are implied. The fields must obey the equations of motion.
On the other hand, Feynman propagators are intrinsically off-shell.
More concretely,
the field $C$ defined in \eqref{eq:b-fields} exists only for $p^2=0$,
whereas the Feynman propagator is of the form $1/p^2$.
In the following we shall explain how to resolve the apparent clash.

First of all, there is nothing that prevents us from performing the calculation
in position space.
For illustration purposes we shall use the example of a scalar field $\phi$
instead of the full-fledged gauge connection on superspace.
We can compute the two-point correlator
of two fields~\eqref{eq:vevphiphi}
\[
\Delta(x-y)=\costate{0}\phi(x)\phi(y)\state{0}
=
-\frac {1}{4 \pi^2} \left(\frac{1}{x^2} + i\pi \sign(x^0)\,\delta(x^2)\right).
\]
The corresponding Feynman propagator~\eqref{eq:feynmanphiphi} can be derived from
the two-point correlator by a simple manipulation~\eqref{eq:feynmanposition}
\begin{align}
i \Delta\indup{F}(x-y)
&=
\costate{0}\mathrm{T}[\phi(x)\phi(y)]\state{0}
=\theta(x^0)\Delta(x)+
\theta(-x^0)\Delta(-x)
\nln
&
=
-\frac {1}{4 \pi^2} \left(\frac{1}{x^2} + i\pi \delta(x^2)\right).
\end{align}
This construction extends without further ado to superspace,
and can be applied to the calculation of the Wilson loop expectation value
$\tvev{W}$.

There is another option at our disposal:
If we blindly replace the Feynman propagator
$i\Delta\indup{F}$ by the two-point correlator $\Delta$
we actually compute $\costate{0}W\state{0}$ which is different from
$\tvev{W}$. The difference between the two is computed via the
difference
\[
i\Delta\indup{F}(x)-\Delta(x)=
-\frac {i}{2 \pi}\, \theta(-x^0)\,\delta(x^2).
\]
This difference is localized to the light cone, and it is purely imaginary.
It is similar to the cut discontinuity of the Feynman propagator
\[
\disc i\Delta\indup{F}(x)
=
-\frac {i}{2 \pi}\, \delta(x^2)
\]
which yields the cut discontinuity of the ordinary Wilson loop
expectation value $\tvev{W}$.
The latter is well-known to be a simpler function
(usually of one degree of transcendentality less).
One can convince oneself that the same applies
to the difference.%
\footnote{The result of $\costate{0}W\state{0}$
depends on the choice of operator ordering in $W$.
The totally symmetrized ordering actually yields
precisely $\disc i\Delta\indup{F}(x)$, hence
$\costate{0}W\state{0}=\Re\tvev{W}$ in this case.}

We will be satisfied with computing the most complicated part
(highest transcendentality)
of the Wilson loop expectation value $\tvev{W}$.
Consequently, we can instead compute $\costate{0}W\state{0}$
by replacing Feynman propagators $i\Delta\indup{F}$
by two-point correlators $\Delta$. Then all the correlators
are perfectly on-shell, and the constraints on the superspace
connection fully apply.
Alternatively, we could decide to compute the discontinuity
$\disc\tvev{W}$.
The cut of the Feynman propagator
$\disc i\Delta\indup{F}$ is another perfectly on-shell quantity.
Eventually $\tvev{W}$ is recovered from a dispersion integral
on $\disc\tvev{W}$.

A final option may be to Fourier transform the obtained
Feynman propagators from position space to momentum space.%
\footnote{Fourier transforms of full superspace are cumbersome
due to superspace torsion:
The fermionic momenta anticommute onto the bosonic momentum,
and momentum space would be non-commutative.
However, the prepotentials $B$ are chiral
and a flat chiral momentum space does exist.}
Here one would have to understand in how far the constraints
on the superspace connection apply and can be used for simplifications.

In this paper we shall mainly adopt the calculation of
$\costate{0}W\state{0}$.  We will then use the mode expansion~\eqref{eq:b-fields}  of the on-shell fields
in terms of the spinor-helicity field $C$.  This will allow us to shortcut the calculation substantially. For a position space calculation see \appref{app:position_spc}.

\subsection{Vertex Correlators}
\label{sec:vertexcorr}

The shape of the Wilson loop is a null polygon in superspace~\cite{Beisert:2012},
i.e.\ a sequence of points $(x_j,\theta_j,\bar\theta_j)$
which are joined by null lines.

For the null line that joins the vertices $j$ and $j+1$ we define
$\lambda_j$, $\bar{\lambda}_j$ by $x^{\dot\alpha\alpha}_{j,j+1} =
\lambda^\alpha_j \bar{\lambda}^{\dot\alpha}_j$, where $x_{j,j+1}$ is
the superspace interval as defined in eq.~\eqref{eq:susy-interval}.
The null line can then be parametrized as follows
\[\label{eq:nullline}
x^{\dot\alpha\alpha}=
x^{\dot\alpha\alpha}_j
+\tau \lambda_{j}^{\alpha}\bar\lambda_{j}^{\dot\alpha}
+2 i\lambda^{\alpha}_{j}\sigma^a \bar\theta_{j,a}^{\dot\alpha}
-2 i\theta_j^{\alpha a}\bar\sigma_a \bar\lambda_{j}^{\dot\alpha},
\quad
\theta^{a\alpha}=\theta_j^{a\alpha}+\lambda_{j}^\alpha\sigma^a,
\quad
\bar\theta^{\dot\alpha}_a=\bar\theta_{j,a}^{\dot\alpha}+\bar\sigma_a\bar\lambda_{j}^{\dot\alpha}.
\]
Here $\tau$ is a bosonic coordinate,
and $\sigma,\bar\sigma$ are 4 additional complex fermionic coordinates.
The null line is ``fat''; it is a $1|8$-dimensional subspace of superspace.
The Wilson line a $1|0$-dimensional curve on the null line.
The restrictions on the gauge field curvature
\eqref{eq:neq4-constraints} imply that the precise choice of curve
does not matter~\cite{Witten:1978xx,Witten:1985nt}.
A Wilson line only depends on the start and end points
$(x_{j},\theta_{j},\bar\theta_{j})$
and $(x_{j+1},\theta_{j+1},\bar\theta_{j+1})$.
We can thus pick any $\sigma(\tau),\bar\sigma(\tau)$
that interpolates between vertices $j$ and $j+1$.
This implies $\sigma(0) = 0$,
$\sigma(1) = \bar\eta_{i}$,
$\bar{\sigma}(0) = 0$,
$\bar{\sigma}(1) = \eta_{i}$.

Correspondingly, the gauge connection $A=A^++A^-+d\Lambda$ \eqref{eq:chiralconn} is
a total derivative
when restricted to the null line \eqref{eq:nullline}%
\footnote{Obviously, this is a classical statement which depends very much on
the classical equations of motion to hold.
In our case we can rely on the linearized classical e.o.m.\ because
the two-point correlator is perfectly on-shell.
(and even the Feynman propagator is on-shell except for coincident points
whose contributions are minute).}
\[
\label{eq:Aexact}
A^+_j=dG^+_j,
\quad
A^-_j=dG^-_j.
\]
Using the definition of $A^+$ and $A^-$ in terms of $B$ and $\bar{B}$ \eqref{eq:prepotentials},
the mode expansion~\eqref{eq:b-fields} and the light cone gauge condition \eqref{eq:scalarC}
a quick computation shows that $G^\pm$ has a solution in the closed form
\begin{align}
G^+_j(x^+,\theta)&= \frac 1 {8 \pi^2} \int d^2 \lambda
\,d^2\bar\lambda\,d^{0|4}\bar\eta\,
\exp\bigbrk{\sfrac i 2\langle\lambda| x^+ |\bar\lambda]+\langle\lambda| \theta |\bar\eta]}\,
\frac{\langle j l\rangle}{\langle \lambda l\rangle \langle \lambda j\rangle}\, C(\lambda, \bar{\lambda}, \bar{\eta}),
\\
G^-_j(x^-,\bar\theta)&= \frac 1 {8 \pi^2} \int d^2 \lambda
\,d^2\bar\lambda\,d^{0|4}\eta\,
\exp\bigbrk{\sfrac i 2\langle\lambda|x^- |\bar\lambda]-\langle\eta| \bar\theta |\bar\lambda]}\,
\frac{[j \bar{l}]}{[\bar{\lambda} \bar{l}] [\bar{\lambda} j]}
\,\bar{C}(\lambda, \bar{\lambda}, \eta).
\end{align}

\begin{figure}\centering
$\displaystyle
\includegraphicsbox[height=4.0cm]{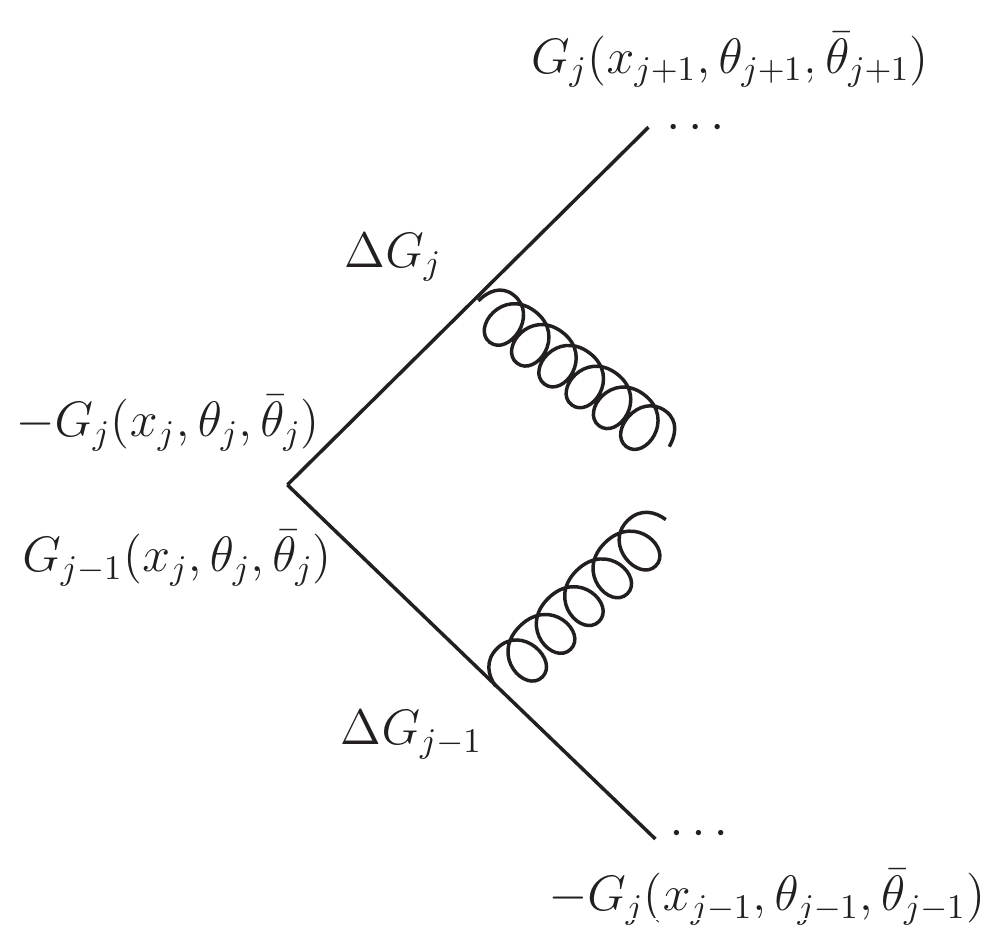}\quad
\Rightarrow \quad \includegraphicsbox[height=4.0cm]{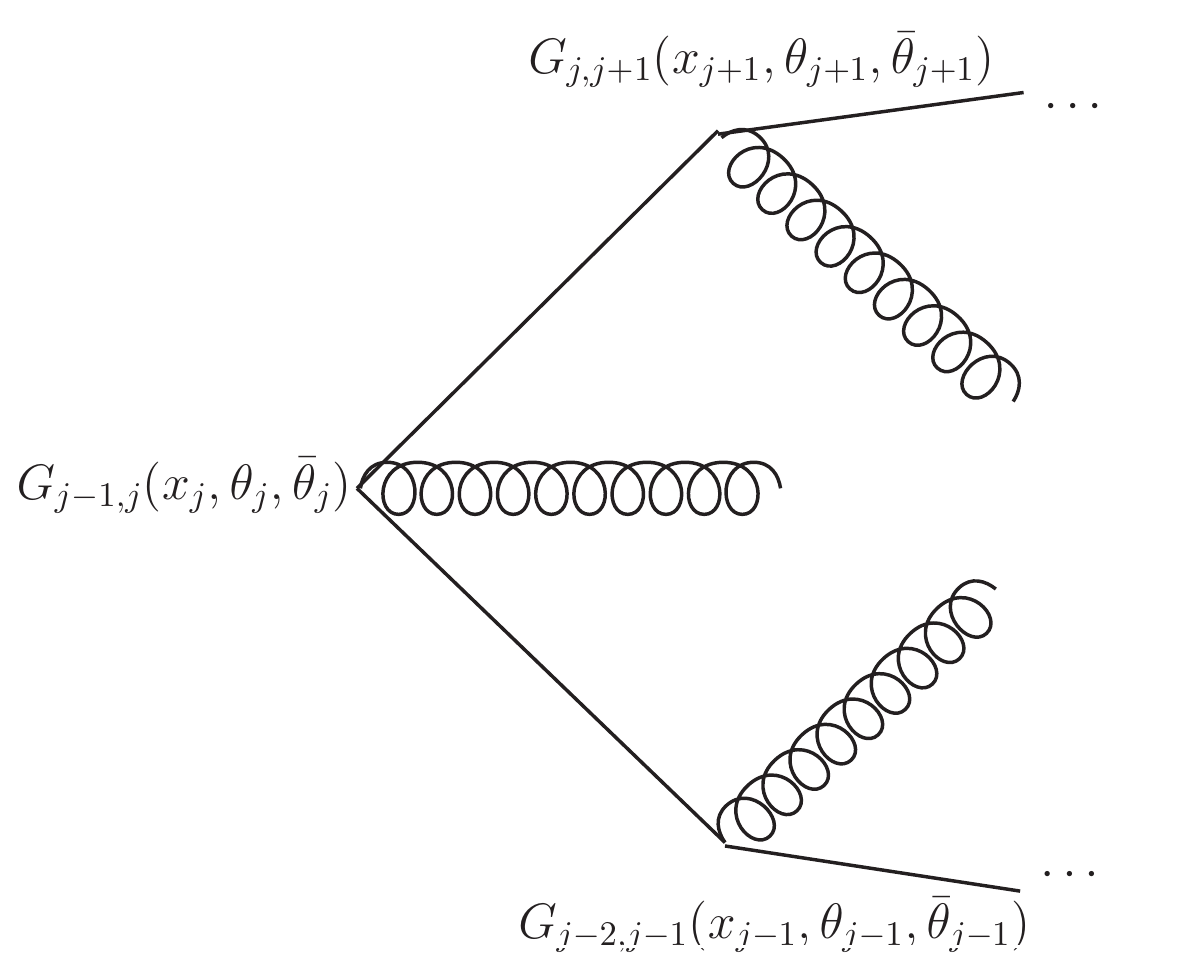} $ \caption{Rearrange the sum of potential shifts over the edges of the
polygon, to a sum over shifts at the vertices.} \label{fig:vertex}
\end{figure}
The Wilson loop integral can now be written
as a sum of potential shifts over the edges of the polygon
\[
\oint A=\sum_{j=1}^n \Delta G_j,
\quad
\Delta G_j=
G_j(x_{j+1},\theta_{j+1},\bar\theta_{j+1})
-G_j(x_{j},\theta_{j},\bar\theta_{j}).
\]
Now there is an interesting rearrangement of the sum
\[
\oint A=\sum_{j=1}^n G_{j-1,j},
\quad
G_{j-1,j}=
G_{j-1}(x_{j},\theta_{j},\bar\theta_{j})
-G_j(x_{j},\theta_{j},\bar\theta_{j}),
\]
which expresses the Wilson loop as a sum over potential
shifts at the vertices, see~\figref{fig:vertex}.
The latter read
\begin{align}
\label{eq:vertexshift}
G^+_{j-1,j}&= \frac 1 {8 \pi^2}\int d^2 \lambda
\,d^2\bar\lambda\,d^{0|4}\bar\eta\,
\exp\bigbrk{\sfrac i 2 \langle\lambda| x_j^+ |\bar\lambda]+\langle\lambda| \theta_j |\bar\eta]}\,
\frac{\langle j{-}1 j\rangle}
     {\langle j{-}1 \lambda\rangle \langle \lambda j\rangle}\, C(\lambda,\bar\lambda,\bar\eta),\nln
G^-_{j-1,j}&= \frac 1 {8 \pi^2}\int d^2 \lambda
\,d^2\bar\lambda\,d^{0|4}\eta\,
\exp\bigbrk{\sfrac i 2 \langle\lambda|x_j^- |\bar\lambda]-\langle\eta| \bar\theta_j |\bar\lambda]}\,
\frac{[j{-}1 j]} {[j{-}1 \bar\lambda] [\bar\lambda j]} \,\bar{C}(\lambda,\bar\lambda,\eta).
\end{align}
At first sight it may be surprising to see that the dependence on
the light cone gauge reference vector $l$ has dropped out from $G_{j-1,j}$. In
fact, the reason is simply that $G_{j-1,j}$ is localized at vertex
$j$, and changes of the gauge must cancel between the contributions
$G_{j-1}$ and $G_j$. It is evident that the early cancellation of
gauge artifacts will substantially simplify the subsequent
calculation.

\begin{figure}\centering
$\displaystyle
M^{(1)}[C] =\half\sum_{j,k}
\left(
\includegraphicsbox[height=3.0cm]{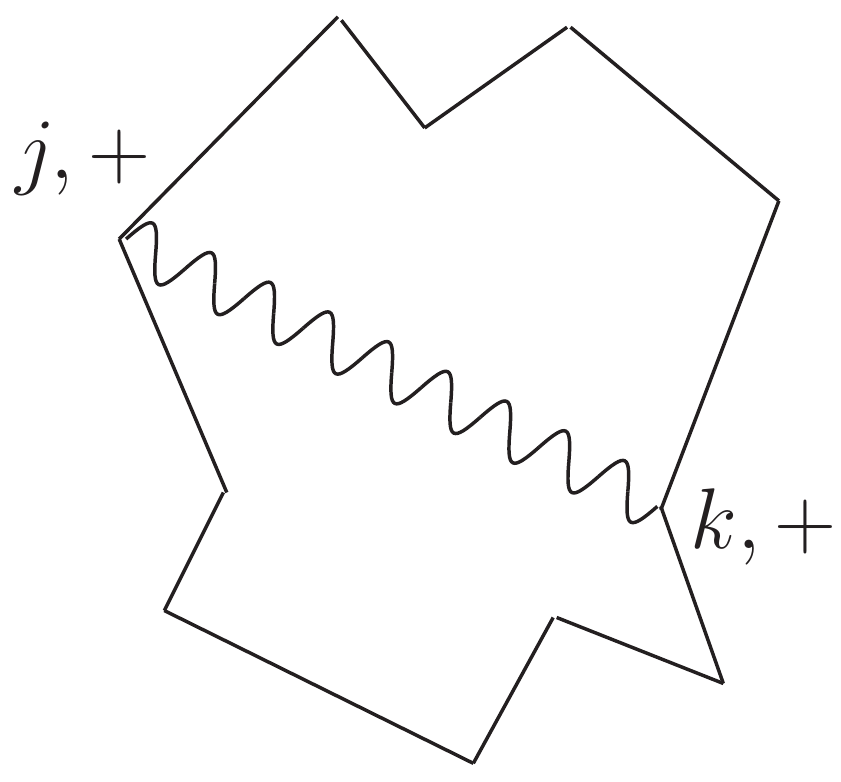}
+\includegraphicsbox[height=3.0cm]{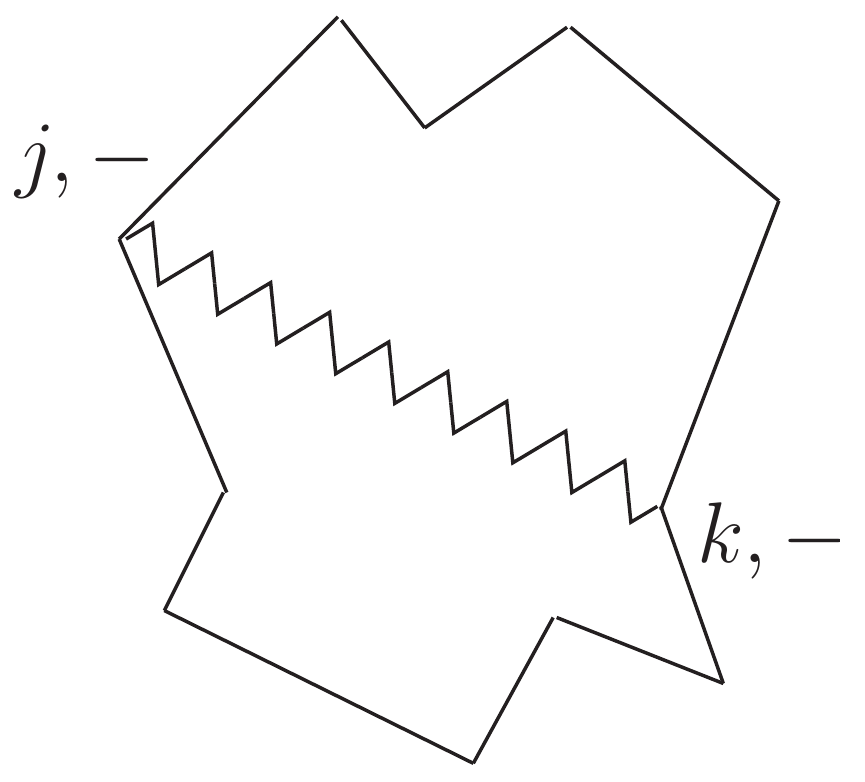} +2\includegraphicsbox[height=3.0cm]{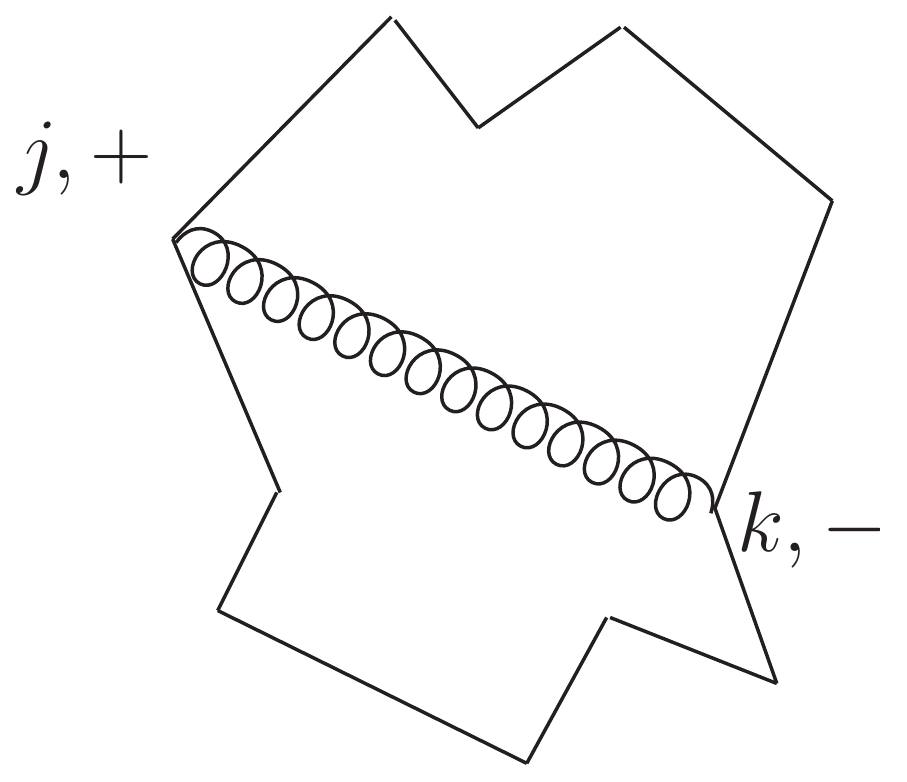}\right)$ \caption{The one-loop expectation value of a
Wilson loop, including sums of chiral, antichiral and mixed-chirality vertex correlators.} \label{fig:WLb}
\end{figure}

For the Wilson loop expectation value \eqref{eq:WLexpval}
we thus have two equivalent representations
\[\label{eq:WLedgevertex}
\frac 1 {64 \pi^{2}} M_n^{(1)}
=\oint \oint' \half\tvev{A A'}
=\sum_{j,k=1}^n \half\tvev{\Delta G_j\, \Delta G_k}
=\sum_{j,k=1}^n \half\tvev{G_{j-1,j}\, G_{k-1,k}}.
\]
The former uses a sum over edge correlators,
the latter a sum over vertex correlators;
the latter will be more convenient to use.

Note that along the lines of the discussion in \secref{sec:onshellcorrelator} we shall replace the expectation value in \eqref{eq:WLedgevertex}
by a vacuum expectation value. This allows us to perform the calculation using on-shell fields in the first place. Secondly, according to
\secref{sec:chiraldecomp}, the expectation values split into three terms of different chirality. The resulting one-loop expectation thus reads
(see \figref{fig:WLb})
\[\label{eq:WLvertex}
\frac 1 {64 \pi^{2}} M_n^{(1)}
=\sum_{j,k=1}^n
\lrbrk{
\half\costate{0}G_{j-1,j}^+\, G_{k-1,k}^+\state{0}
+\half\costate{0}G_{j-1,j}^-\, G_{k-1,k}^-\state{0}
+\costate{0}G_{j-1,j}^+\, G_{k-1,k}^-\state{0}
}.
\]
In the following we shall consider the chiral and the mixed chiral contributions by substituting the vertex gauge potential shifts $G_{j-1,j}$,
evaluating the correlators and performing the integrals.

Note that the vertex correlators also play an important role in twistor space calculations.
As we shall see in~\secref{sec:ambi}, in twistor
space, each vertex corresponds to an edge connecting two adjacent ambitwistors.

\subsection{Chiral Correlator}
\label{sec:chircorr}

In this section we compute the expectation values of the chiral-chiral (or equivalently antichiral-antichiral) vertex shifts of eq.~\eqref{eq:WLvertex}.  The remaining mixed chiral expectation values are computed in the next section.

Using the two-point function of two $C$ fields, we obtain the following result for the the two-point function of two $G^{+}$ fields
\begin{multline}
  \langle 0 \vert G_{j-1,j}^{+}\, G_{k-1,k}^{+}\vert 0 \rangle =  \frac 1 {64 \pi^{4}} \int_+ d^{2} \lambda\, d^{2} \bar{\lambda}\, d^{0\vert 4} \bar{\eta} \times
\\
 \exp \left(\frac i 2 \langle \lambda \vert x_{k,j}^{+} \vert \bar{\lambda}] + \langle \lambda \vert \theta_{k,j} \vert \bar{\eta}]\right) \frac {\langle j-1 j\rangle}{\langle j-1 \lambda \rangle \langle j \lambda\rangle} \frac {\langle k-1 k\rangle}{\langle k-1 \lambda \rangle \langle k \lambda\rangle}\,.
\end{multline}  Then, we multiply the numerator and the denominator of the integrand by $[\bar{\lambda} \bar{\rho}]^{4}$ and make a change of variable $\bar{\eta} = \bar{\zeta} [\bar{\lambda} \bar{\rho}]$ to get
\begin{multline}
    \langle 0 \vert G_{j-1,j}^{+}(x_{j}^{+}, \theta_{j}) G_{k-1,k}^{+}(x_{k}^{+}, \theta_{k})\vert 0 \rangle =  \frac 1 {64 \pi^{4}} \int_+ d^{2} \lambda \,d^{2} \bar{\lambda}\, d^{0\vert 4} \bar{\zeta} \times
\\
 \exp \left(\frac i 2 \langle \lambda \vert x_{k,j}^{+} \vert \bar{\lambda}] - \langle \lambda \vert \theta_{k,j} \vert \bar{\zeta}] [\bar{\rho} \bar{\lambda}]\right)
\frac {\langle j-1 j\rangle}{\langle j-1 \lambda \rangle \langle j \lambda\rangle [\bar{\lambda} \bar{\rho}]^{2}}\,
 \frac {\langle k-1 k\rangle}{\langle k-1 \lambda \rangle \langle k \lambda\rangle [\bar{\lambda} \bar{\rho}]^{2}}\,.
\end{multline}

Now consider the following differential operators
\begin{equation}
  \label{eq:chiralchiral1}
  \mathcal{D}_{\ell} = -i \langle \ell \vert \sigma^{\mu} \vert \bar{\rho}] \frac \partial {\partial x_{k,j}^{+}{}^{\mu}}\,.
\end{equation}  These differential operators have been designed to cancel the $\lambda$ and $\bar{\lambda}$ dependence in the denominator of the integrand in eq.~\eqref{eq:chiralchiral1}.
Since
\begin{equation}
  \mathcal{D}_{\ell} \left(\frac i 2 \langle \lambda \vert x_{k,j}^{+} \vert \bar{\lambda}]\right)
= \langle \ell \lambda\rangle [\bar{\rho} \bar{\lambda}],
\end{equation} we have that
\begin{multline}
  \mathcal{D}_{j-1} \mathcal{D}_{j} \mathcal{D}_{k-1} \mathcal{D}_{k}   \langle 0 \vert G_{j-1,j}^{+}\, G_{k-1,k}^{+}\vert 0 \rangle =\\
\langle j-1 j\rangle \langle k-1 k\rangle \int d^{0\vert 4} \bar{\zeta} \left(-\frac 1 {4 \pi^{2}}\right) \frac 1 {(x_{k,j}^{+} + 2 i \theta_{k,j} \vert \bar{\zeta}] [\bar{\rho})^{2}}\,.
\end{multline}
The integral over $\bar{\zeta}$ can be done as follows
\begin{equation}
  \int d^{0\vert 4} \bar{\zeta} \frac 1 {(x_{k,j}^+ + [\bar{\rho} \vert \bar{\sigma} \theta_{k,j} \vert
\bar{\zeta}])^2} = \int  d^{0\vert 4} \bar{\zeta} \exp(i [\bar{\rho}\vert \bar{\sigma}^\mu\theta_{k,j} \vert \bar{\zeta}]\partial_\mu ) \frac 1 {(x_{k,j}^{+})^{2}} = 2^{4} 4! \frac {\delta^{0\vert 4}(\theta_{k,j} \vert x_{k,j}^{+} \vert \bar{\rho}])}{\left((x_{k,j}^{+})^{2}\right)^{5}}\,,
\end{equation}
where for the first equality we have used a translation operator applied to $1/{(x_{k,j}^{+})^{2}}$
and for the second equality we have expanded the exponential.

Now we want to find another expression which gives the same result when acted upon by the
product $\mathcal{D}_{j-1} \mathcal{D}_{j} \mathcal{D}_{k-1} \mathcal{D}_{k}$ of differential operators.
This seems to be very hard, but consider the action on $1/{(x_{k,j}^{+})^{2}}$.  It is straightforward to show that
\begin{equation}
  \mathcal{D}_{j-1} \mathcal{D}_{j} \mathcal{D}_{k-1} \mathcal{D}_{k} \frac 1 {(x_{k,j}^{+})^{2}} = 2^{4} 4! \frac {\langle j-1\vert x_{k,j}^{+}\vert \bar{\rho}] \langle j\vert x_{k,j}^{+}\vert \bar{\rho}] \langle k-1\vert x_{k,j}^{+}\vert \bar{\rho}] \langle k\vert x_{k,j}^{+}\vert \bar{\rho}]}{\left((x_{k,j}^{+})^{2}\right)^{5}}\,.
\end{equation}

If we use the fact that $\mathcal{D}_{\ell} \langle \lambda \vert x_{k,j}^{+}\vert \bar{\rho}] = 0$ and $\mathcal{D}_{\ell} \theta_{k,j} \vert x_{k,j}^{+} \vert \bar{\rho}] = 0$ for any $\lambda$ and $\ell$, we get that
\begin{multline}
  \mathcal{D}_{j-1} \mathcal{D}_{j} \mathcal{D}_{k-1} \mathcal{D}_{k} \left(-\frac 1 {4 \pi^{2}} \frac {\langle j-1 j\rangle \langle k-1 k\rangle \delta^{0\vert 4}(\theta_{k,j} \vert x_{k,j}^{+} \vert \bar{\rho}])}{(x_{k,j}^{+})^{2} \langle j-1\vert x_{k,j}^{+}\vert \bar{\rho}] \langle j\vert x_{k,j}^{+}\vert \bar{\rho}] \langle k-1\vert x_{k,j}^{+}\vert \bar{\rho}] \langle k\vert x_{k,j}^{+}\vert \bar{\rho}]}\right) =\\
  \mathcal{D}_{j-1} \mathcal{D}_{j} \mathcal{D}_{k-1} \mathcal{D}_{k}   \langle 0 \vert G_{j-1,j}^{+}\, G_{k-1,k}^{+}\vert 0 \rangle.
\end{multline}

In conclusion, up to terms which vanish under the action of $\mathcal{D}_{j-1} \mathcal{D}_{j} \mathcal{D}_{k-1} \mathcal{D}_{k}$, we have
\begin{equation}
  \label{eq:space-timeR}
  \langle 0 \vert G_{j-1,j}^{+}\, G_{k-1,k}^{+}\vert 0 \rangle = -\frac 1 {4 \pi^{2}} \frac {\langle j-1 j\rangle \langle k-1 k\rangle \delta^{0\vert 4}(\theta_{k,j} \vert x_{k,j}^{+} \vert \bar{\rho}])}{(x_{k,j}^{+})^{2} \langle j-1\vert x_{k,j}^{+}\vert \bar{\rho}] \langle j\vert x_{k,j}^{+}\vert \bar{\rho}] \langle k-1\vert x_{k,j}^{+}\vert \bar{\rho}] \langle k\vert x_{k,j}^{+}\vert \bar{\rho}]}\,.
\end{equation}
The right-hand side of eq.~\eqref{eq:space-timeR} is the space-time
form of the $R$-invariant of the points $(x_{j-1}^{+}, \theta_{j-1})$, $(x_{j}^{+}, \theta_{j})$,
$(x_{k-1}^{+}, \theta_{k-1})$, $(x_{k}^{+}, \theta_{k})$
and a reference spinor $\bar{\rho}$.
The equivalence of this space-time form of the $R$-invariant
and the twistor form is explicitly shown in \appref{app:rinvar}
(a proof can also be found in ref.~\cite{Bullimore:2010pj}).

One may be surprised by the appearance of $\bar{\rho}$ in the right-hand side, when there is no such dependence in the left-hand side of eq.~\eqref{eq:space-timeR}.  However, the dependence on $\bar{\rho}$ can be interpreted as an integration constant.  Indeed, using identities between $R$-invariants, one can show that the difference between the expressions in the right-hand side of eq.~\eqref{eq:space-timeR} for two different values of $\bar{\rho}$ is annihilated by $\mathcal{D}_{j-1} \mathcal{D}_{j} \mathcal{D}_{k-1} \mathcal{D}_{k}$.%
\footnote{The difference is a linear combination of four $R$-invariants, each depending on only three of the four points $j-1$, $j$, $k-1$, $k$.}

If we now sum up the contributions of all the chiral-chiral vertex correlators,
with the same reference spinor $\bar{\rho}$, we find the NMHV scattering amplitude
in the form obtained by Mason and Skinner in ref.~\cite{Mason:2010yk}.
In \secref{sec:twistor} we will see that the analogous
computation in twistor space is more straightforward.

The antichiral correlator between vertices $j$ and $k$, $\langle
G^-_{j-1,j} G^-_{k-1,k}\rangle$, is given by the conjugate $R$
invariant, which depends on a conjugate
reference spinor $\rho$.

\subsection{Mixed Chirality Correlator}

Using the two-point function of a $C$ and a $\bar{C}$ field, we find, after performing some trivial integrations, that
\begin{multline}
  \label{eq:mixedcorrint}
  \langle 0 \vert G_{j-1,j}^{+}\, G_{k-1,k}^{-}\vert 0\rangle = -\frac 1{256 \pi^{4}} \int_+ d^{2} \lambda \,d^{2} \bar{\lambda}\, \times\\ \exp \left(-\frac i 2 \langle \lambda \vert x_{j,k}^{+-} \vert \bar{\lambda}]\right) \frac {\langle j-1 j\rangle [k-1 k]}{\langle j-1 \lambda\rangle \langle \lambda j\rangle [k-1 \bar{\lambda}] [\bar{\lambda} k]}\,,
\end{multline}
where $x_{j,k}^{+-} = x_{k}^{-} - x_{j}^{+} + 4 i \theta_{j} \bar{\theta}_{k}$
\eqref{eq:mixedinterval}.

It is not obvious how to compute these integrals directly, but we can use the fact that the answer satisfies differential equations with simple
source terms.
The solution to these differential equations is not unique,
but there is a discrete symmetry that fixes the coefficient
of the homogeneous solution.


It is convenient to use \eqref{eq:mixedcorrint} and find
differential operators with respect to the components of $x^{+-}_{j,k}$ in the basis
$(\lambda_{j-1},\lambda_j;\bar\lambda_{k-1},\bar\lambda_k)$,
\[a=\langle j{-}1|x_{j,k}^{+-}| k{-}1],\quad b= \langle j|x_{j,k}^{+-}| k{-}1],\quad c=\langle
j{-}1|x_{j,k}^{+-}| k], \quad d=\langle j |x_{j,k}^{+-}| k], \]
then
\[
\left(x_{j,k}^{+-}\right)^2=\frac 1 2\frac{ad-bc}{\langle j{-}1 j\rangle [k{-}1 k]}\,. \]
We also have that
\begin{equation}
  \langle \lambda \vert x_{j,k}^{+-} \vert \bar{\lambda}] =
\frac {\langle \lambda j\rangle [\bar{\lambda} k] a}{\langle j{-}1 j\rangle [k{-}1 k]}
+ \frac {\langle j{-}1 \lambda\rangle [\bar{\lambda} k] b}{\langle j{-}1 j\rangle [k{-}1 k]}
+ \frac {\langle \lambda j\rangle [k{-}1 \bar{\lambda}] c}{\langle j{-}1 j\rangle [k{-}1 k]}
+ \frac {\langle j{-}1 \lambda \rangle [k{-}1 \bar{\lambda}] d}{\langle j{-}1 j\rangle [k{-}1 k]}\,.
\end{equation}

The integral \eqref{eq:mixedcorrint} only depends on $x_{j,k}^{+-}$ through
$\exp(-\tfrac i 2 \langle \lambda \vert x_{j,k}^{+-} \vert \bar\lambda])$,
and differentiating with respect
to $a$ gives a factor $-\tfrac i 2 {\langle\lambda j\rangle[\bar\lambda k]}/{\langle j-1 j\rangle [k-1 k]}$ etc. Therefore, the second-order differential operator with
respect to $a,d$ or $b,c$ removes all brackets in the denominator
\begin{multline}
\frac{\partial^2 }{\partial a\,\partial d}\langle 0\vert G^+_{j-1,j}
G^-_{k-1,k}\vert 0\rangle = \frac{\partial^2 }{\partial b\,\partial c}
\langle 0\vert G^+_{j-1,j} G^-_{k-1,k}\vert 0\rangle =\\= \frac 1 {1024 \pi^{4}} \frac 1 {\langle j{-}1
j\rangle [ k{-}1 k]} \int_+ d^{2}\lambda \,d^2\bar\lambda\, \exp\left(\frac i 2 \langle \lambda|
x^{+-}_{j,k}| \bar\lambda]\right).
\end{multline}

The differential operators reduce the integral to a simpler one,
which is nothing but the momentum representation of the scalar
propagator
\[ \int_+
d^{2} \lambda \,d^2\bar\lambda\, \exp\left(\frac
i 2 \langle \lambda| x| \bar\lambda]\right) = -16 \pi^{2} \frac{1}{x^2}\,.\]
 Therefore, we have
\begin{equation}
  \frac {\partial^{2}}{\partial a \partial d} \langle 0\vert G^+_{j-1,j} G^-_{k-1,k}\vert 0\rangle =   \frac {\partial^{2}}{\partial b \partial c} \langle 0\vert G^+_{j-1,j} G^-_{k-1,k}\vert 0\rangle = -\frac 1 {64 \pi^{2}} \frac 1 {a d - b c}\,.
\end{equation}

A solution to these two differential equations is easily found by integration in terms of dilogarithms and logarithms
\[
\frac{\partial^2 }{\partial a\,\partial d}\left(-\Li_2\frac{b c}{a d}+\half\log (a d)\log\frac{b c}{a d}\right)=\frac{\partial^2
}{\partial b\,\partial c} \left(-\Li_2\frac{b c}{a d}+\half\log (a d)\log\frac{b c}{a d}\right)=\frac{1}{bc-ad}\,.
\]

Besides an additive constant which is not very interesting, there is
an ambiguity in solving the equations in the class of
transcendentality two functions. It corresponds to adding a factor
of $\log (a d) \log (b c)$ times a rational number.  Such a term is
annihilated by both second-order differential operators we
considered.
To fix the coefficient it suffices to demand that the expression is
antisymmetric with respect to interchanges of $\lambda_{j-1}$ and $\lambda_j$ or $\bar\lambda_{k-1}$ and $\bar\lambda_k$.
This obvious symmetry of the integral \eqref{eq:mixedcorrint}
should be reflected in the integrand (up to shifts by constants which we neglect).

In conclusion, the resulting integral reads
\begin{multline}\label{eq:mixedcorr}
64 \pi^{2} \langle 0\vert G^+_{j-1,j} G^-_{k-1,k}\vert 0\rangle = -\Li_2\left(\frac{\langle j{-}1| x^{+-}_{j,k} |k]\langle j| x^{+-}_{j,k} |
k{-}1]} {\langle j{-}1| x^{+-}_{j,k} |k{-}1]\langle j | x^{+-}_{j,k}
|k]}\right)+\\ \half\log \left(\langle j{-}1| x^{+-}_{j,k}
|k{-}1]\langle j | x^{+-}_{j,k} |k]\right) \log \left(\frac{\langle
j{-}1| x^{+-}_{j,k} |k]\langle j| x^{+-}_{j,k} | k{-}1]} {\langle
j{-}1| x^{+-}_{j,k} |k{-}1]\langle j | x^{+-}_{j,k} |k]}\right).
\end{multline}

We should note that if the points $j$ and $k$ become too close,
then the answer in eq.~\eqref{eq:mixedcorr} becomes divergent.
This is how the UV divergences of the Wilson loop manifest themselves.
In \secref{sec:regularization} we will discuss some ways to regularize these divergences.

It ought to be mentioned that the above expression is not invariant under
rescaling of the spinor variables.
It is however reassuring to observe that in the sum over all vertices
this dependence drops out.
This cancellation depends crucially on the correct choice of coefficient
for the homogeneous solution of the above differential equations.
\section{Twistor Space Calculation}
\label{sec:twistor}

The above results for the vertex correlators have
convenient expressions in terms of twistor variables.
Here we present our twistor conventions,
translate our above results,
and show how the calculations can be cut short
if performed directly in twistor space.

\subsection{Ambitwistors}
\label{sec:ambi}

\begin{figure}\centering
$\displaystyle
\includegraphicsbox[height=3.5cm]{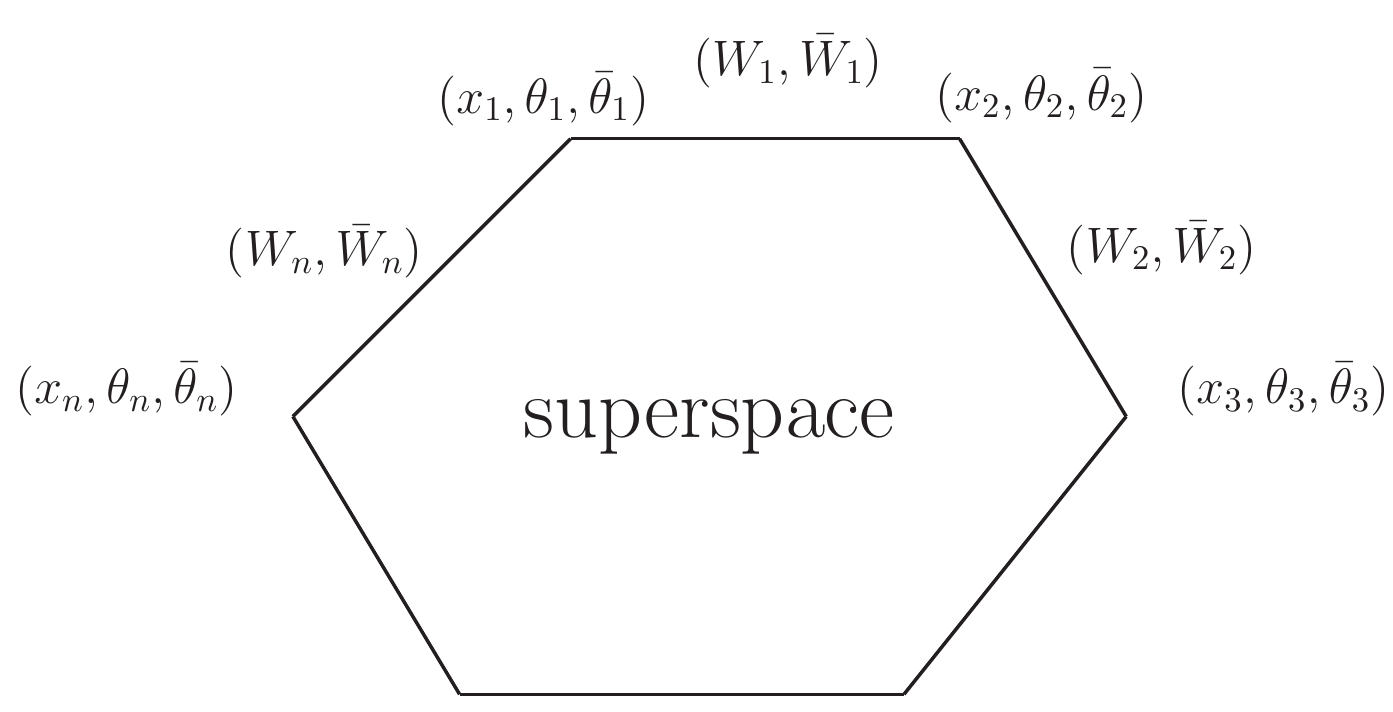}\quad \Leftrightarrow \quad \includegraphicsbox[height=3.5cm]{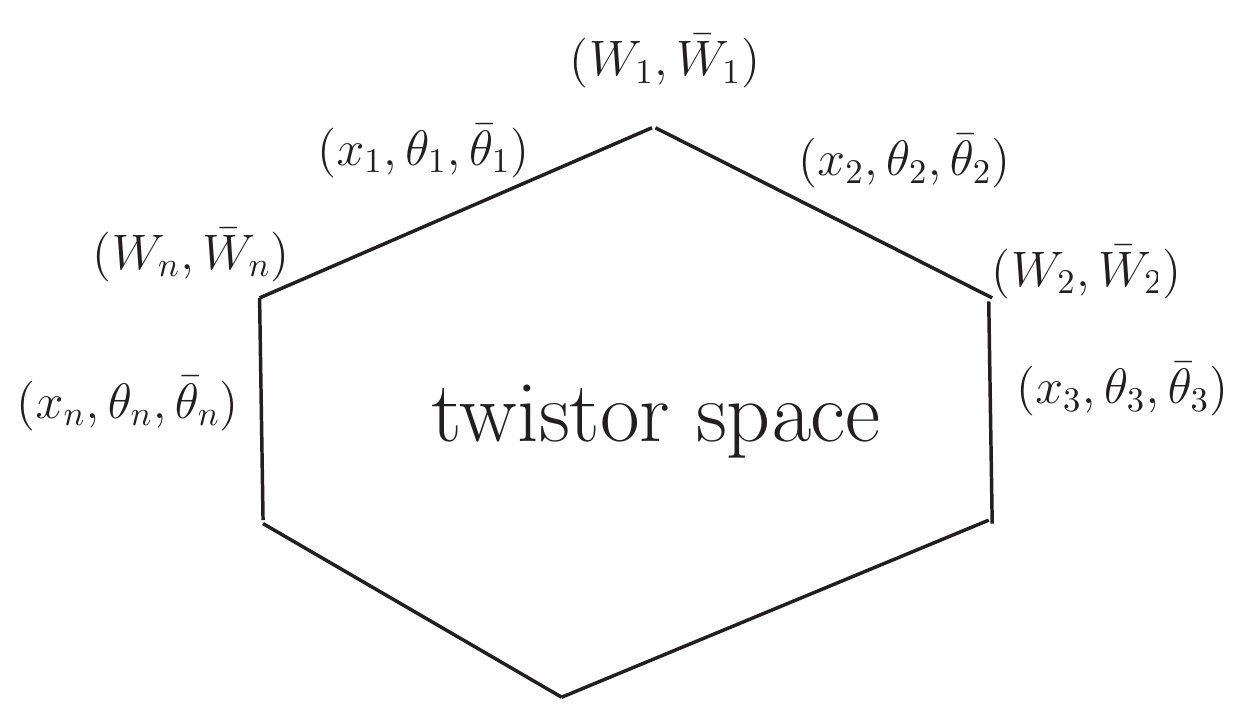}$
\caption{A null polygon in superspace and the dual polygon in twistor space.} \label{fig:twistorpolygon}
\end{figure}

The Wilson loop is a sequence of null lines.
In \secref{sec:vertexcorr} we specified
these null lines through the polygon vertices.
A useful alternative description of a null line is through
an ambitwistor $(W,\bar W)$ \cite{Mason:2005kn}.
Consequently the Wilson loop contour is also specified
through a sequence of ambitwistors, see \figref{fig:twistorpolygon}.
We will now specify these twistor variables for the polygon
and spell out their relations.
See ref.~\cite{Beisert:2012} for more details of the construction.

The twistor equations
$\mu_{\dot{\alpha}} = \frac 1 4 \lambda^\alpha x^+_{\alpha \dot{\alpha}}$,
$\chi^a = \lambda^\alpha {\theta}_\alpha^a$ and
$\bar{\mu}_{\alpha} = \sfrac 1 4 x^-_{\alpha \dot{\alpha}} \bar{\lambda}^{\dot{\alpha}}$,
$\bar{\chi}_{a} = \bar{\theta}_{\dot{\alpha} a} \bar{\lambda}^{\dot{\alpha}}$
for $(x,\theta,\bar\theta)$ define a null line.
They are solved precisely by the explicit parametrization
of null lines given in \eqref{eq:nullline}.
It then makes sense to collect the quantities $\mu_{\dot{\alpha}}$,
$\chi^a$, $\bar{\mu}_{\alpha}$, $\bar{\chi}_{a}$
in twistor variables
$W=(-i\lambda^\alpha,\mu_{\dot\alpha},\chi^a)$ and
$\bar W=(\bar\mu_{\alpha},i\bar\lambda^{\dot\alpha},\bar\chi_a)$
which transform nicely as projective vectors under the superconformal
group $\grp{PSL}(2,2|4)$.

For the null segment connecting vertices $j$ and $j+1$,
we shall use the ambitwistor $(W_j,\bar W_j)$. More explicitly,
we define $\lambda_j$, $\bar{\lambda}_j$ by $x_{j,j+1} = \lambda_j \bar{\lambda}_j$,
where $x_{j,j+1}$ is a superspace interval as defined in eq.~\eqref{eq:susy-interval}.
Also, we set
\begin{alignat}{2}
\label{eq:shiftindex}
  \mu_{j} &:= \sfrac 1 4 \langle j\vert x_j^+= \sfrac 1 4 \langle j\vert x_{j+1}^+, &\qquad
  \bar{\mu}_{j} &:= \sfrac 1 4 x_j^- \vert j ]= \sfrac 1 4 x_{j+1}^- \vert j ],\nln
\chi_j &:= \langle j\vert \theta_j= \langle j\vert \theta_{j+1}, &\qquad
\bar{\chi}_{j} &:= \bar{\theta}_{j} \vert j]= \bar{\theta}_{j+1} \vert j],
\end{alignat}
then the twistors and dual twistors have components
\begin{equation}
  W_i = (-i \lambda_i^\alpha, \mu_{i \dot{\alpha}}, \chi_i^a), \qquad
  \bar{W}_i = (\bar{\mu}_{i \alpha}, i \bar{\lambda}_i^{\dot{\alpha}}, \bar{\chi}_{i a}).
\end{equation}
The scalar product $\langle j,k]:=W_j \cdot \bar{W}_k$ between a twistor and dual twistor is defined by
\begin{equation}
 \langle j,k] = -i \lambda_j^\alpha \bar{\mu}_{k \alpha} + i \mu_{j \dot{\alpha}} \bar{\lambda}_k^{\dot{\alpha}} + \chi_j^a \bar{\chi}_{k a}
= -\sfrac i 4 \langle j \vert (x_k^- - x_j^+ + 4 i \theta_j \bar{\theta}_k )\vert k ]
= - \sfrac i 4 \langle j \vert x_{j,k}^{+-} \vert k ].
\end{equation}
The relation $x^+ - x^- = 4 i \theta \bar{\theta} $
along with the relations in \eqref{eq:shiftindex}
implies the following identities
\[\label{eq:zerobrackets}
\langle j,j-1]
=\langle j,j]
=\langle j,j+1]
=0.
\]

In general, the answers become simpler in twistor language.  
For instance, the mixed chiral vertex correlator \eqref{eq:mixedcorr}
has the following simple expression in terms of twistor products
\[
\label{eq:mixedcorrtwistor}
64 \pi^{2} \costate{0} G^+_{j-1,j} G^-_{k-1,k}\state{0}
= -\Li_2 X_{j,k} +\half\log \langle j{-}1, k{-}1]\langle
j, k] \log X_{j,k},
\]
where we have defined the twistor cross-ratios
\begin{equation}
  \label{eq:twistorCR}
  X_{j,k} = \frac{\langle j{-}1, k]\langle j, k{-}1]}{\langle
j{-}1, k{-}1]\langle j, k]}\,,
\end{equation}
which are invariant under rescaling of any of the involved twistors
as well as under superconformal transformations.

\subsection{Correlators}
\label{sec:twistor2}

Next we transform the on-shell momentum space fields $C(\lambda,\bar\lambda,\bar\eta)$
and $\bar{C}(\lambda,\bar\lambda,\eta)$ to twistor space.
There are several reasons to do this.
First of all, we hope to obtain the tree-level NMHV amplitude,
which is most naturally expressed in twistor space.
Moreover, the one-loop amplitudes also have a simple form in twistor space.
Another reason to study the transformation to twistor space
is the fact that the superconformal symmetry becomes more obvious in this language.

An immediate drawback of the twistor transformation is that it
is hard to define properly in Minkowski signature.
Typically, one Wick rotates to ${+}{+}{-}{-}$ signature or
complexifies spacetime altogether.
The resulting expressions remain meaningful after this transformation.
Unfortunately, integration contours are not obvious anymore,
and would have to be specified in order to make sense of most integrals.
We will not elaborate on the choice (or existence) of contours in
this paper.

We use the following definitions for the twistor transforms of the mode expansions $C$ and $\bar{C}$%
\footnote{The dimension of $\lambda$ (used to represent null momenta)
is not the same as the dimension of $\lambda_j$ (used to represent null distances).
Due to the projective nature of twistors, this difference stays without consequences.}
\begin{align}
  C(\lambda,\mu,\chi) &\equiv \int d^2 \bar{\lambda}\, d^{0 \vert 4} \bar{\eta} \,
\exp \left(-2 i [\mu \bar{\lambda}] + \chi \bar{\eta}\right) C(\lambda, \bar{\lambda}, \bar{\eta}),
\\
  \bar{C}(\bar\lambda,\bar\mu,\bar\chi) &\equiv \int d^2 \lambda \, d^{0 \vert 4} \eta \,
\exp \left(-2 i \langle \lambda \bar{\mu}\rangle - \eta \bar{\chi}\right) \bar{C}(\lambda, \bar{\lambda},\eta).
\end{align}
Relation \eqref{eq:fermfourier}
translates to a relation between the twistor fields
\begin{equation}
\label{eq:conjugatetwistor}
  \bar{C}(\bar{W}) = \frac 1 {(2 \pi)^2} \int d^{4 \vert 4} W \exp \left(2 W \cdot \bar{W}\right) C(W).
\end{equation}
The prepotentials \eqref{eq:b-fields} in light cone gauge~\eqref{eq:scalarC}
also find a simple expression in terms of the twistor fields
\begin{align}
B^{\alpha\beta}(x^+,\theta)&= \frac 1 {8 \pi^2}\int
\frac{\langle\lambda d\lambda\rangle\,l^{\alpha}l^{\beta}}{\langle \lambda l \rangle^2}\,
C(\lambda,\quarter\lambda x^+,\lambda\theta),
\nln
\bar B^{\dot\alpha\dot\beta}(x^-,\bar\theta)&= \frac 1 {8 \pi^2} \int
\frac{[\bar\lambda d\bar\lambda]\,\bar l^{\dot\alpha}\bar l^{\dot\beta}}{[ \bar l \bar\lambda ]^2}\,
\bar{C}(\bar\lambda,\quarter x^-\bar\lambda,\bar\theta\bar\lambda).
\end{align}
The above expressions are integrals over
a contour in $\mathbb{CP}^1$'s which are the twistor duals of the points
$(x^{+},\theta)$, $(x^{-},\bar\theta)$ in chiral or antichiral superspace.  As described in more detail in~\cite{Beisert:2012},
a point in full superspace corresponds in complexified ambitwistor space
to a $\mathbb{CP}^{1} \times \mathbb{CP}^{1}$.
Each of the $\mathbb{CP}^{1}$ factors can be seen as the twistor (or conjugate twistor)
associated to the points in chiral (or antichiral space).

Finally, we need to transform the two-point correlator
\eqref{eq:cc-propagator} to twistor space.
Here, the main complication is the restriction to positive energies
in integrals $\int_+$,
which makes sense only in Minkowski signature,
but not in split signature or complexified spacetime.
Simply dropping the step function is not an option because
in integrals the negative energy contributions typically
cancel most of the positive energy contributions,
and the result would almost vanish.%
\footnote{According to the discussion in \secref{sec:onshellcorrelator},
dropping the step function amounts to computing the
discontinuity on the expectation value.
For Wilson loops the discontinuity usually
has one degree of transcendentality less.}

We cast the step function to the form of a Fourier integral
\[
\theta(x)=\frac{1}{2\pi i}\int \frac{dt}{t-i\epsilon}\,\exp(ixt),
\]
which can be taken to a different signature up to a suitable choice of
integration contour.
Furthermore, the energy $E(\lambda,\bar\lambda)$ is not
a convenient expression in twistor space.
As we are only interested in
distinguishing the positive from the negative light cone,
we can safely replace the energy $E$ by a light-cone energy
given by $E\indup{l.c.}=-2\langle \lambda \rho\rangle[\bar\rho\bar\lambda]$
where the spinors $\rho,\bar\rho$ describe a reference null direction.
In other words we replace
\[
\theta(E(\lambda,\bar\lambda))\to
\frac{1}{2\pi i}\int \frac{dt}{t}\,\exp\bigbrk{-2it\langle \lambda \rho\rangle[\bar\rho\bar\lambda]}.
\]
and obtain for the twistor space correlation function
\[
\costate{0} C(\lambda,\mu,\chi)\,C(\lambda',\mu',\chi')\state{0}
=
-\frac{1}{4}
\int \frac{dt}{t}\, \frac{d z}{ z}\,
\delta^2(\lambda+z\lambda')\,
\delta^2(\mu+z\mu'+t\langle \lambda \rho\rangle \bar\rho)\,
\delta^{0 \vert 4}(\chi+z\chi').
\]
By rescaling the integration variables we end up with a neat twistor space expression
for the chiral and antichiral correlators%
\footnote{It is tempting to scale away $\langle \lambda \rho\rangle\simeq W\cdot\bar W_\star$ as well,
but such a rescaling would obscure the conjugation relation \eqref{eq:conjugatetwistor}
between $C$ and $\bar C$, and may have other undesired side-effects.}
\begin{align}
  \label{eq:cc-twistor}
  \costate{0} C(W)\, C(W')\state{0}
&=
-\frac{1}{4} \int \frac {d s}{s}\, \frac {d t}{t}\,\delta^{4 \vert 4}(s W + t W' + (W\cdot\bar W_\star)W_\star),
\nln
  \costate{0} \bar{C}(\bar W)\, \bar{C}(\bar W')\state{0}
&=
-\frac{1}{4} \int \frac {d s}{s}\, \frac {d t}{t}\,\delta^{4 \vert 4}(s \bar W + t \bar W' + (W_\star\cdot \bar W)\bar W_\star),
\end{align}
where $W_\star = (0, \bar{\rho}, 0)$
$\bar W_\star = (\rho, 0, 0)$
are reference twistors.
Remarkably, this is the propagator of the twistor field in the axial gauge,
as shown by Mason and Skinner in ref.~\cite{Mason:2010yk}.
It has support when the twistors $W$, $W'$ and $W_\star$ lie
on a common projective line.

The mixed chiral correlator in twistor space reads
\[
\costate{0}   C(W)\, \bar{C}(\bar W') \state{0}
=
-\frac{1}{16\pi^2}
\int \frac{ds}{s}\,\frac{dt}{t}\,
\exp \left(s W\cdot \bar W'+t (W\cdot \bar W_\star)(W_\star\cdot \bar W')\right)
\]
Corresponding to the above observation, this expression might serve
as the mixed chiral propagator in an ambitwistor theory.

\subsection{Vertex Correlators}

Here we will compute the vertex correlators directly in twistor space.
First we transform the shift of gauge potential at a vertex
\eqref{eq:vertexshift}
\begin{align}
G^+_{j-1,j}&= \frac 1 {8 \pi^2}\int
\frac{\langle\lambda d\lambda\rangle\,\langle j{-}1 j\rangle}{\langle j{-}1 \lambda\rangle \langle \lambda j\rangle}\,
C(\lambda,\quarter\lambda x^+_j,\lambda\theta_j),
\nln
G^-_{j-1,j}&= -\frac 1 {8 \pi^2}\int
\frac{[\bar\lambda d\bar\lambda]\,[j{-}1 j]} {[j{-}1 \bar\lambda] [\bar\lambda j]} \,
\bar{C}(\bar\lambda,\quarter x^-_j\bar\lambda, \bar\theta_j\bar\lambda).
\end{align}
We expand $\lambda,\bar\lambda$
as $\lambda=\lambda_{j-1}+u\lambda_j$,
$\bar\lambda=\bar\lambda_{j-1}+v\bar\lambda_j$
and use the identities \eqref{eq:shiftindex}
to find the following twistor space representation
\[
G^+_{j-1,j}= \frac 1 {8 \pi^2}
\int \frac{du}{u}\, C(W_{j-1}+uW_j),
\quad
G^-_{j-1,j}= -\frac 1 {8 \pi^2}
\int \frac{dv}{v}\, \bar{C}(\bar W_{j-1}+v\bar W_j).
\]
Now it is straight-forward to compute the chiral correlator between two vertices $j$ and $k$
which reads after some rescaling of integration variables
\begin{align}
\costate{0} G^+_{j-1,j} G^+_{k-1,k}\state{0}
=- \frac{1}{256\pi^4}
\int \frac{d\alpha}{\alpha}\,\frac{d \beta}{\beta}\,\frac{d \gamma}{\gamma}\,\frac{d \delta}{\delta}\,
\delta^{4|4}(\alpha W_{j-1} + \beta W_j+ \gamma W_{k-1} + \delta W_k + W_\star).
\end{align}
The antichiral correlator is simply the conjugate expression. We recognize this as the correlator between two edges $j$ and $k$ in twistor
space. This is expected since each vertex in space-time corresponds to a line in twistor space, and the $W_\star$ we have here corresponds to
the reference twistor in the axial gauge form of the propagator in twistor space.

Now we proceed to the mixed chirality correlator between two
vertices. It reads simply
\[\costate{0} G^+_{j-1,j}\, G^-_{k-1,k}\state{0}=
\frac{1}{1024\pi^6}
\int \frac{d u}{u}\,\frac{d v}{v}\,\frac{d s}{s}\,\frac{d t}{t}\,
\exp\bigbrk{s W_u\cdot\bar W_v+t (W_u\cdot\bar W_\star)
(W_\star\cdot\bar W_v)},
\]
where $W_u:= W_{j-1}+u W_j$ and $\bar W_v:= \bar W_{k-1}+ v \bar W_k$.
It would be desirable to show that this multiple
integral evaluates to \eqref{eq:mixedcorr}.
We have not made serious attempts in this direction,
but it appears that a careful consideration of integration contours
may be required to prove the equivalence.

\section{Regularizations}
\label{sec:regularization}

From now on we will consider only the mixed chirality contributions since
the purely chiral contributions are rational, finite
and equal to the well-known counterparts in the chiral Wilson loop.

\begin{figure}\centering
\includegraphics[height=4cm]{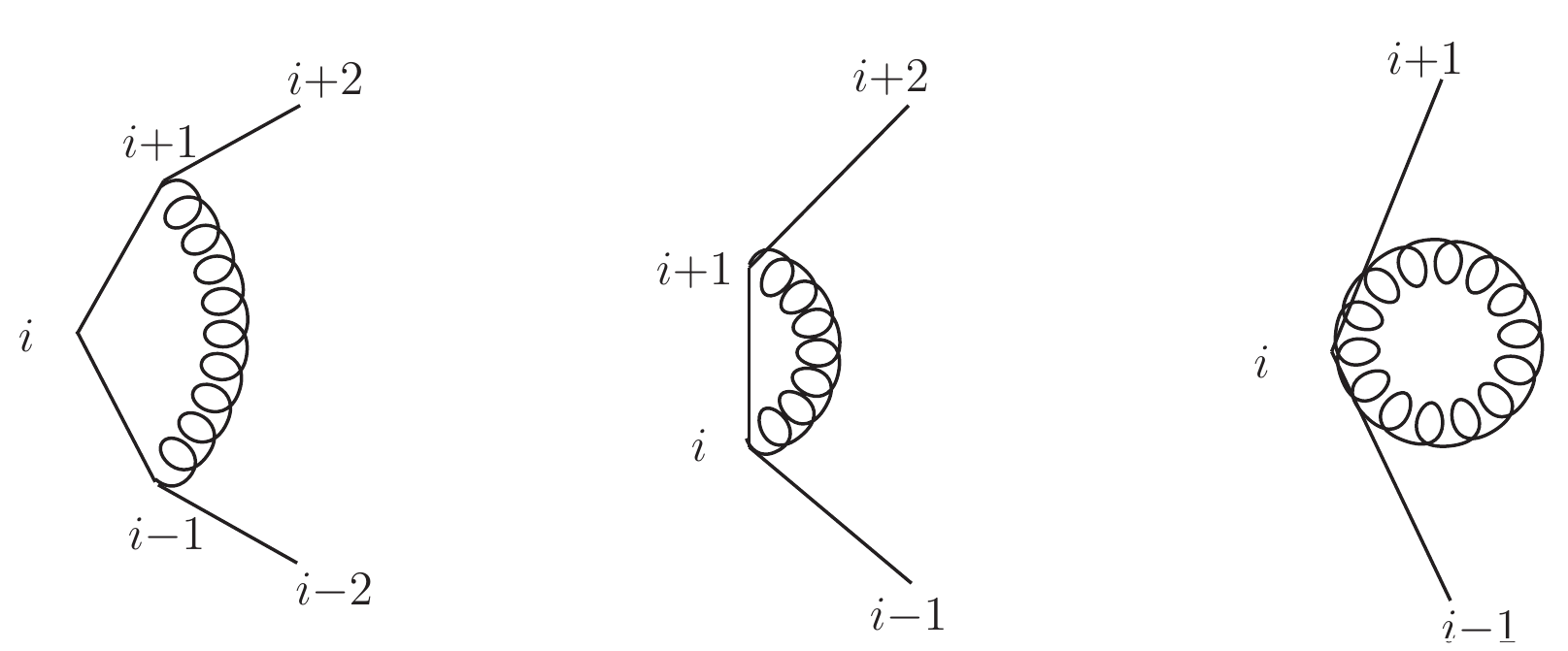}
\caption{Diagrams for divergent mixed-chirality correlators.} \label{fig:divergence}
\end{figure}

Now that we have the vertex correlator, we need to sum over all
pairs of vertices as in eq.~\eqref{eq:WLvertex},
\[
\frac 1 {64 \pi^{2}} M^{(1)}_{n,+-}=\sum_{j,k=1}^n
\langle G^+_{j-1,j}G^-_{k-1,k}\rangle.
\]
However, it is easy to see the $\Li_2$ term diverges for $|k-j|<2$, so does the $\log\log$ term for $|k-j|<3$, see
\figref{fig:divergence}. In these cases, either a regularization, or a finite quantity to be extracted from the full answer, is needed, and
there are various ways to do it as we discuss now.

\subsection{Framing}

\begin{figure}\centering
\includegraphics[width=.8\textwidth]{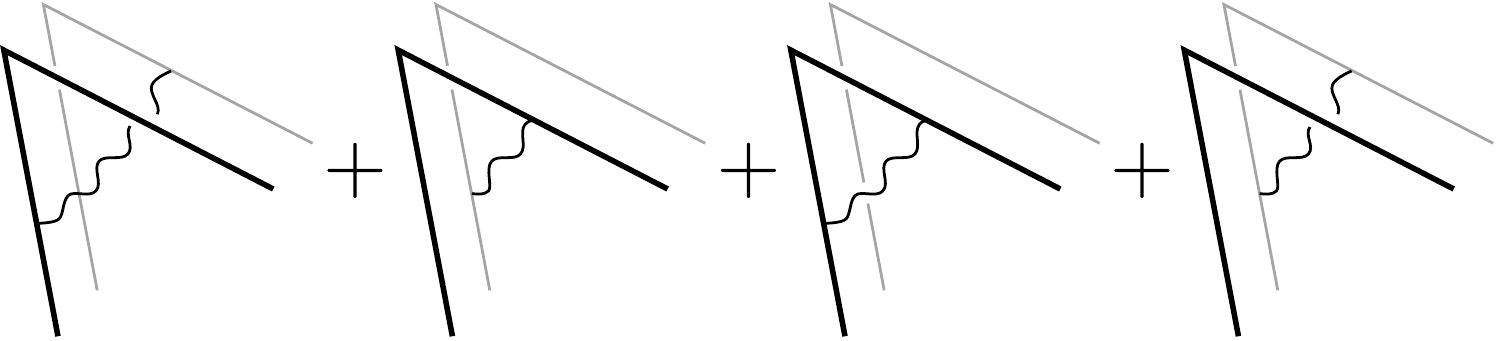}
\caption{The one-loop expectation value of two Wilson loops.
The initial Wilson loop contour $C$ is represented by a solid line,
while the contour of its displaced copy $C'$ is represented by a gray line.
There are four kinds of contributions to the one-loop expectation
value of $\langle \mathcal{W}[C] \mathcal{W}[C']\rangle$,
but when dividing by $\langle \mathcal{W}[C]\rangle \langle \mathcal{W}[C']\rangle$,
the contribution of the last two diagrams is canceled.}
\label{fig:framing}
\end{figure}

One way to regularize the one-loop result is to frame the Wilson loop. By
shifting each vertex of the null polygon $C$ by any vector,%
\footnote{In order to obtain the right branch cut structure the safest option is to take the shift to be space-like.}
which preserves the null condition, we have a shifted null
polygon, $C'$, and we consider the ratio
\[\frac{\langle \mathcal{W}[C]\mathcal{W}[C']\rangle}{\langle \mathcal{W}[C]\rangle \langle \mathcal{W}[C']\rangle}\,.
\]
At one loop, it is equivalent to the sum
\[M\indup{framing}^{(1)}[C,C']:=\half M^{(1)}[C,C']
-\half M^{(1)}[C]- \half M^{(1)}[C'],\]
which is given by (one half) the sum of correlators between edges (or vertices) of $C$ and edges (or vertices) of $C'$ (see
\figref{fig:framing}),
\begin{multline}\label{eq:framingGG} \frac 1 {64 \pi^{2}} M\indup{framing}^{(1)}[C,C']=
\half \oint_C\oint_{C'}\bigbrk{\langle A^+A^-\rangle+\langle A^-A^+\rangle}
=\\=\half\sum_{j\in C,k\in C'}\bigbrk{\langle G^+_{j-1,j}G^-_{k-1,k}\rangle+\langle
G^-_{j-1,j}G^+_{k-1,k}\rangle}.\end{multline}
Note that this is symmetric under the exchange of $C$ and $C'$.
The contributions of chiral-antichiral and antichiral-chiral vertex-vertex correlators in eq.~\eqref{eq:framingGG} differ by quantities which vanish when the contours $C$ and $C'$ become coincident.  If we discard such vanishing terms in the following we can use
\begin{equation}
  \frac 1 {64 \pi^{2}} M\indup{framing}^{(1)}[C,C']= \sum_{j\in C,k\in C'} \langle G^+_{j-1,j}G^-_{k-1,k}\rangle.
\end{equation}

Since the contour $C$ is specified by a sequence of momentum
ambitwistors $W_k, \bar W_k$, the contour $C'$ can be specified by shifted
momentum ambitwistors $W'_k, \bar W'_k$. A particular choice is to
shift all twistors (conjugates) along the same direction of a
reference twistor $W_*$ (conjugate $\bar W_*$), with
$\tprod{*}{*}\neq 0$
\[ W'_k=W_k+i\epsilon \frac{\tprod{k}{*}}{\tprod{*}{*}} W_*, \qquad \bar W'_k=\bar W_k-i\epsilon \frac{\tprod{*}{k}}{\tprod{*}{*}}\bar W_*,\]
for which, up to $O(\epsilon^2)$ terms, indeed we have $W'_k \bar
W'_k=W'_k \bar W'_{k\pm 1}=0$.

The correlators between well-separated points have a finite limit as the framing goes away ($\epsilon\rightarrow 0$). The divergent terms are regularized simply
by replacing \[\label{eq:axialreplacement}\tprod{j}{k} \to i \epsilon
\frac{\tprod{j}{*}\tprod{*}{k}}{\tprod{*}{*}}:=\epsilon \tprod{j}{k}^*\] for
$k=j,j\pm 1$.
Then, the superconformal cross-ratios defined in eq.~\eqref{eq:twistorCR} are regularized as follows
\begin{equation}
  X_{j,j\pm 2} \to \epsilon X_{j,j\pm2}^*, \quad
  X_{j,j\pm1} \to \epsilon^{-1} X_{j,j\pm1}^*, \quad
  X_{j,j} \to 1.
\end{equation}

Explicitly, the regularized one-loop expectation value using
$W_*$-$\bar W_*$ framing is given by%
\footnote{Just as for the symbol
described in ref.~\cite{Goncharov:2010jf}, from here on we shall not
be careful with the signs of the arguments of logarithm functions.
This amounts to a choice of branch cuts.}
\begin{align}
\label{eq:axialframing}
M^{(1)}_{n,*}=&
\sum_{j=1}^n\biggl\{
-\log^2 \epsilon
+\log\epsilon\,\log \frac{\tprod{j+1}{j-1}\tprod{j-1}{j+1}}{\tprod{j+1}{j}^*\tprod{j}{j+1}^*}
\nln & \qquad
+\sum_{k=j{+}3}^{j{-}3}
  \left(-\Li_2 X_{j,k}+\half\log \bigbrk{\tprod{j{-}1}{k{-}1}\tprod{j}{k}} \log X_{j,k}\right)
\nln & \qquad
+\sum_{k=j\pm2}\half\log \bigbrk{\tprod{j{-}1}{k{-}1}\tprod{j}{k}}\log {X^*_{j,k}}
\nln & \qquad
+\sum_{k=j\pm1}
\half \log\bigbrk{\tprod{j}{k{-}1}^{(*)}\tprod{j{-}1}{k}^{(*)}}\log X_{j,k}^*
\biggr\}
+\mathcal{O}(\epsilon),
\end{align}
where we have neglected terms with $k=j$ since they simply give constants
like $\zeta(2)$ which we are not careful about. We have checked that
the weight in each of the twistors and conjugates vanishes. Roughly
speaking, the $W_*$-$\bar W_*$ framing can be viewed as an axial
regularization, which breaks superconformal symmetry explicitly by
the axial directions, $W_*$ and $\bar W_*$.

\subsection{Super-Poincar\'e}

Instead of reference twistors $W_*$ and $\bar W_*$, we can try to use the matrix corresponding
to the infinity twistor $I$ for regularization purposes.
The antisymmetric matrix $I$ projects any twistor (or conjugate twistor)
to its $\lambda$ (or $\bar \lambda$) component, thus
\[W I W'=\langle \lambda \lambda'\rangle, \qquad \bar W I \bar
W'=[\bar\lambda \bar\lambda'].
\]
The matrix $I$ breaks superconformal symmetry
down to super-Poincar\'e symmetry.

A motivation for introducing the spinor brackets for regularization
is that they arise naturally in the dimensional reduction scheme
which preserves super-Poincar\'e symmetry.
By fully supersymmetrizing the bosonic result~\cite{Bern:1994zx}
using twistor/spinor brackets, we can propose a super-Poincar\'e invariant
expression for the regularized one-loop expectation value.
At the moment we have no first principles derivation for the following
expression, it remains a guess%
\footnote{%
We should note that the divergent part is similar to the one in framing regularization~\eqref{eq:axialframing}.
Also, the divergent part is a bit more complicated than in the bosonic case.
In particular, it depends on odd variables as well as
next-to-adjacent twistors.}
\begin{align}\label{eq:superpoincare}
M^{(1)}_{n,I}=&
- \sum_{j=1}^n
\left(
\frac{2}{\epsilon^2}
+\frac{2}{\epsilon}\log\frac{\tprod{j{-}1}{j{+}1}\tprod{j{+}1}{j{-}1}}
                            {\mu^2\sprod{j}{j{+}1}\cprod{j}{j{+}1}}
\right)
\nln&
-\sum_{j=1}^n\sum_{k=j+3}^{j-3}\Li_2 X_{j,k}
-\half\sum_{j=1}^n \left(\log^2\frac{\sprod{j}{j{+}1}}{\cprod{j}{j{+}1}}-\log^2\frac{\sprod{j{-}1}{j}}{\cprod{j}{j{+}1}}-\log^2\frac{\cprod{j{-}1}{j{-}2}}{\sprod{j}{j{-}1}}\right)
\nln&
- \half\sum_{j=1}^n
\left(
\log^2\frac{\tprod{j{-}1}{j{+}1}}{\mu\cprod{j}{j{+}1}}
+\log^2\frac{\mu\cprod{j{-}1}{j{-}2}}{\tprod{j}{j{-}2}}
+\log^2\frac{\tprod{j}{j{+}2}}{\mu\sprod{j}{j{+}1}}+
\log^2\frac{\mu\sprod{j}{j{-}1}}{\tprod{j}{j{-}2}}\right)
\nln&
-\half\sum_{j=1}^n
\left(\sum_{k=j+2}^{j-3}\log^2\frac{\tprod{j{-}1}{k}}{\tprod{j}{k}}+\sum_{k=j+3}^{j-2}
\log^2\frac{\tprod{j}{k}}{\tprod{j}{k{-}1}}-\sum_{k=j+2}^{j-2}
\log^2\frac{\tprod{j{-}1}{k{-}1}}{\tprod{j}{k}}\right)
\nln&
+\half\gamma\sum_{j=1}^n
\log^2\frac{\tprod{j{-}1}{j{+}1}\sprod{j}{j{+}1}\cprod{j{-}1}{j}}{\tprod{j{+}1}{j{-}1}\sprod{j{-}1}{j}\cprod{j}{j{+}1}}\,.
\end{align}
Note that there is some freedom in supersymmetrizing the bosonic result.
Requiring proper scaling for all twistors and for all conjugate twistors
yields some constraints that guided us to the above result.
We note that the structure multiplied by the coefficient $\gamma$
has proper twistor scaling,
reduces to zero when dropping fermionic coordinates
and obeys some discrete symmetries.
It also does not modify the well-defined finite correlator to be obtained
in \secref{sec:boxing}. Hence we have no means to fix the coefficient $\gamma$,
but for simplicity we will subsequently set it to zero.

We can also take the derivative of \eqref{eq:superpoincare}, which is
essentially its polylogarithm symbol
\begin{align}\label{eq:MHVder}
\delta M^{(1)}_{n,I}=&
\mathbin{\phantom{+}} \sum_{j=1}^n\lrbrk{
+\frac{2}{\epsilon}
+\log \frac{\cprod{j}{j{+}1}\tprod{j{+}1}{j{-}1}\tprod{j}{j{+}2}}{\mu^2\sprod{j}{j{+}1}\cprod{j}{j{-}1}\cprod{j{+}1}{j{+}2}}}
\delta\log\sprod{j}{j{+}1}
\nln &
+ \sum_{j=1}^n \lrbrk{
+\frac{2}{\epsilon}
+\log \frac{\sprod{j}{j{+}1}\tprod{j{-}1}{j{+}1}\tprod{j{+}2}{j}}{\mu^2\cprod{j}{j{+}1}\sprod{j{-}1}{j}\sprod{j{+}2}{j{+}1}}}
\delta\log\cprod{j}{j{+}1}
\nln &
+\sum_{j=1}^n\lrbrk{
-\frac{2}{\epsilon}
+\log\frac{\mu^2\tprod{j{-}1}{j{+}2}\sprod{j}{j{+}1}\cprod{j{+}1}{j{+}2}\tprod{j}{j{+}3}}
         {\tprod{j{-}1}{j{+}1}\tprod{j}{j{+}2}\tprod{j}{j{+}2}\tprod{j{+}1}{j{+}3}}}
\delta\log\tprod{j}{j{+}2}
\nln &
+ \sum_{j=1}^n\lrbrk{
-\frac{2}{\epsilon}
+\log\frac{\mu^2\tprod{j{+}1}{j{-}2}\sprod{j}{j{-}1}\cprod{j{-}1}{j{-}2}\tprod{j}{j{-}3}}
         {\tprod{j+1}{j{-}1}\tprod{j}{j{-}2}\tprod{j}{j{-}2}\tprod{j{-}1}{j{-}3}}}
\delta\log\tprod{j}{j{-}2}
\nln &
+ \sum_{j=1}^n \sum_{k=j+3}^{j-3}
\log \frac{X_{j,k}X_{j+1,k+1}(1-X_{j+1,k})(1-X_{j,k+1})}{(1-X_{j,k})(1-X_{j+1,k+1})}\,\delta\log\tprod{j}{k}.
\end{align}
It is straightforward to see that the result is a
supersymmetrization of the regularized bosonic one-loop expectation value.
For $|j-k|\geq 3$, by discarding all fermionic components,
the combination
\[
\frac{(1-X_{j+1,k})(1-X_{j,k+1})X_{j,k}X_{j+1,k+1}}{(1-X_{j,k})(1-X_{j+1,k+1})}
=
\frac{(x^{+-}_{j+1,k})^2(x^{+-}_{j,k+1})^2}{(x^{+-}_{j,k})^2(x^{+-}_{j+1,k+1})^2}:=u^{+-}_{j,k}
\]
reduces to the bosonic cross ratio
$u_{j,k}:=(x^2_{j+1,k}x^2_{j,k+1})/(x^2_{j+1,k+1}x^2_{j,k})$, and
$\tprod{j}{k}\rightarrow {\langle j k{-}1 k k{+}1 \rangle}/{\langle k{-}1 k\rangle \langle k k{+}1\rangle}$,
thus the term reduces to the derivative
of the finite correlator between the edges $j$ and $k$
\[
\log u_{j,k}\,\delta\log\frac{\langle j k{-}1 k k{+}1 \rangle}{\langle k{-}1 k\rangle \langle k k{+}1\rangle}\,.
\]
Terms with
$k=j\pm 1, j\pm 2$ depend on $I$, and they also reduce to the
derivative of regularized terms in the bosonic result.
Note that $\tprod{j{\mp}1}{j{\pm}1}\to\sprod{j{\mp}1}{j}\cprod{j}{j{\pm}1}$.

\subsection{Boxing}
\label{sec:boxing}

\begin{figure}\centering
\includegraphics[width=\textwidth]{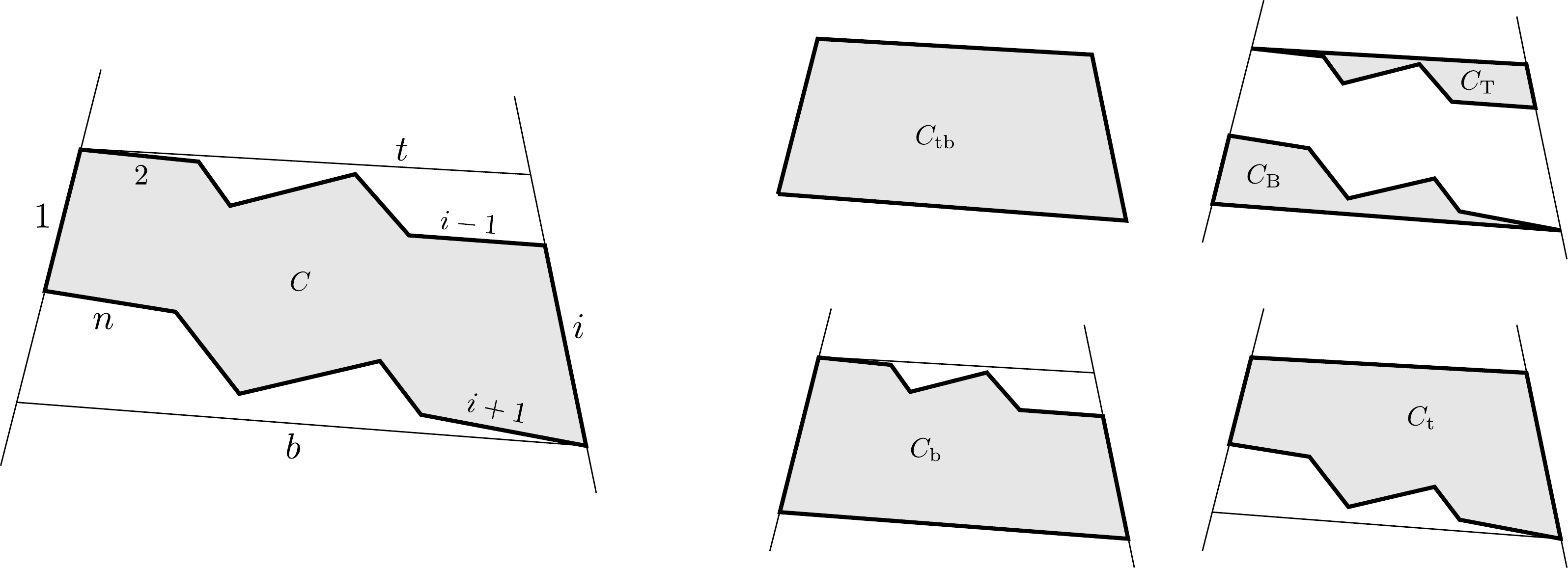}
\caption{The boxing regularization. We have labeled the momenta of
the initial polygonal contour from $1$ to $n$.  The vertices $v_{i}$
are at the intersection of the edges labeled by $i-1$ and $i$ and
have coordinates $(x_{i}, \theta_{i}, \bar{\theta}_{i})$.
The new sides $\mathrm{t}$ and $\mathrm{b}$ are light-like.
We represent the six relevant contours $C$, $C_{\mathrm{t}}$, $C_{\mathrm{b}}$, $C_{\mathrm{tb}}$, $C_{\mathrm{T}}$, $C_{\mathrm{B}}$.}
\label{fig:boxing}
\end{figure}

Finally, similar to~\cite{Gaiotto:2011dt} in the bosonic case, we can use the following ``boxing'' procedure to extract a finite and
superconformal quantity of the one-loop expectation value, as shown in \figref{fig:boxing}. It is a prescription to compute a finite object,
which is not a simple Wilson loop, and we call it the ``boxed Wilson loop''. This prescription, when applied to Wilson loops in any
regularization scheme, should yield the same answer, as we will confirm below.

First we pick two edges, say $1$ and $i$, and extend them from $v_1$ and $v_i$ to two new vertices, which are then connected to $v_{i+1}$ and
$v_2$ by two additional null edges $\mathrm{b}$ and $\mathrm{t}$, respectively (see \figref{fig:boxing}).
Then the boxed Wilson loop is defined as a combination of four Wilson loops expectation values
\[
\frac{\vev{\mathcal{W}[C]}\vev{\mathcal{W}[C\indup{tb}]}}{\vev{\mathcal{W}[C\indup{t}]}\vev{\mathcal{W}[C\indup{b}]}}\,,
\]
where we have specified the four
polygons, $C$, $C\indup{tb}$, $C\indup{t}$, $C\indup{b}$,
by listing the twistors, including
\[\label{eq:edges} W\indup{t}
=W_1-\frac{\langle 1,i]}{\langle 2,i]}
\,W_2,\qquad W\indup{b}
=W_i-\frac{\langle i,1]}{\langle i{+}1,1]}\,
W_{i+1},\]
and similarly for conjugate twistors.\footnote{
A light-like line in space-time is dual to
a twistor $W$ and a conjugate twistor $\bar{W}$ such that $W \cdot
\bar{W} = 0$.  Two light-like lines, represented by two twistor
pairs $(W, \bar{W})$ and $(W', \bar{W}')$, intersect if and only if
$W \cdot \bar{W}' = W' \cdot \bar{W} = 0$.  The incidence relations
in \figref{fig:boxing} imply that $W\indup{t} \cdot \bar{W}_{i} = W\indup{t}
\cdot \bar{W}_{1} = W\indup{t} \cdot \bar{W}_{2} = 0$, which are solved
by the first equality in~\eqref{eq:edges}.}
At one-loop the combination reduces to the following remainder function
\begin{align}\label{eq:defremainder}
r:=&M^{(1)}[C]+M^{(1)}[C\indup{tb}]-M^{(1)}[C\indup{t}]-M^{(1)}[C\indup{b}]\nln
=&M^{(1)}(1, \ldots, n)
+M^{(1)}(1, \mathrm{t}, i, \mathrm{b})
-M^{(1)}(1, \mathrm{t}, i, \ldots)
-M^{(1)}(1, \ldots, i, \mathrm{b}).
\end{align}

By \eqref{eq:oneloop}, the one-loop mixed chirality expectation
value is given by a double integral along the null polygonal contour
$C$
\[\frac{1}{64\pi^2}M^{(1)}[C]=\oint_{C}\oint_{C}\langle A^+ A^-\rangle,\]
thus the boxed Wilson loop at one loop is given by%
\footnote{Since
$\oint_{C} + \oint_{C\indup{tb}} - \oint_{C\indup{t}} - \oint_{C\indup{b}} =
0$, it is easy to show $\oint_{C} \oint_{C} +
\oint_{C\indup{tb}} \oint_{C\indup{tb}} -
\oint_{C\indup{t}} \oint_{C\indup{t}} - \oint_{C\indup{b}}
\oint_{C\indup{b}}= \oint_{C\indup{B}} \oint_{C\indup{T}} + \oint_{C\indup{T}} \oint_{C\indup{B}}$.
}
\[\label{eq:remainder}
r=\oint_{C\indup{T}}\oint_{C\indup{B}}\bigbrk{\langle A^+
A^-\rangle+\langle A^- A^+\rangle}=\sum_{\substack{j\in C\indup{T}\\k\in C\indup{B}}}
\bigbrk{\langle G^+_{j{-}1,j}G^-_{k{-}1,k}\rangle+\langle
G^-_{j{-}1,j}G^+_{k{-}1,k}\rangle}.\]
The sum is over pairs of edges
(or vertices), $j$ of the top null polygon $C\indup{T}$, and $k$ of
the bottom one $C\indup{B}$.
In terms of twistors the two contours are $C\indup{T}=\{2,\ldots,i{-}1,i,\mathrm{t}\}$ and
$C\indup{B}=\{i{+}1,\ldots,n,1,\mathrm{b}\}$, respectively.

A generic edge (or vertex) of the top polygon is well separated from
one of the bottom one, yielding finite correlators for the remainder
function. There are special cases when some correlators naively
diverge, because e.g.\ the vertex $v_2$ lies on the null line $1$.
However, similar to the bosonic case shown in the appendix C
of~\cite{Gaiotto:2011dt}, if we carefully take the limit when $v_2$
approaches line $1$, we find $r=r^+ + r^-$ with
\begin{align} r^+=&\sum^i_{j=4}\sum^n_{k=i{+}2}\left(-\Li_2
X_{j,k}+\frac1 2\log\langle j{-}1,k{-}1]\langle j,k]\log
X_{j,k}\right)
\nln&
+\frac1 2\log^2\frac{\langle 3, n]}{\langle 3,
1]}+\frac 1 2\log^2\frac{\langle 2, \mathrm{b}]}{\langle 3, \mathrm{b}]}
-\frac1
2\log^2\frac{\langle i, n]}{\langle i, 1]}-\frac1
2\log^2\frac{\langle 2, n]}{\langle 3, n]}
\nln&
+\frac 1 2\log\langle i,1]
\langle \mathrm{t}, \mathrm{b}]\log X'_{\mathrm{t},\mathrm{b}}+\frac1 2\log \langle 2, n] \langle 3,
1]\log X'_{3,\mathrm{b}},
\end{align}
where $X'_{i,\mathrm{b}}:={\langle i-1, \mathrm{b}]\langle i, n]}/{\langle i, \mathrm{b}]\langle i-1, n]}$,
and similarly for its conjugate $r^-$.
Thus we confirm that the boxed Wilson loop \eqref{eq:remainder} is indeed finite and superconformal, and its explicit expression agrees with that
of~\cite{Gaiotto:2011dt}, if we replace supersymmetric products $\tprod{j}{k}$ by bosonic ones ${\langle j k{-}1 k k{+}1 \rangle}/{\langle
k-1 k\rangle \langle k k+1\rangle}$.

As an important consistency check, we have explicitly used regularized expectation values, the axial-framing and the super-Poincar\'e forms, to
calculate the boxed Wilson loop. By plugging \eqref{eq:axialframing} and \eqref{eq:superpoincare} into \eqref{eq:defremainder}, we find the same
one-loop result as above in both cases.  In particular all reference twistors $W_*,\bar W_*$ or infinity twistors $I_{A B}$, $I^{A B}$, as well as all divergent contributions neatly cancel.

\section{Yangian symmetry and anomalies}
\label{sec:yangian}

Let us now turn to the definition of the Yangian in full superspace and to the analysis of its anomalies.

\subsection{Yangian generators in ambitwistor space}

The space of functions of ambitwistor space variables $(W^A,\bar W_B)$ admits a representation of the generators $\gen{J}^{A}{}_{B}$ of
the unitary superalgebra $\alg{u}(2,2|4)$ by single derivative operators%
\begin{align}\label{eq:chirachir}
{\gen{J}^{A}}{}_B = \sum_{i=1}^n \gen{J}^{A}_{i, C}= \sum_{i=1}^n (-1)^A W^A_i \partial_{i, B} - (-1)^{A B}\bar W_{i, B}\bar\partial_i^A
\end{align}
 where the sum is taken over the sites of the Wilson loop. The central charge $\gen{C}$ and the hypercharge $\gen{B}$
are obtained by taking the supertrace and trace of $\gen{J}^A{}_B$ respectively.

The level-one generators $\genY{J}^A{}_B$ of the Yangian~\cite{arXiv:0902.2987} transform in the adjoint representation
under the level-zero generators
\[ \label{eq:levone-alg} [\gen{J}^A{}_B, \genY{J}^C{}_D] = (-1)^C \delta^C_B \genY{J}^A{}_D - (-1)^{C+(A+B)(C+D)}\delta^A_D \genY{J}^C{}_B.\]
They are represented by a bilocal formula\footnote{The sign factor $(-1)^A$ was included to eliminate a corresponding factor in the definition of the Yangian charges, see below.}
\[\label{eq:yangrep}{\genY{J}^A}{}_B = \sum_{i,j=1}^n\ \sign(j-i)\ {\gen{J}^A_i}_{C}{\gen{J}^C_j}_{B} = \sum_{i<j}\ {\gen{J}^A_i}_{C}{\gen{J}^C_j}_{B} - (i \leftrightarrow j)\] where we use the sign function to rewrite the ordered sums on the far right hand side of the equation above in terms of sums taken over all sites of the Wilson loop.

Yangian invariance of a function of ambitwistor variables $F(W^A,\bar W_B)$ is achieved when \[j F(W^A,\bar W_B) = 0\] holds for all $j= \gen{J}$ or $j = \genY{J}$.

\paragraph{Superconformal invariance.}
\label{sec:func}
All generators \[\{ \gen{P},\gen{L},\bar{\gen{L}},\gen{K},\gen{D}\vert\gen{Q},\bar{\gen{Q}},\gen{S},\bar{\gen{S}}\}\] of the superconformal algebra $\alg{psu}(2,2|4)$ are neatly represented by $\gen{J}^A{}_B$ so that we can treat them all at once.

The ambitwistor brackets $\langle k, l]$  defined in \secref{sec:ambi} are superconformal invariants \[\gen{J}^A{}_B \langle i, j] = 0,\qquad |i-j|\geq 2.\]
Since the generators $\gen{J}^A{}_B$ are represented by single derivative operators on ambitwistor space  any function $F(\langle k, l])$ of finite ambitwistor brackets is a superconformal invariant, too
\begin{align}
  \label{eq:invar-eq}
\gen{J}^A{}_B F(\langle k, l]) = 0.
\end{align} It is important to note that the dual Coxeter number of $\alg{psu}(2,2|4)$ is zero.  This fact is very helpful during calculations where we often encounter terms proportional to $(-1)^A\delta^A_A $.  Further comments about superconformal invariants can be found in~\appref{app:invariants}.

 Due to regularization (see \secref{sec:regularization}) a wider class of functions $F_{\rm reg}$ with additional dependencies on auxiliary twistors ${W}_\ast$ as in the framing regularization or explicitly non-superconformally invariant combinations of the twistor data like the angled and square brackets \[\langle a, b\rangle = W_a^A I_{AB} W_b^B,\qquad [a,b]= \bar W_{a, A}I^{AB}\bar W_{b, B}\] in supersymmetric regularization has to be considered. These do not in general satisfy superconformal invariance.  We expect therefore an anomalous remainder $\mathcal{A}$ of the invariance equations \[\gen{J}^A{}_B F_{\rm reg} = \mathcal{A}.\]  This has implications for Yangian invariance.

\paragraph{Yangian invariance.}

The generators of the first level in the Yangian ${\rm Y}[\alg{psu}(2,2|4)]$ are given by second order derivatives. This requires any superconformally invariant function of ambitwistors $(W^A, \bar W_B)$ to satisfy an additional second order differential equation
\begin{align}
  \label{yanginvar-eq}
  \genY{J}^A{}_B F(W^A, \bar W_B) = 0.
\end{align}
It is easily checked that a single ambitwistor bracket $\langle k, l]$ on its own is also invariant under the first level generators of ${\rm Y}[\alg{psu}(2,2|4)]$.
However, a generic function $F(\langle k,l])$ of brackets is in general not an invariant as~\eqref{yanginvar-eq}\footnote{The occurring derivative is defined by $\partial_{k,l} = \partial/\partial \langle k, l]$. The function $S$ is a factor defined by $S_{kl,ij} = \sign(k-i)-\sign(k-j)-\sign(l-i)+\sign(l-j)$.}
\begin{align}
  \label{eq:universal}
{\genY{J}^A}{}_B F(\langle m, n]) = (-1)^A&\sum_{i,j,k,l=1}^n S_{kl,ij}W_i^A\bar{W}_{l,B}\langle k,j] \partial_{k,l}\partial_{i,j} F(\langle m, n])\notag\\ &\qquad\qquad-\delta^A_B \sum_{k,l=1}^n \sign(k-l)\langle k,l ] \partial_{k,l}F(\langle  m, n]).
\end{align} is a non-trivial second order partial differential equation.
The trace term proportional to $\delta^A_B$ in~\eqref{eq:universal} only appears when considering the level one hypercharge $\genY{B}$ of the Yangian ${\rm Y}[\alg{u}(2,2|4)]$.  This generator was shown to be an additional symmetry of the scattering amplitudes of $\mathcal{N}=4$ SYM~\cite{Beisert:2011pn} not contained in the Yangian ${\rm Y}[\alg{psu}(2,2|4)]$.  This trace term contains a single derivative with respect to the brackets as can be seen above.  Thus ambitwistor brackets transform covariantly under $\genY{B}$ \[\genY{B}\langle k, l] = 8\ {\rm sign}(l-k) \langle k, l].\] Furthermore it is worth mentioning that the twistor constraints~\eqref{eq:zerobrackets} are even invariant under the full ${\rm Y}[\alg{u}(2,2|4)]$.

\subsection{Anomaly of Yangian symmetry}
\label{sec:anomnotion}

It has been shown that one-loop corrections to the chiral supersymmetric Wilson loop~\cite{arXiv:1010.1167,Mason:2010yk} break the chiral $\mathcal{N}=4$ supersymmetry transformations~\cite{arXiv:1103.3008}.  Its conformal anomaly has been investigated most recently in~\cite{Belitsky:2012nu}.

Also the non-chiral supersymmetric $n$-polygonal Wilson loop $\langle\mathcal{W}\rangle$ presented in this paper suffers from ultraviolet divergences in the regions close to the cusps. These need to be regularized which in turn breaks Yangian invariance \[j F_n \neq 0\] for $j \in {\rm Y}[\alg{psu}(2,2|4)]$.  In contradistinction to the chiral super Wilson loop however it should be possible to find a regularization for the non-chiral Wilson loop that at least preserves super-Poincar\'e symmetry. A very promising guess for such a regularization was given in~\secref{sec:regularization}.

In the following we treat the anomalies  \[j M_n^{(1)} = \mathcal{A}_{n,j}\] for the non-chiral MHV one-loop expectation value in different regularizations.  We investigate not only the anomalies of the symmetry generators $\gen{J}^A{}_B$ of the superconformal algebra $\alg{psu}(2,2|4)$
but also the anomalies \[\label{eq:yanganomeq}\genY{J}^A{}_B M_n^{(1)} = \widehat{\mathcal{A}}^A_{n, B}\] of the Yangian generators.

Naturally, it would be better to check explicitly finite, regularization independent quantities for superconformal and Yangian invariance.  An interesting class of such quantities is provided by $r = M^{(1)}[C] + M^{(1)}[C\indup{tb}] - M^{(1)}[C\indup{t}] - M^{(1)}[C\indup{b}]$ in~\eqref{eq:defremainder} that is obtained by the boxing procedure in~\secref{sec:boxing}.
We find that these are clearly superconformally invariant \[\gen{J}^A{}_B\ r = 0\] as they have no dependence on the regulators.  On the other hand this does not extend to Yangian symmetries which remain broken even when used on these finite quantities.

\subsection{Vertex correlators}
\label{sec:fincor}

 We begin by inspecting finite mixed correlators~\eqref{eq:mixedcorr}
$\costate{0} G^+_{j-1,j}G^-_{k-1,k}\state{0}$ with $j$ and $k$ well separated.
These are obviously invariant under superconformal
transformations as they are functions of ambitwistor brackets alone.

How do the Yangian level one generators fare?
When simply acting with ${\genY{J}^A}{}_B$ on the vertex correlators in~\eqref{eq:mixedcorr} we find
\begin{align}
\label{eq:yangCorr}
\genY{J}^A{}_B \costate{0} G^+_{j-1,j}G^-_{k-1,k}\state{0} &=
64\pi^2(-1)^A\left[\frac{W^A_{j-1}\bar W_{k,B}}{\langle j-1, k]} - \frac{W^A_{j}\bar W_{k-1,B}}{\langle j, k-1]}\right]\nln &\qquad\qquad\qquad + \delta^A_B\log\Bigg(\frac{\langle j-1,k]\langle j,k-1]}{\langle j-1,k-1]\langle j,k]}\Bigg)
\end{align}
so they are not Yangian invariant on their own.
Nevertheless, the anomaly is of the form $f_{j-1,k}-f_{j,k-1}$
(the trace term is slightly different, but the conclusion is the same)
which naively telescopes in the sum over all vertices
\[
\sum_{j,k=1}^n \lrbrk{f_{j-1,k}-f_{j,k-1}}
=
\sum_{j,k=1}^n \lrbrk{f_{j,k}-f_{j,k}}=0.
\]
The trouble is that \eqref{eq:yangCorr} holds only for the finite
vertex correlators with $|j-k|\geq 3$.
The divergent correlators for $|j-k|\leq 2$ need to be regularized.
This turns out to inevitably break superconformal and Yangian invariance.
Therefore it is fair to say that the one-loop Wilson loop expectation value
is perfectly superconformal and Yangian invariant except for the effects
of regularization.
Only the divergent correlators of nearby vertices call for regularization
and break both symmetries in an analogous fashion.
These anomaly terms are computed in the subsequent subsections.

It is worth mentioning that the expression in \eqref{eq:yangCorr}
makes no reference to the vertex
which defines the ordering in the Yangian action \eqref{eq:yangrep}.
This is because the function is also superconformally invariant
in which case the Yangian action respects cyclic symmetry \cite{arXiv:0902.2987}.
However, the regularized vertex correlators for $|j-k|\leq 2$
break superconformal symmetry and consequently introduce dependence on the
reference vertex.

It is helpful to cast $\costate{0} G^+_{j-1,j}G^-_{k-1,k}\state{0}$ into the form of a symbol
\begin{align}
\label{eq:corrsymb}
\mathcal{S}\costate{0} G^+_{j-1,j}G^-_{k-1,k}\state{0} =
\sum_{\stackrel{\scriptstyle i= j-1, j }{l= k-1, k}} R_{i,l} \otimes \langle i, l].
\end{align}
It is remarkable that there are only single brackets in the second entry.  A very similar observation for the form of the symbols of scattering amplitudes has been made in~\cite{CaronHuot:2011ky}.  The $R_{i,l}$ represent the rational functions which appear as the first entry of the symbol for a given second entry $\tprod{i}{l}$.
A generator of $\alg{u}(2,2|4)$ acts like a logarithmic derivative on the last entry of a symbol thus lowering the transcendentality by one.  As can be seen from~\eqref{eq:corrsymb} this is just the bracket $\langle i, l]$.  However, the Yangian level-one generators $\genY{J}^A{}_B$ generically act on both parts of the symbol thus producing rational terms when acting on a finite correlator.

On inspection of~\eqref{eq:universal} it is evident that the only generator acting twice on the second part of a symbol of the form~\eqref{eq:corrsymb} is the level-one hypercharge generator $\genY{B}$. This explains the logarithmic terms in \eqref{eq:yangCorr} proportional to the trace $\delta^A_B$.  The anomaly of $\genY{B}$ therefore suffers from additional single logarithm contributions.

Correlators that need regularization can be inspected in the same way.  Supersymmetric and axial regularization also have symbols with only one bracket in the second entry for the divergent propagators $|j-k| < 3$:
\begin{align}
  \label{eq:regulSus}
\mathcal{S}\langle G^+_{j-1,j}G^-_{k-1,k}\rangle_{\rm susy}& =
\sum_{\stackrel{\scriptstyle i= j-1, j }{l= k-1, k}} \Big(R^{(1)}_{i,l} \otimes \langle i, l] + R^{(2)}_{i,l} \otimes \langle i, l\rangle + R^{(3)}_{i,l} \otimes [ i, l] \Big),
\\
\mathcal{S}\langle G^+_{j-1,j}G^-_{k-1,k}\rangle_{\rm axial} &=
\sum_{\stackrel{\scriptstyle i= j-1, j }{l= k-1, k}} \Big( {\cal R}^{(1)}_{i,l} \otimes \langle i, l] + {\cal R}^{(2)}_{i,l} \otimes \langle i, \bar\ast] + {\cal R}^{(3)}_{i,l} \otimes \langle \ast, l]  + {\cal R}^{(4)}_{i,l} \otimes \langle \ast, \bar\ast] \Big).
\end{align}
The functions $R^{(i)}$ and ${\cal R}^{(i)}$ in~\eqref{eq:regulSus} are all rational
and they differ in both schemes.
The presence of non-invariant brackets $\langle\cdot,\cdot\rangle$ and $[\cdot,\cdot]$ in super-Poincar\'e regularization or $\langle \ast,\cdot]$ and $\langle\cdot,\bar\ast]$ in axial regularization in the second entries break superconformal invariance. Similarly we expect further contributions to the anomalies of all $j \in {\rm Y}(\alg{u}(2,2|4))$

\subsection{Super-Poincar\'e regularization}

\paragraph{The superconformal anomaly.}
From the variation of $M_n^{(1)}$ given in~\eqref{eq:MHVder} follows that it is only necessary to know the action of the generators of $\alg{psu}(2,2|4)$ on spinor brackets.  We write any superconformal generator ${\gen{J}^A}_B$ acting on a function \[F = F(\langle k,\ l], \langle k,k+1\rangle, [k, k+1])\] as a function of derivatives with respect to brackets
\begin{align}
{\gen{J}^A}_B  F =\ (-1)^A I_{BC} &\sum_{i=1}^n (W_i^A W_{i+1}^C -W_{i+1}^A W^C_{i})\partial_i F\nln
 -& (-1)^A  I^{AC}\sum_{i=1}^n(\bar{W}_{i,B}\bar{W}_{i+1,C} -\bar{W}_{i+1,B}\bar{W}_{i,C})\bar\partial_i F
\end{align}
where $\partial_i = {\partial}/{\partial \langle i, i+1\rangle}$, similarly for $\bar\partial_i$.  For $F = M_n^{(1)}$ in supersymmetric regularization, the right hand side is
\begin{align}\label{eq:psuaxanom}
{\gen{J}^A}_B M^{(1)}_n =&\ (-1)^AI_{BC}\sum_{i} \left( \frac{W_i^A W_{i+1}^C -W_{i+1}^A W^C_{i}}{\sprod{i}{i+1}} \right)\nln  &\qquad\qquad\qquad\times\left[\frac{2}{\epsilon} + \log\left(\frac{\cprod{i}{i+1} \langle i,i+2] \langle i+1,i-1] }{\mu^2\sprod{i}{i+1}\cprod{i+1}{i+2}\cprod{i}{i-1}}\right)\right]\nln &-I^{AC} \sum_i \left(\frac{\bar{W}_{i,B}\bar{W}_{i+1,C} -\bar{W}_{i+1,B}\bar{W}_{i,C}}{\cprod{i}{i+1}}\right)\nln
&\qquad\qquad\qquad\times\left[\frac{2}{\epsilon} + \log\left(\frac{\sprod{i}{i+1} \langle i-1, i+1] \langle i+2, i] }{\mu^2\sprod{i-1}{i}\cprod{i}{i+1}\sprod{i+2}{i+1}}\right)\right]
\end{align}
where $I_{AB}$ and $I^{AB}$ are infinity (bi-)twistors. The right hand side of~\eqref{eq:psuaxanom} is zero for any of the Poincar\'e generators as well as supersymmetry and $R$-symmetry thus realizing full super-Poincar\'e symmetry free of anomalies. We are left with the conformal anomaly of the Wilson loop.

When comparing this anomaly to the literature,
e.g.\ \cite{arXiv:0712.1223}, note that the bosonic
result is often split
\[\langle \mathcal{W}[C_n]\rangle = Z_n F_n\]
into a divergent part $Z_n$ and a finite part $F_n$.
The divergent part $Z_{n}$ is defined such that it contains
the full dependence on the renormalization scale $\mu$.
Ref.~\cite{arXiv:0712.1223}, computed the anomaly of the conformal group,
when acting on $\log F_{n}$. This fact must be taken into account
when comparing to the above anomaly of the whole answer,
including the contribution of the divergent part $Z_{n}$.

\paragraph{The Yangian anomaly.}

The calculation of the Yangian anomaly \[\genY{J}^A{}_B M_n^{(1)} = \widehat{\mathcal{A}}^A_{n, B}.\] of $M_n^{(1)}$ can be done in a similar fashion.

As an example we will give the form of the anomaly of the level-one hypercharge $\genY{B}$. Its form is especially nice compared to the anomalies of the other first level generators $\genY{J}^A{}_B$ which can be deduced using~\eqref{eq:levone-alg}.  Just as before we can find the action of $\genY{B}$ on a function $F$ in terms of derivatives with respect to brackets.
The result of acting on $F=M_n^{(1)}$ is
\begin{align}
\label{eq:spyanom}
\genY{B} M_n^{(1)} &= 2 \sum_{j=1}^n \Bigg[\frac{\langle j-1,j+2](-1)^A W_j^A\bar{W}_{j+1,A}}{\langle j-1, j+1]\langle j,j+2]} - \frac{\langle j+2,j-1] (-1)^A W_{j+1}^A\bar{W}_{j,A}}{\langle j+1, j-1]\langle j+2,j]}\notag\\ &\ + 2 \left(\frac{(-1)^A W_{j+2}^A\bar{W}_{j,A}}{\langle j+2,j]}-\frac{(-1)^A W_{j}^A\bar{W}_{j+2,A}}{\langle j,j+2]}\right)\Bigg] + 16 \sum_{j=1}^{n-2}\log\left(\frac{\langle j+2,j]}{\langle j,j+2]}\right)\notag\\ &\ + 16 \log\left(\frac{\langle 1,2\rangle[n-1,n]}{\langle n-1,n\rangle[1,2]}\right)
\end{align}
where the regularization dependent part of the anomaly is fully contained in the terms proportional to $W_i \bar{W}_{i+2}$ and $W_{i+2} \bar{W}_i$.  The last term is a contribution from the $1,n$ boundary.

For Yangian level one generators $\genY{J}^A{}_B$ invariance under cyclic shifts $i \to i+1$ needs to be checked explicitly.  This is done by calculating the difference between a Yangian generator ${\genY{J}^A_{1,n\ B}}$ between site $1$ and site $n$ and a Yangian generator which is shifted by one site ${\genY{J}^A_{2,n+1\ B}}$.  For $\alg{psu}(2,2|4)$ one finds \[{\genY{J}^A_{2,n+1\ B}} - {\genY{J}^A_{1,n\ B}} = 2 (-1)^{(A+C)(C+B)} \gen{J}^C_{1, B}\gen{J}^A{}_{C} - 2 \gen{J}^A_{1,C}{\gen{J}^C}_{B}.\]  Superconformally as well as cyclically invariant functions will be annihilated by the right hand side, proving the compatibility of the Yangian with cyclic shifts.  In the anomalous case presented here, the right hand side is non-vanishing which is the reason for the cyclic asymmetry of the last term in~\eqref{eq:spyanom}.

\subsection{Axial regularization}

\paragraph{The superconformal anomaly.}

Now consider framing as described in \secref{sec:regularization}. When acting with $\gen{J}^A{}_B$ on a function \[F = F(\langle k,l], \langle k,\bar\ast], \langle\ast,k],\langle \ast,\bar\ast])\] in axial regularization the invariance equation is no longer trivially satisfied \[{\gen{J}^A}_B F = (-1)^A\sum_{j=1}^n\Big[ W^A_j \bar{W}_{\bar\ast, B}\frac{\partial F}{\partial \langle j, \bar\ast]}- W^A_{\ast} \bar{W}_{j, B} \frac{\partial F}{\partial \langle \ast, j]}\Big].\]
Setting $F = M^{(1)}_{n,\ast}$ we find
\begin{align}
\label{eq:frameanom}
{\gen{J}^A}_B M^{(1)}_{n,\ast} = (-1)^A\sum_{i=1}^n \left[\frac{W^A_{\ast} \bar{W}_{i, B}}{\langle \ast, i]} - \frac{W^A_i \bar{W}_{\bar\ast, B}}{\langle i, \bar\ast]}\right]\log\left(\epsilon^2\frac{\langle i-1,i+1]^\ast\langle i+1, i-1]^\ast}{\langle i-1,i+1]\langle i+1, i-1]}\right)
\end{align}
employing the notation~\eqref{eq:axialreplacement} introduced in~\secref{sec:regularization}.

This compares nicely with \eqref{eq:psuaxanom}.  In both cases there are single logarithmic terms weighted by rational functions depending on the symmetry breaking brackets.

The twistors $W_\ast$ and $\bar W_{\bar\ast}$ do not get transformed under the action of the generators of $\alg{psu}(2,2|4)$.   Hence, the brackets $\langle i, \bar\ast]$ and $\langle\ast, i]$ are not invariant. Obviously, if the auxiliary twistors $W_\ast$ and $W_{\bar\ast}$ were to be transformed under superconformal transformations we would find the expectation value~\eqref{eq:axialframing} $M^{(1)}_{n,\ast}$ to be an invariant ${\gen{J'}^A}_B M^{(1)}_{n,\ast} = 0$.

\paragraph{The Yangian anomaly.}
In the following we will use some additional notation to shorten the expression for the Yangian anomaly.  We write\footnote{When restricted to bosonic components this denotes the intersection point between a line $(jk)$ and the plane $\bar{W}_k$.} \[(i\ j)\cap k := W_i \langle j, k] - W_j\langle i, k].\] This resembles the notation used in~\cite{ArkaniHamed:2010kv}.  Similarly, for antichiral twistor variables, we use \[k\cap (i\ j) := \bar{W}_i \langle k, j]  - \bar{W}_k \langle k,i].\]  They satisfy the relation \[\langle (i\ j)\cap k, m] = \langle j, i\cap(k\ m)].\]
 Finally, to write the Yangian anomaly in a more compact form we will make use of the notation \[([i\ j] k)\cap(l\ m) = W_i\langle (j\ k)\cap l,m] - W_j \langle (i\ k)\cap l,m ]\,.\]  When restricted to bosonic components this quantity indicates that the points $(j\ k)\cap l, (i\ k)\cap l$ and $(i\ j)\cap l$ are linearly related which enables us via a Pl\"ucker identity to replace this expression by a simpler one.  However on inclusion of the fermionic directions there are additional sign factors from the fermions that prevent us from doing so.

The Yangian anomaly can be straightforwardly calculated. It is given by
\begin{align}
\genY{J}^A{}_B M^{(1),\ast}_n
&= \sum_{i=1}^{n-1}\Bigg(2\,\frac{\langle([i-1\ \ast] i)\cap(i+2\ \bar\ast),i+1]}{\langle i-1, i+1]\langle i, i+2] \langle \ast, \bar\ast]} - 1 \Bigg) \frac{(-1)^A W_i^A \bar{W}_{i+1,B}}{\langle i, i+1]^\ast}\notag\\
& \qquad+\Bigg(2\,\frac{\langle i+1, (\ast\ i+2)\cap([\bar\ast\ i-1] i)]}{\langle i+1, i-1]\langle i+2, i] \langle \ast, \bar\ast]} - 1 \Bigg) \frac{(-1)^A W_{i+1}^A \bar{W}_{i,B}}{\langle i+1,i]^\ast}\notag\\
&-\sum_{i=1}^n\Bigg(\frac{(-1)^A W_i^A \bar{W}_{i+2, B}}{\langle i, i+2]}-\frac{(-1)^A W_{i+2}^A \bar{W}_{i, B}}{\langle i+2, i]}\Bigg)\notag\\
&+2\,\frac{(-1)^A [(n-1\ \ast)\cap 1]^A \bar W_{n,B}}{\langle n-1,1]\langle \ast, n]} - 2\frac{(-1)^A W_n^A [1\cap (n-1\ \bar\ast)]_B}{\langle 1, n-1]\langle n, \bar\ast]}\notag\\
&+ 2\,\frac{(-1)^A W_i^A [n\cap (2\ \bar\ast)]_B }{\langle 1, \bar\ast]\langle n, 2]}-2\,\frac{(-1)^A [(2\ast)\cap n]^A\bar{W}_{1, B}}{\langle 2, n]\langle \ast, 1]}\notag\\
&+ \left(\frac{2\langle n-1, 2]\langle n,1]}{\langle n-1, 1]\langle n, 2]}- 1 \right)\frac{(-1)^A W_n^A \bar{W}_{1,B}}{\langle n,1]^\ast}\notag\\ &- \left(\frac{2\langle 2, n-1] \langle 1,n]}{\langle 1, n-1]\langle 2, n]}-1\right)\frac{(-1)^A W_1^A \bar{W}_{n,B}}{\langle 1,n]^\ast} + 2 \delta^A_B \sum_{j=1}^{n-2}\log\left(\frac{\langle j+2, j]}{\langle j, j+2]}\right).
\end{align}

Despite the fact that we could make superconformal symmetry exact by transforming the auxiliary twistors $W_\ast$ and $\bar W_{\bar\ast}$, too, the same trick does not cure the Yangian anomaly $\widehat{\mathcal A}^A{}_B$.  The bilocal structure of the Yangian generators distinguishes the auxiliary sites as we need to insert these into the chain $1\rightarrow \ldots \rightarrow n \rightarrow 1$.  Putting them between $n$ and $1$ the new level-one generators $\genY{J'}{}^{A}{}_B$ are defined by $\genY{J}^A{}_B$ and additional pieces from the new sites
\[{\genY{J'}{}^A}{}_B = {\genY{J}^A}{}_B + {\gen{J}^A}_C {\gen{J}^C_\ast}_{B} + {\gen{J}^A}_C {\gen{J}^C_{\bar\ast}}_{B} - {\gen{J}^A_{\bar\ast}}_C {\gen{J}^C}_{B} - {\gen{J}^A_{\ast}}_C {\gen{J}^C}_{B} + {\gen{J}^A_{\ast}}_C {\gen{J}^C_{\bar\ast}}_B - {\gen{J}^A_{\bar\ast}}_C {\gen{J}^C_\ast}_B. \]  Their action on $M^{(1),\ast}_n$ is given by
 \[{\genY{J'}{}^A}{}_B M^{(1),\ast}_n = {\widehat{\mathcal{A}}_{n, B}^A} - \gen{J'}^A{}_C\mathcal{A}^C_{n,B} + f^{A}{}_E{}^{F}{}_D{}^{G}{}_{B} \gen{J}^E_{\ast,F}{\gen{J}^D_{\bar\ast, G}} M^{(1),\ast}_n\] with $f^{A}{}_E{}^{F}{}_D{}^{G}{}_{B} = (-1)^A \delta^A_E\delta^F_D\delta^G_B-(-1)^{A+(A+G)(G+F)}\delta^A_D\delta^F_B\delta^G_E$.  In particular, cyclic symmetry remains broken after the inclusion of the auxiliary points into the superconformal generators.

\subsection{Boxing the Wilson loop}

We saw that the above two regularized Wilson loop expectation values break
parts of superconformal and Yangian symmetry. Moreover, the anomaly terms
are different in both cases.
This is particularly inconvenient when the aim is to construct the result
from unbroken or anomalous symmetry consideration.
This is, however, not very surprising because both results
are divergent when the regulator is removed, $\epsilon\to0$.
In other words, the above Wilson loops are regularized but not renormalized,
and therefore all answers certainly depend on the regularization scheme.
It only makes sense to consider the symmetries of a regularized but not renormalized
quantity within any given regularization scheme.

Let us take a look at correlators of local operators in a
conformal theory. Naively they are also divergent and need to be regularized.
In addition, local operators are renormalized, and when the regulator is
removed, the correlation functions are not only perfectly finite, but also
transform nicely under superconformal symmetry (albeit with quantum corrections
to scaling dimensions).

The boxed Wilson loop introduced in \secref{sec:boxing} can be regarded as
such a renormalization of a Wilson loop.
The quantity~\eqref{eq:defremainder} and the ones obtained
by choosing different reference twistors $i$ and $j$
do not depend on the regularization scheme,
they are finite and manifestly superconformally invariant.
However, when inspecting $r^+$ in~\secref{sec:boxing}
we notice the occurrence of brackets like
\[\langle k, t] = \langle k, 1] - \frac{\langle i,1]}{\langle i,2]} \langle k,2].\]
Their occurrence breaks Yangian invariance. This is easily seen when
considering the symbols $\mathcal{S}r$ of these quantities. We find
terms like
\[R_{i,j,k,l}\otimes\left(1-\frac{\langle i,k]\langle
j,l]}{\langle i,l]\langle j, k]}\right).\]
The Yangian acts twice
on the second entry of the symbol leaving behind additional
logarithmic terms on the right hand side of the anomalous invariance
equations of Yangian level-one generators.

The boxed Wilson loop is finite and respects superconformal symmetry,
but it does not respect Yangian symmetry.
Naively this seems to imply that superconformal symmetry is exact
while Yangian symmetry is broken or anomalous.
However one has to bear in mind that the boxed
Wilson loop is not a simple planar Wilson loop expectation value anymore.
For instance, at the one-loop level, the boxed Wilson loop
is equivalent to the correlator of two Wilson loops
\[r = \frac{\langle W[C]\rangle \langle W[C\indup{tb}]\rangle}{\langle W[C\indup{t}]\rangle\langle W[C\indup{b}]\rangle}
= \frac{\langle W[C\indup{T}] W[C\indup{B}]\rangle}{\langle W[C\indup{T}]\rangle \langle W[C\indup{B}]\rangle}
+\order{g^4},
\]
where $C\indup{T}$, $C\indup{B}$ refer to the top and bottom polygons
enclosed by the edges $(\mathrm{t},2,\ldots,i)$ and $(\mathrm{b},i+1,\ldots,n,1)$ in \figref{fig:boxing}.
In the string worldsheet picture, the simple planar Wilson loop has the topology of a disk
while the correlator has annulus topology.
Yangian invariance is expected only for disc topology,
because a loop surrounding the disc which represents a Yangian generator
can be contracted to a point,
see the discussions in \cite{Beisert:2009cs}.
Hence it is not surprising that we find no Yangian invariance
from the quantities obtained through boxing
despite the fact that they are finite and superconformally invariant.

Once again from experience with local operators  we know that two-point functions of local operators do not exhibit Yangian invariance.  Hence it is not surprising that we find no Yangian invariance from the quantities obtained through boxing despite the fact that they are finite and superconformally invariant.

\section{Conclusions}
\label{sec:conclusions}

In conclusion, we have computed the one-loop expectation value of polygonal light-like Wilson loops in full superspace.
The answer we obtained has two pieces: one rational of Grassmann
weight four, and one transcendental of transcendentality degree
two.

For the rational part, the computation in full superspace is
identical to the computation in chiral superspace, in the sense that
they are both computed by integrating the end points of a propagator
along the sides of the Wilson loop.

However, for the transcendental
part the computation looks different.  Both in the
twistor~\cite{Mason:2010yk} and space-time~\cite{arXiv:1010.1167}
version of the chiral Wilson loop, the one-loop computation uses a
quadratic interaction vertex, besides the integration along the sides.
The extra interaction vertex gives rise to the integrand of the Wilson loop,
which is the same as the integrand of the corresponding scattering
amplitude.
In contrast, the corresponding computation in non-chiral superspace directly
yields the integrated result, and does not employ any
interaction vertices. It would be interesting to see if there is
a useful notion of integrand for the non-chiral Wilson loop.

We have presented several computations: in momentum space, in space-time
and in momentum twistor space.  In order to regularize the divergences,
we have used the framing regularization.
We have also presented a guess for the finite part of the answer which preserves Poincar\'e
supersymmetry.

Another way to deal with divergences is to construct finite
quantities.  Inspired by~\cite{Alday:2010ku}, we considered a finite
combination of Wilson loops called the ``boxed Wilson loop'', which
depends on a choice of two reference edges.

In the chiral case, the Poincar\'e supersymmetry generators are $\partial/{\partial \theta}$ and
$\theta \partial/{\partial x}$.
The first chiral-half of Poincar\'e supersymmetry is not anomalous,
but the second one is.
However, if we use a non-chiral formalism, the generators and the answer
are modified in such a way that $\gen{Q}$ and $\gen{\bar{Q}}$ symmetries are both exact.

If we expand the transcendental part in powers of $\theta \bar\theta$,
the second chiral-half anomaly of the chiral result,
i.e.\ the term at zeroth order, can be interpreted as
coming from the generator acting on higher order terms, since
the full result is invariant (see \cite{CaronHuot:2011ky}).
At zeroth order in the $\theta \bar{\theta}$ expansion the answer is
identical to the answer obtained for the chiral Wilson loop.

Finally, we have investigated the superconformal and Yangian anomaly of
several regularized one-loop Wilson loop expectation values.
It turned out that the result is superconformally invariant whenever it is finite.
Conversely, no regularization turned out to be exactly invariant under the Yangian.
In fact, it is quite common for integrable models that the Yangian
is not an exact symmetry.
For example, the Hamiltonian of an integrable spin chain is typically not Yangian invariant.
Instead, the Yangian action converts the bulk Hamiltonian to a telescoping sum.
The resulting boundary terms usually remain and break exact Yangian invariance.
Gladly, this behavior turns out to be sufficient for integrability.
Here the situation is very similar:
The Yangian action \eqref{eq:yangCorr} leaves behind some
terms which telescope in a sum.
Naively, we thus have Yangian invariance.
Unfortunately, some (boundary) terms require regularization
and spoil exact invariance.
Nevertheless, the cancellation of the majority of terms is very remarkable
and should be taken as a consequence of integrability of the problem.

We have computed the Yangian anomaly for
different types of regularizations.
In particular, we have seen that the transcendentality of the
Yangian anomaly is reduced by two degrees compared to the Wilson loop
expectation value.
This seems to imply that it would be substantially simpler to compute
in practice.
It would therefore be good to be able to calculate or quantify
this anomaly in more general terms.
Along the lines of \cite{CaronHuot:2011ky,CaronHuot:2011kk} this
could give easy access to yet higher loop orders.

\medskip

Obviously one would like to compute this Wilson loop in full superspace to higher loops.
Beyond one-loop level, one would have to use
interaction vertices and work with non-abelian gauge fields.
Presumably, the two-loop answer will contain the tree level $\text{N}^2\text{MHV}$, the
one-loop NMHV and the two-loop MHV answers, with a similar pattern for higher loops. This is in line with the recent findings that, in some
sense, a measure of the difficulty of a computation is given by NMHV level plus the loop order.

We believe the results of this work will contribute towards
understanding the perturbation theory of $\mathcal{N}=4$ SYM in
ambitwistor space.  This ambitwistor theory is very elegant but it has
proven hard to quantize.  We hope that availability of results in a
non-chiral formulation will contribute to the understanding of the
quantization of this theory.

Finally, let us comment on the duality with scattering amplitudes.
As we have already mentioned and as discussed in more detail in
ref.~\cite{Beisert:2012}, there is no straightforward
correspondence with scattering amplitudes.  This happens because
supersymmetric intervals in full superspace contain terms quadratic in
the fermionic variables and therefore are not given by differences
$x_{i} - x_{j}$.  This implies that a direct identification of the
particle momenta with the supersymmetric intervals will violate
momentum conservation.

Instead, one could attempt to identify particle momenta with
differences $x_{i} - x_{j}$ of the bosonic superspace coordinates.
This satisfies momentum conservation, but the particles are not
massless anymore since $x_{ij}^{2} \neq 0$.  If we want to take this
proposal seriously, we need to explain the discrepancy in the number
of degrees of freedom;
a massless $\mathcal{N}=4$ multiplet containing
states with helicities between $1$ and $-1$ has $2^4=16$ states
(corresponding to a superfield in 4 $\bar\eta$'s)
while a massive multiplet has $2^8=256$ states
(corresponding to a superfield in 4 $\bar\eta$'s and 4 $\eta$'s).

\subsection{Acknowledgments}

We have benefited from discussions with
Simon Caron-Huot,
Tristan McLoughlin,
David Mesterhazy,
Matteo Rosso
and David Skinner.

The work of NB and BUWS is partially supported by grant no.\
200021-137616 from the Swiss National Science Foundation.
The work of NB is partially supported
by grant no.\ 962 by the German-Israeli Foundation (GIF).
While he was at Brown University, CV was supported by the US Department of
Energy under contract DE-FG02-91ER40688 and the US National Science
Foundation under grant PHY-0643150. NB and CV are grateful to the
Kavli Institute for Theoretical Physics and the organizers of the
``Harmony of Scattering Amplitudes'' workshop for hospitality during
part of this work. The research at Kavli Institute for Theoretical
Physics was supported in part by the U.S.\ National Science
Foundation under grant no.\ NSF PHY05-51164.
We are also grateful to
Perimeter Institute for hospitality during the ``Integrability in
Gauge and String Theory'' conference. SH and BUWS' stay at Perimeter
Institute was partially funded by the EC ``Unify'' grant with the
number PIRSES-GA-2010-269217.

\appendix

\section{Conventions}
\label{app:conventions}

Our convention for the metric signature is ${+}{-}{-}{-}$.

We are raising and lowering spinor indices with the $SL(2)$ invariant antisymmetric matrices $\epsilon$, $\psi_\alpha = \varepsilon_{\alpha \beta} \psi^\beta$, $\psi^\alpha = \varepsilon^{\alpha \beta} \psi_\beta$, $\bar{\psi}_{\dot{\alpha}} = \varepsilon_{\dot{\alpha} \dot{\beta}} \bar{\psi}^{\dot{\beta}}$, $\bar{\psi}^{\dot{\alpha}} = \varepsilon^{\dot{\alpha} \dot{\beta}} \bar{\psi}_{\dot{\beta}}$.  We have $\varepsilon_{\alpha \beta} \varepsilon^{\beta \gamma} = \delta_\alpha^\gamma$, $\varepsilon_{\dot{\alpha} \dot{\beta}} \varepsilon^{\dot{\beta} \dot{\gamma}} = \delta_{\dot{\alpha}}^{\dot{\gamma}}$.

We use a shorthand notation for index contractions: $\langle \psi \chi\rangle  = \psi^\alpha \chi_\alpha$, $[\bar{\psi} \bar{\chi}] = \bar{\psi}_{\dot{\alpha}} \bar{\chi}^{\dot{\alpha}}$.  Complex conjugation changes the order of the Grassmann variables $\langle\psi \chi\rangle^* = [\bar{\chi} \bar{\psi}]$.

We also use the $2 \times 2$ matrices $\sigma^\mu_{\alpha \dot{\alpha}} = (\mathbf{1}, \vec{\sigma})_{\alpha \dot{\alpha}}$ and $\left(\bar{\sigma}^\mu\right)^{\dot{\alpha} \alpha} = (\mathbf{1},-\vec{\sigma})^{\dot{\alpha} \alpha}$ where $\vec{\sigma}$ are the three $3$-dimensional Pauli matrices.  The $\sigma^\mu$ and $\bar{\sigma}^\mu$ matrices are related by
\begin{equation}
  \left(\bar{\sigma}^\mu\right)^{\dot{\alpha} \alpha} = \varepsilon^{\alpha \beta} \varepsilon^{\dot{\alpha} \dot{\beta}} \sigma^\mu_{\beta \dot{\beta}}.
\end{equation}  The $\sigma^\mu$ and $\bar{\sigma}^\mu$ matrices also satisfy the following relations
\begin{align}
  \sigma^\mu \bar{\sigma}^\nu + \sigma^\nu \bar{\sigma}^\mu &= 2 \eta^{\mu \nu} \mathbf{1},\\
  \bar{\sigma}^\mu \sigma^\nu + \bar{\sigma}^\nu \sigma^\mu &= 2 \eta^{\mu \nu} \mathbf{1}.
\end{align}
Finally, some relations which are useful in calculations are
\begin{equation}
  \sigma^\mu_{\alpha \dot{\beta}} \bar{\sigma}_\mu^{\dot{\alpha} \beta} = 2 \delta_\alpha^\beta \delta_{\dot{\alpha}}^{\dot{\beta}}, \quad
  \sigma^\mu_{\alpha \dot{\alpha}} \bar{\sigma}_\nu^{\dot{\alpha} \alpha} = 2 \delta_\nu^\mu.
\end{equation}
Throughout the text we use the notations:
\begin{equation}
  x_{\alpha \dot{\alpha}} = x_\mu \sigma^\mu_{\alpha \dot{\alpha}}, \quad
  x^{\dot{\alpha} \alpha} = x_\mu \left(\bar{\sigma}^\mu\right)^{\dot{\alpha} \alpha}, \quad
  \partial_{\alpha \dot{\alpha}} = \sigma_{\alpha \dot{\alpha}}^\mu \partial_\mu.
\end{equation}
Note that with this convention we have $\partial_{\alpha \dot{\alpha}} x^{\dot{\beta} \beta} = 2 \delta_\alpha^\beta \delta^{\dot{\beta}}_{\dot{\alpha}}$.

\section{Bosonic Prepotentials}
\label{app:bos_reduction}

In this appendix we discuss the bosonic reduction of the supersymmetric calculations presented above.  Since the meaning of the different fields appearing in the supersymmetric computation may be unclear, we aim to understand them better by considering only their lowest components in the Grassmann expansion.

The construction described below is valid for an abelian gauge theory, or for a non-abelian theory at the linearized level.  This is sufficient for computing one loop corrections to the Wilson loops.  It is not clear how to extend this construction to make it work at the non-linear level.

We start by imposing the Lorentz gauge condition $\partial^\mu A_\mu = 0$.  This gauge condition can be solved by $A_\mu = \partial^\nu H_{\nu \mu}$ where $H_{\mu \nu}$ is an antisymmetric rank two tensor.  The field $H$ is known as the Hertz potential.

The ``prepotential'' $H_{\mu \nu}$ has its own gauge symmetry under which the potential $A$ is unchanged $\delta H_{\mu \nu} = \partial^\rho K_{\mu \nu \rho}$, where $K_{\mu \nu \rho}$ is a rank three completely antisymmetric tensor.  The rank three antisymmetric tensor $K_{\mu \nu \rho}$ can be dualised to a vector so the gauge transformation of $H_{\mu \nu}$ can be alternatively written: $\delta H_{\mu \nu} = \varepsilon_{\mu \nu \rho \sigma} \partial^\rho K^\sigma$.

Note that we could also add a piece proportional to $\eta_{\mu \nu} H$ to $H_{\mu \nu}$.  Then $H_{\mu \nu}$ will not be antisymmetric anymore and a variation of $H$ would produce a gauge transformation of $A$, $\delta A_\mu = \partial_\mu \delta H$.  Therefore, we can alternatively describe a $U(1)$ theory by a rank two tensor $H_{\mu \nu}$ of a special kind, which can be decomposed to an antisymmetric tensor with a gauge symmetry and a scalar.

Let us write the tensor $H_{\mu \nu}$ in spinor language:
\begin{equation}
  H_{\mu \nu} \sigma^\mu_{\alpha \dot{\alpha}} \sigma^\nu_{\beta \dot{\beta}} \equiv  \varepsilon_{\dot{\alpha} \dot{\beta}} H_{\alpha \beta} - \varepsilon_{\alpha \beta} \bar{H}_{\dot{\alpha} \dot{\beta}}.
\end{equation}  If $A$ is antihermitian then $H_{\mu \nu}$ is antihermitian and $H_{\alpha \beta}$ and $\bar{H}_{\dot{\alpha} \dot{\beta}}$ are related by hermitian conjugation.

If $H_{\mu \nu} = -H_{\nu \mu}$, then $H_{\alpha \beta} = H_{\beta \alpha}$ and $\bar{H}_{\dot{\alpha} \dot{\beta}} = \bar{H}_{\dot{\beta} \dot{\alpha}}$.  This is the usual decomposition of an antisymmetric tensor in selfdual and anti-selfdual parts.  If $H_{\mu \nu}$
is not antisymmetric but it can be decomposed to an antisymmetric tensor and a scalar,
then the same spinor decomposition holds but $H_{\alpha \beta}$ and $\bar{H}_{\dot{\alpha} \dot{\beta}}$ are not symmetric anymore.

Now we can identify up to a factor the fields $H_{\alpha \beta}$ and $\bar{H}_{\dot{\alpha} \dot{\beta}}$ with the fields $B_{\alpha \beta}$ and $\bar{B}_{\dot{\alpha} \dot{\beta}}$ defined before, since the equation $A_\mu = \partial^\nu H_{\nu \mu}$ translates to
\begin{equation}
A_{\alpha \dot{\alpha}} = -\frac 1 2 \partial_{\alpha \dot{\beta}} \bar{H}^{\dot{\beta}}_{\hphantom{\dot{\beta}} \dot{\alpha}} + \frac 1 2 \partial_{\beta \dot{\alpha}} H^\beta_{\hphantom{\beta} \alpha}
\end{equation} in spinor language.  This is to be compared with eq.~\eqref{eq:a-bosonic}.  If we take $B$, $\bar{B}$, $H$ and $\bar{H}$ to be symmetric, we have that
\begin{equation}
  B_{\alpha \beta} = \frac 1 2 H_{\alpha \beta}, \quad
  \bar{B}_{\dot{\alpha} \dot{\beta}} = \frac 1 2 \bar{H}_{\dot{\alpha} \dot{\beta}}.
\end{equation}

Let us now impose a light-cone gauge condition $l^\mu H_{\mu \nu} = 0$ with $l^2 = 0$ on the field $H_{\mu \nu}$, just like in the supersymmetric computation.  This implies that $l^\mu A_\mu = 0$ as well.  In momentum space, this constraint and the Lorentz gauge constraint, read
\begin{equation}
  l^\mu A_\mu(p) = 0, \quad \text{with $l^2 = 0$}, \qquad
  p^\mu A_\mu(p) = 0, \quad \text{for all $p$}.
\end{equation}  The propagator in this gauge reads
\begin{equation}
  \langle A_\mu(p) A_\nu(q)\rangle = \delta^4(p+q) \frac {-i}{p^2} \left(\eta_{\mu \nu} - \frac {l_\mu p_\nu + p_\mu l_\nu}{p \cdot l} + p^2 \frac {l_\mu l_\nu}{(p \cdot l)^2}\right).
\end{equation}

Now we can obtain the $\langle H_{\mu \rho}(p) H_{\nu \sigma}(q)\rangle$ from $\langle A_\mu(p) A_\nu(q)\rangle = p^\rho p^\sigma \langle H_{\mu \rho}(p) H_{\nu \sigma}(q)\rangle$ and the symmetries of the fields $H$.  This yields
\begin{equation}
  \langle H_{\mu \rho}(p) H_{\nu \sigma}(q)\rangle = \delta^4(p+q) \frac {-i (\eta_{\mu \nu} l_\rho l_\sigma - \eta_{\rho \nu} l_\mu l_\sigma - \eta_{\mu \sigma} l_\rho l_\nu + \eta_{\rho \sigma} l_\mu l_\nu)}{p^2 (p \cdot l)^2}.
\end{equation}

In spinor language, this reads
\begin{gather}
  \langle H_{\alpha \beta}(p) H_{\gamma \delta}(q)\rangle = 0, \qquad
  \langle \bar{H}_{\dot{\alpha} \dot{\beta}}(p) \bar{H}_{\dot{\gamma} \dot{\delta}}(q)\rangle = 0,\\
  \langle H_{\alpha \beta}(p) \bar{H}_{\dot{\alpha} \dot{\beta}}(q)\rangle = -2 i \delta^4(p+q) \frac {l_\alpha l_\beta \bar{l}_{\dot{\alpha}} \bar{l}_{\dot{\beta}}} {p^2 (p \cdot l)^2}.
\end{gather}  Of course, when we do the same computation in $\mathcal{N} = 4$ super Yang--Mills, all the fields are on-shell and the naive computation of the right-hand side yields $\tfrac 1 0$.  This infinity is regularized by putting in the Feynman $+i \epsilon$ prescription for the propagator.  We will use the identity
\begin{equation}
  \frac 1 {x + 0 i} = \text{p.v.} \frac 1 x - i \pi \delta (x).
\end{equation}

We can put the fields $H$ and $\bar{H}$ on-shell by setting
\begin{equation}
  H_{\alpha \beta}(p) = \delta (p^2) \frac {l_\alpha l_\beta}{\langle \lambda l\rangle^2} H(\lambda, \bar{\lambda}), \quad
  \bar{H}_{\dot{\alpha} \dot{\beta}} (q) = \delta (q^2) \frac {\bar{l}_{\dot{\alpha}} \bar{l}_{\dot{\beta}}}{[\bar{\lambda}' \bar{l}]^2} \bar{H}(\lambda', \bar{\lambda}'),
\end{equation}
where we have set $p_{\alpha \dot{\alpha}} = \lambda_\alpha \bar{\lambda}_{\dot{\alpha}}$
on the support of $\delta (p^2)$ and $q_{\alpha \dot{\alpha}} = \lambda'_\alpha \bar{\lambda}'_{\dot{\alpha}}$
on the support of $\delta (q^2)$.
The fields $H(\lambda, \bar{\lambda})$ and $\bar{H}(\lambda', \bar{\lambda}')$  are,
up to a factor $\tfrac 1 2$, the same as the fields $C$ and $\bar{C}$,
truncated to their lowest component in the Grassmann expansion.
Note that the fact that the $\langle C C\rangle$ two-point function
is of Grassmann weight four is consistent with the vanishing
of the $\langle H H\rangle$ two-point function.

Using the on-shell version of the fields and the fact that the propagator with the Feynman $+i \epsilon$ prescription reduces to $-i \pi \delta (p^2)$ on-shell, we get
\begin{equation}
  \delta (q^2) \langle H(\lambda, \bar{\lambda}) \bar{H}(\lambda', \bar{\lambda}')\rangle \frac {[\bar{\lambda} \bar{l}]^2}{[\bar{\lambda}' \bar{l}]^2} = -8 \pi \delta^4(p + q).
\end{equation}

It is not hard to show that
\begin{equation}
  \delta^4 (p + q) = \frac 1 4 \delta (q^2) \left\lvert \frac {\lambda_1}{\lambda_1'}\right\rvert^2 \int \frac {d s}s \delta^2 (\lambda - s \lambda') \delta^2 (s \bar{\lambda} + \bar{\lambda}'),
\end{equation} so we finally get
\begin{equation}
   \label{eq:hh-propagator}
   \langle H(\lambda, \bar{\lambda}) \bar{H}(\lambda', \bar{\lambda}')\rangle = -2 \pi \int d s s^3 \delta^2(\lambda - s \lambda') \delta^2 (s \bar{\lambda} + \bar{\lambda}').
\end{equation}

One can check that the scaling constraints
\begin{equation}
  H(z \lambda, z^{-1} \bar{\lambda}) = z^2 H(\lambda, \bar{\lambda}), \quad
  \bar{H}(z \lambda, z^{-1} \bar{\lambda}) = z^{-2} \bar{H}(\lambda, \bar{\lambda})
\end{equation} are satisfied by the two-point function in eq.~\eqref{eq:hh-propagator}.  Note that, because of the absence of Grassmann variables, the exponent of $s$ in eq.~\eqref{eq:hh-propagator} is different from the exponent of $s$ in eq.~\eqref{eq:cc-propagator}.

\section{Position Space Calculations}
\label{app:position_spc}

\subsection{Prepotential Correlators}

With the propagators \eqref{eq:cc-propagator} ff.\ it is possible to compute correlators of the prepotentials
\begin{align}
\label{eq:superprop}
\costate{0}B^{\alpha\beta}(x^+_1,\theta_1)\,B^{\gamma\delta}(x^+_2,\theta_2)\state{0}
&=\Delta^{\alpha\beta\gamma\delta}(x^+_{12},\theta_{12}), \nln
\costate{0}B^{\alpha\beta}(x^+_1,\theta_1)\,\bar
B^{\dot\gamma\dot\delta}(x^-_2,\bar\theta_2)\state{0}
&=\Delta^{\alpha\beta\dot\gamma\dot\delta}(x^{+-}_{12}),
\nln \costate{0}\bar B^{\dot\alpha\dot\beta}(x^-_1,\bar\theta_1)\,\bar
B^{\dot\gamma\dot\delta}(x^-_2,\bar\theta_2)\state{0}
&=\Delta^{\dot\alpha\dot\beta\dot\gamma\dot\delta}(x^{-}_{12},\bar\theta_{12}),
\end{align}
where, after some initial trivial integrations,
\begin{align}
\label{eq:superpropmom}
\Delta^{\alpha\beta\gamma\delta}(x^+_{12},\theta_{12})&=\frac{1}{64\pi^4}
\int_+ d^2\lambda \,d^2\bar\lambda\,
\exp\left(-\frac{i}{2}\langle\lambda|x^+_{12}|\bar\lambda]\right)\,
\frac{l^{\alpha}l^{\beta}l^{\gamma}l^{\delta}}
     {\langle\lambda\, l\rangle^4}\, \delta^{0|4}(\langle\lambda|\theta_{12}), \nln
\Delta^{\alpha\beta\dot\gamma\dot\delta}(x^{+-}_{12})&=\frac{1}{256\pi^4}
\int_+ d^2\lambda \,d^2\bar\lambda\,
\exp\left(-\frac{i}{2}\langle\lambda| x^{+-}_{12}|\bar\lambda]\right)\,
\frac{l^{\alpha} l^{\beta}\bar l^{\dot\gamma}\bar l^{\dot\delta}}
     {\langle\lambda\, l\rangle^2[\bar l\, \bar\lambda]^2}\,,
\nln \Delta^{\dot\alpha\dot\beta\dot\gamma\dot\delta}(x^{-}_{12},\bar\theta_{12})&=\frac{1}{64\pi^4} \int_+ d^2\lambda
\,d^2\bar\lambda\, \exp\left(-\frac{i}{2}\langle\lambda | x^-_{12} | \bar\lambda]\right)\,
\frac{\bar l^{\dot\alpha}\bar l^{\dot\beta}\bar l^{\dot\gamma}\bar l^{\dot\delta}}
     {[\bar l\, \bar\lambda]^4}\, \delta^{0|4}(\bar \theta_{12} |\bar \lambda]).
\end{align}
To solve the remaining integrals we have to do a calculation very similar to the calculation of the vertex correlators $\costate{0}G^{\pm}_{j-1,j}G^{\pm}_{k-1,k}\state{0}$ presented in \secref{sec:chircorr}.  The remaining instances of $\lambda$ and $\bar\lambda$ can be represented through $(\partial/\partial x)$ contracted with $\bar l$ and $l$, respectively.
This leads to derivative operators
\begin{align}
\mathcal{D}^+ = i\langle l | \sigma^\mu | \bar l ] \frac{\partial}{\partial x^+_{12}{}^\mu}\,,\quad
\mathcal{D}^{\pm} = i\langle l | \sigma^\mu | \bar l ] \frac{\partial}{\partial x^{+-}_{12}{}^\mu}\,,\quad
\mathcal{D}^- = i\langle l | \sigma^\mu | \bar l ] \frac{\partial}{\partial x^-_{12}{}^\mu}\,,
\end{align}
for the three different correlators.  Making use of the momentum representation of the scalar propagator
\[
\int_+ d^2\lambda \,d^2\bar\lambda\,
\exp\bigbrk{-\sfrac{i}{2}\langle\lambda| x |\bar\lambda]}=-16\pi^2 \frac{1}{x^2}\,.
\]
it is possible to perform the corresponding integrations and we obtain the solutions
\begin{align}
  \Delta^{\alpha\beta\gamma\delta}(x^+_{12},\theta_{12})&=- \frac{1}{4\pi^2}
\frac{l^{\alpha}l^{\beta}l^{\gamma}l^{\delta}\,
      \delta^{0|4}(\theta_{12}| x^+_{12} | \bar l])}
     {\langle l | x^+_{12} |\bar l ]^4(x^+_{12})^2}\,,\nln
\Delta^{\alpha\beta\dot\gamma\dot\delta}(x^{+-}_{12})&=-\frac{1}{64\pi^2}
\frac{l^{\alpha}l^{\beta}\bar l^{\dot\gamma}\bar l^{\dot\delta}
(x^{+-}_{12})^2(\alpha + \log (x^{+-}_{12})^2)}
     {\langle l | x^{+-}_{12}| \bar l]^2} -\frac{1}{64\pi^2} \frac{l^{\{\alpha}
(x^{+-}_{12})^{\beta\}\{\dot\gamma}\bar l^{\dot\delta\}}}
          {\langle l| x^{+-}_{12}| \bar l]}\,,
\nln
\Delta^{\dot\alpha\dot\beta\dot\gamma\dot\delta}(x^-_{12},\bar\theta_{12})&=-\frac{1}{4\pi^2}
\frac{\bar l^{\dot\alpha}\bar l^{\dot\beta}\bar l^{\dot\gamma}\bar l^{\dot\delta}\,
      \delta^{0|4}(\langle l| x^-_{12} |\bar\theta_{12})}
     {\langle l | x^-_{12} | \bar l]^4(x^-_{12})^2}\,.
\end{align}
where $\alpha$ is an unspecified integration constant.

One can confirm that these propagators are harmonic functions~\eqref{eq:constraintharm}: Contractions of the second derivatives in
$x$ and $\theta$ all vanish.
Furthermore, the duality constraint~\eqref{eq:constraintpre} is fulfilled. To see this, it is possible to evaluate the equation
\begin{align}
\label{eq:correxam}
D^{(1)}_{a \alpha} D^{(1)}_{b \beta} \Delta^{\alpha\beta\dot\gamma\dot\delta}(x^{+-}_{12}) = -\frac12 \varepsilon_{abcd}{\bar D^{(1)}}{}^c_{\dot\alpha}{\bar D^{(1)}}{}^d_{\dot\beta}\Delta^{\dot\alpha\dot\beta\dot\gamma\dot\delta}(x_{12}^-,\bar\theta_{12})
\end{align}
at $\theta_1 = 0$ using a supersymmetry translation.  The left hand side is given by
\[
\left. D^{(1)}_{a \alpha} D^{(1)}_{b \beta} \Delta^{\alpha\beta\dot\gamma\dot\delta}(x^{+-}_{12})\right| = \frac{1}{4\pi^2}\frac{\bar l^{\dot\gamma}\bar l^{\dot\delta} \langle l | x_{12}^{-} | \bar \theta_{12}]_a \langle l | x_{12}^{-} | \bar \theta_{12}]_b}{\langle l| x_{12}^-|\bar l]^2 (x_{12}^-)^2}
\]
which is easily seen to be equal to the right hand side of~\eqref{eq:correxam} when $\theta_1 = 0$.  Similar considerations work for~\eqref{eq:constraintpre} between chiral-chiral and mixed chirality correlators.

\subsection{Mixed Edge Correlator}

Let us consider the mixed chirality correlator $\costate{0}A^+_j A^-_k\state{0}$
in the chiral decomposition \eqref{eq:chiralconn} and the propagator
\eqref{eq:superprop}. Importantly, the mixed propagator depends only
on the mixed chirality interval
$x^{+-}_{jk}=-x^+_j+x^-_k+4 i\theta_j\bar\theta_k$. There is no
explicit dependence on the fermionic coordinates, they merely enter
through $x^{+-}_{jk}$. This fact simplifies the result somewhat
\begin{align}
\costate{0}A^+_j\, A^-_k \state{0} =&
-\left(-\frac 1 2(dx^+_j)^{\dot\beta\gamma}+2i(d\theta_j)^{\gamma b}
(\bar\theta_k)_{b}{}^{\dot\beta}\right)
\left(\frac 1 2 (dx^-_k)^{\dot\gamma\beta}-2i(d\bar\theta_k)_{b}{}^{\dot\gamma} (\theta_j)^{\beta
b}\right)\nln&\qquad \times \partial_{\delta\dot\beta}
\partial_{\beta\dot\delta} {\Delta^{\delta}}_{\gamma}{{}^{\dot\delta}}_{\dot\gamma}(x^{+-}_{jk})-2 i(d\theta_j)^{\gamma b} (d\bar\theta_k)_{b}{}^{\dot\gamma} \partial_{\delta\dot\delta}{\Delta^{\delta}}_{\gamma}{{}^{\dot\delta}}_{\dot\gamma}(x^{+-}_{jk}).
\end{align}
Moreover, the propagator is a harmonic function and thus one can
exchange the indices $\beta$ and $\delta$ in the second derivative.
Effectively, the propagator appears only in the combination
\[
\partial_{\gamma\dot\gamma}
\Delta^{\gamma\delta\dot\gamma\dot\delta}(x)=-\frac{1}{32\pi^2}\,
\frac{l^{\delta}\bar{l}^{\dot\delta} (\alpha' + \log x^2)}
     {\langle l \vert x\vert\bar l]}\,,
\qquad \alpha'= \alpha + 3.
\]

Now consider the case where point $j$ is restricted to a twistor and point $k$ to a conjugate twistor.  Using the parametrization in eq.~\eqref{eq:nullline} we have
\begin{align}
x^+_j(\bar\kappa,\sigma) &= x^+_j + \lambda_j\bar\kappa, &
\theta_j(\bar\kappa,\sigma) &= \theta_j + \lambda_j\sigma, \nln
x^-_k(\kappa,\bar\sigma) &= x^-_k + \kappa\bar\lambda_k, &
\bar\theta_k(\kappa,\bar\sigma) &= \bar\theta_k +
\bar\sigma\bar\lambda_k.
\end{align}
Since the gauge connections are exact on the sides of the Wilson loop (see eq.~\eqref{eq:Aexact}), we can make the ansatz
\[
\costate{0}A^+_j\,A^-_k\state{0}=\frac 1 {64 \pi^{2}} d_j d_k I^{+-}_{jk}(x^{+-}_{jk}).
\]
Comparing both sides one finds the following two differential
equations on the integral function $I$
\begin{align}
\frac 1 {64 \pi^{2}} \lambda_j^{\delta} \bar\lambda_k^{\dot\delta}
\partial_{\delta\dot\delta} \, I^{+-}_{jk}(x) &=
-\lambda_j^\alpha \bar\lambda_k^{\dot\alpha}
\partial_{\delta\dot\delta}\, \Delta^{\delta}{}_\alpha{}^{\dot\delta}{}_{\dot\alpha}(x), \nln \frac 1 {64 \pi^{2}} \lambda_j^{\delta}
\bar\lambda_k^{\dot\delta} \partial_{\beta\dot\delta} \partial_{\delta\dot\beta}
\, I^{+-}_{jk}(x) &= -\lambda_j^\gamma \bar\lambda_k^{\dot\gamma}
\partial_{\delta\dot\beta}
\partial_{\beta\dot\delta} \Delta^{\delta}{}_{\gamma}{}^{\dot\delta}{}_{\dot\gamma}(x).
\end{align}
Now in general we can assume that the spinors $l$ (or $\bar{l}$) are not collinear to
the $\lambda_{j}$ (or $\bar{\lambda}_{k}$), and thus they form a basis for spinors. We
decompose the coordinate $x$ in this basis
\[
x=\frac{\langle l\vert x\vert\bar l]\,\lambda_j\bar\lambda_k
      -\langle j\vert x\vert\bar l]\,l\bar\lambda_k
      -\langle l\vert x\vert\bar k]\,\lambda_j\bar l
      +\langle j \vert x\vert\bar k]\,l\bar l}
{\langle l j\rangle [k \bar l]}\,,
\] which implies that
\[
\qquad x^2 =
\frac{\langle l\vert x\vert\bar l]\, \langle j \vert x\vert\bar k]-\langle l\vert x\vert\bar k]\, \langle j\vert x\vert\bar l]}
      {\langle l j\rangle [k \bar l]}\,.
\]
We rewrite the differential equations in these coordinates
\begin{align}
\frac{\partial I^{+-}_{jk}(x)}{\partial \langle l\vert x\vert\bar l]} \, &=
\frac{1}{\langle l\vert x\vert\bar l]} \log
\frac{\langle l\vert x\vert\bar l]\, \langle j \vert x\vert\bar k]-\langle l\vert x\vert\bar k]\, \langle j\vert x\vert\bar l]}
      {\langle l j\rangle [k \bar l]}
+\frac{\alpha'}{\langle l\vert x\vert\bar l]} \,, \\
\frac{\partial^2I^{+-}_{jk}(x)}{\partial \langle j\vert x\vert\bar l] \partial
\langle l\vert x\vert\bar k]}
&=\frac{1}{\langle l\vert x\vert\bar l]\, \langle j \vert x\vert\bar k]-\langle l\vert x\vert\bar k]\, \langle j\vert x\vert\bar l]}\,.
\end{align}
Up to functions of $(\langle l\vert x\vert\bar k],\langle j\vert x\vert\bar k])$
and of $(\langle j\vert x\vert\bar l],\langle j\vert x\vert\bar k])$ the solution reads
\begin{equation}
I^{+-}_{jk}(x)
=\Li_2\left(\frac{\langle l\vert x\vert\bar k]\, \langle j\vert x\vert\bar l]}{\langle l\vert x\vert\bar l]\, \langle j \vert x\vert\bar k]}\right)
+\half
\log^2\left(\frac{\langle \rho j\rangle [\bar k \bar\rho]}{\langle l\vert x\vert\bar l]\, \langle j \vert x\vert\bar k]}\right)
+\alpha'\log \langle l\vert x\vert\bar l]\,.
\end{equation}
Note that $\langle j \vert x\vert\bar k]$ is constant along the edges
since
\[\langle j \vert x_{jk}^{+-}\vert\bar k]=\langle j \vert x_{j+1,k}^{+-}\vert\bar k]=\langle j \vert x_{j,k+1}^{+-}\vert\bar k]=\langle j \vert x_{j+1,k+1}^{+-}\vert\bar k]=4 i W_j\bar W_k.\]

Now we can straightforwardly integrate $\langle A_{j}^{+} A_{k}^{-}\rangle$ along the two edges $j$ and $k$ which interpolate between the points $x_{j}$ and $x_{j+1}$ and between the points $x_{k}$ and $x_{k+1}$.  Because the gauge fields are exact differentials on these edges, we only have contributions from the boundary terms:
\[
\sum_{j,k} \lrbrk{ I^{+-}_{jk}(x^{+-}_{j+1,k+1})
-I^{+-}_{jk}(x^{+-}_{j,k+1}) -I^{+-}_{jk}(x^{+-}_{j+1,k})
+I^{+-}_{jk}(x^{+-}_{j,k}) }.
\]

Just like for the other computations, here also we need a regularization.  This can be done as in \secref{sec:regularization}.

The sum should also be independent of $l$ and $\bar{l}$.
This is not obvious since dilogarithm identities are necessary to show it.
It is easy to see that the dependence on the integration constant
$\alpha'$ cancels in the sum.
We have also checked the independence on $l$ and $\bar{l}$ at the level of
the symbol and we have confirmed that the answer obtained in
this way agrees with the answer obtained by
the other methods described in this paper.


\section{\texorpdfstring{$R$}{R}-invariants}
\label{app:rinvar}

Let us show that the $R$-invariant
\begin{equation}
  [j-1 j k-1 k \star] = \frac {\delta^{0\vert 4}(\langle j-1 j k-1
k\rangle \chi_{\star} + \text{cycle})}{\langle j-1 j k-1 k\rangle
\langle j k-1 k \star\rangle \langle k-1 k \star j-1\rangle \langle k
\star j-1 j\rangle \langle \star j-1 j k-1\rangle},
\end{equation} is, up to a factor, the same as the one in
eq.~\eqref{eq:space-timeR}, for $W_{\star} = (0, \bar{\rho} \vert 0)$.

First, we need to compute $\langle a a+1 b \star\rangle$.  We have
\begin{multline}
  \langle a a+1 b \star\rangle =
  \begin{vmatrix}
    \lambda_{a} & \lambda_{a+1} & \lambda_{b} & 0\\
    \frac 1 4 \langle a \vert x_{a} & \frac 1 4 \langle a+1 \vert
x_{a+1} & \frac 1 4 \langle b \vert x_{b} & \bar{\rho}
  \end{vmatrix} =\\
  \begin{vmatrix}
    \lambda_{a} & \lambda_{a+1} & \lambda_{b} & 0\\
    0 & 0 & \frac 1 4 \langle b \vert (x_{b} - x_{a+1}) & \bar{\rho}
  \end{vmatrix} = \frac 1 4 \langle a a+1\rangle \langle b \vert
x_{b}-x_{a+1}\vert \bar{\rho}],
\end{multline} where we have used eq.~\eqref{eq:shiftindex} to rewrite
the expressions.

Similarly, we can compute $\langle a a+1 b b+1\rangle$.  We find
\begin{multline}
  \langle a a+1 b b+1\rangle =
  \begin{vmatrix}
    \lambda_{a} & \lambda_{a+1} & \lambda_{b} & \lambda_{b+1}\\
    \frac 1 4 \langle a \vert x_{a} & \frac 1 4 \langle a+1 \vert
x_{a+1} & \frac 1 4 \langle b \vert x_{b} & \frac 1 4 \langle b+1 \vert
x_{b+1}
  \end{vmatrix} =\\
  \begin{vmatrix}
    \lambda_{a} & \lambda_{a+1} & \lambda_{b} & \lambda_{b+1}\\
    0 & 0 & \frac 1 4 \langle b \vert (x_{b} - x_{a+1}) & \frac 1 4
\langle b \vert (x_{b+1} - x_{a+1})
  \end{vmatrix} =\\ \frac 1 {16} \langle a a+1\rangle \langle b
b+1\rangle (x_{a+1}-x_{b+1})^{2}.
\end{multline}

After some computation using $\chi_{\star} = 0$ and $\chi_{j} = \langle
\lambda_{j} \vert \theta_{j}$, we find that
\begin{equation}
  \delta^{0\vert 4}(\langle j-1 j k-1 k\rangle \chi_{\star} +
\text{cycle}) = \frac 1 {4^{4}} \langle j-1 j\rangle^{4} \langle k-1
k\rangle^{4} \delta^{4}(\theta_{j,k} \vert x_{j,k}^{+} \vert \bar{\rho}]).
\end{equation}

Putting everything together, we find
\begin{equation}
  [j-1 j k-1 k \star] = 16 \frac {\langle j-1 j\rangle \langle k-1
k\rangle \delta^{0\vert 4}(\theta_{k,j} \vert x_{k,j}^{+} \vert
\bar{\rho}])}{(x_{k,j}^{+})^{2} \langle j-1\vert x_{k,j}^{+}\vert
\bar{\rho}] \langle j\vert x_{k,j}^{+}\vert \bar{\rho}] \langle k-1\vert
x_{k,j}^{+}\vert \bar{\rho}] \langle k\vert x_{k,j}^{+}\vert \bar{\rho}]},
\end{equation} which, up to a constant factor, is the same as the
right-hand side of eq.~\eqref{eq:space-timeR}.

\section{Invariants and cross-ratios}
\label{app:invariants}

In this appendix we contemplate about superconformal invariants
which can be constructed from the ambitwistor variables $W_j$ and $\bar W_j$.
Finite superconformal observables should be functions of these.

In the absence of fermionic degrees of freedom,
there are essentially two types of invariants:
mixed and chiral brackets
\[
\langle jk]=W_j^a\bar W_{k,a},
\qquad
\langle jkmn\rangle=\varepsilon_{abcd}W_j^a W_k^b W_m^c W_n^d.
\]
Due to momentum ambitwistor constraints,
the two types of brackets are even related $\langle jk]\sim \langle j,k-1,k,k+1\rangle$,
and it suffices to consider only the chiral brackets.
Superconformal invariants are constructed from these quantities,
taking care that the overall twistor weights vanish.

In the full superspace, the picture is similar, but there are also important differences.
First of all, the totally antisymmetric tensor $\varepsilon_{abcd}$
is not superconformally invariant,
and there is no replacement in the form of a tensor.
Nevertheless there is a generalization of the chiral bracket which exists for superalgebras%
\footnote{Note that one can write
$\langle 1234\rangle$ as the integral $\int d^{0|4}\sigma\,\delta^{0|4}(\sigma\cdot W)$
over 4 fermionic variables $\sigma_k$.}
\[
\int d^ns\, f(s)\, \delta^{4|4}(s\cdot W),
\]
where $f(s)$ is some function of $n$ variables $s_k$.
The Grassmannian integrals
(see refs.~\cite{arXiv:0907.5418, arXiv:1001.3348, arXiv:1002.4625})
yielding the N$^k$MHV tree-level scattering amplitudes
are precisely of this kind.
The essential feature of this integral is that it merely depends on chiral data.

On the other hand, the mixed chiral bracket $\tprod{j}{k}$ generalizes
straight-forwardly to the supersymmetric case.
In (finite) loop corrections to scattering amplitudes these brackets
usually appear in cross-ratios with balanced twistor weights
\[
X_{j,k}:=\frac{\tprod{j{-}1}{k}\tprod{j}{k-1}}{\tprod{j{-}1}{k-1}\tprod{j}{k}}\,.
\]

The main difference between the bosonic and supersymmetric case
is that the in the presence of fermions there exist additional conformal cross-ratios.
Let us briefly consider the cases of $n=4,5,6$ edges. In the bosonic case,
it is well-known that there exist no conformal cross-ratios for $n=4,5$,
and there are three cross-ratios for $n=6$.
In the supersymmetric case, there are also no superconformal invariants for $n=4$.
However, for $n=5$, there is the following superconformal invariant%
\footnote{Note that this invariant cannot be written as a
rational function of the above cross-ratios.}
\[
\frac{\tprods{1}{3}\tprods{3}{5}\tprods{5}{2}\tprods{2}{4}\tprods{4}{1}}
     {\tprods{1}{4}\tprods{4}{2}\tprods{2}{5}\tprods{5}{3}\tprods{3}{1}}
=1+\order{\chi\bar\chi}.
\]
When the fermions are discarded,
there are some relations to show that this expression is exactly $1$,
and hence cannot be considered an independent invariant.
In the presence of fermions, however, the invariant receives
non-trivial corrections in $\chi\bar\chi$.

For $n=6$ the picture is similar. We find $6+1$ independent superconformal invariants
\[
X_{k,k+3}
=\frac{\tprod{k-1}{k+3}\tprod{k}{k+2}}{\tprod{k-1}{k+2}\tprod{k}{k+3}}
\quad\mbox{and}\quad
\frac{\tprods{1}{3}\tprods{2}{4}\tprods{3}{5}\tprods{4}{6}\tprods{5}{1}\tprods{6}{2}}
     {\tprods{1}{4}\tprods{2}{5}\tprods{3}{6}\tprods{4}{1}\tprods{5}{2}\tprods{6}{3}}\,.
\]
In the bosonic case these are also invariant, but there are four constraints
which leave behind the well-known $3$ independent conformal cross-ratios.

It would be interesting to investigate further the number of independent
superconformal cross-ratios. In the bosonic case some general considerations
of the dimension of the (little) group and number of degrees of freedom
yield the answer. However, the presence of fermionic variables obscures the counting somewhat.

\phantomsection
\addcontentsline{toc}{section}{\refname}
\bibliography{superwilson1loop}

\begin{thebibliography}{10}
\providecommand{\href}[2]{#2}
\providecommand{\arxivref}[2]{\href{http://arxiv.org/abs/#1}{#2}}
\providecommand{\doiref}[2]{\href{http://dx.doi.org/#1}{#2}}
\providecommand{\nbbstauthor}[1]{#1}
\providecommand{\nbbstjournal}[1]{\textsf{#1}}
\providecommand{\nbbsttitle}[1]{\textit{#1}}
\providecommand{\nbbsturl}[1]{\texttt{#1}}
\providecommand{\nbbsteprint}[1]{\texttt{#1}}
\providecommand{\nbbststyle}{\raggedright\small\parskip0pt}
\nbbststyle

\bibitem{Beisert:2010jr}
\nbbstauthor{N.~Beisert et~al.},
\nbbsttitle{``{Review of AdS/CFT Integrability: An Overview}''},
\nbbstjournal{\doiref{10.1007/s11005-011-0529-2}{Lett.~Math.~Phys.~99,~3~(2012)}},
\nbbsteprint{\arxivref{1012.3982}{arxiv:1012.3982}}.

\bibitem{arXiv:1007.3246}
\nbbstauthor{B.~Eden, G.~P.~Korchemsky and E.~Sokatchev},
\nbbsttitle{``{From correlation functions to scattering amplitudes}''},
\nbbstjournal{\doiref{10.1007/JHEP12(2011)002}{JHEP~1112,~002~(2011)}},
\nbbsteprint{\arxivref{1007.3246}{arxiv:1007.3246}}.

\bibitem{arXiv:1007.3243}
\nbbstauthor{L.~F.~Alday, B.~Eden, G.~P.~Korchemsky, J.~Maldacena and
  E.~Sokatchev},
\nbbsttitle{``{From correlation functions to Wilson loops}''},
\nbbstjournal{\doiref{10.1007/JHEP09(2011)123}{JHEP~1109,~123~(2011)}},
\nbbsteprint{\arxivref{1007.3243}{arxiv:1007.3243}}.

\bibitem{arXiv:1103.3714}
\nbbstauthor{B.~Eden, P.~Heslop, G.~P.~Korchemsky and E.~Sokatchev},
\nbbsttitle{``{The super-correlator/super-amplitude duality: Part I}''},
\nbbsteprint{\arxivref{1103.3714}{arxiv:1103.3714}}.

\bibitem{arXiv:1103.4353}
\nbbstauthor{B.~Eden, P.~Heslop, G.~P.~Korchemsky and E.~Sokatchev},
\nbbsttitle{``{The super-correlator/super-amplitude duality: Part II}''},
\nbbsteprint{\arxivref{1103.4353}{arxiv:1103.4353}}.

\bibitem{arXiv:0705.0303}
\nbbstauthor{L.~F.~Alday and J.~M.~Maldacena},
\nbbsttitle{``{Gluon scattering amplitudes at strong coupling}''},
\nbbstjournal{\doiref{10.1088/1126-6708/2007/06/064}{JHEP~0706,~064~(2007)}},
\nbbsteprint{\arxivref{0705.0303}{arxiv:0705.0303}}.

\bibitem{Alday:2009zm}
\nbbstauthor{L.~F.~Alday, J.~M.~Henn, J.~Plefka and T.~Schuster},
\nbbsttitle{``{Scattering into the fifth dimension of $\mathcal{N}$ = 4 super
  Yang-Mills}''},
\nbbstjournal{\doiref{10.1007/JHEP01(2010)077}{JHEP~1001,~077~(2010)}},
\nbbsteprint{\arxivref{0908.0684}{arxiv:0908.0684}}.

\bibitem{arXiv:1103.3008}
\nbbstauthor{A.~Belitsky, G.~Korchemsky and E.~Sokatchev},
\nbbsttitle{``{Are scattering amplitudes dual to super Wilson loops?}''},
\nbbstjournal{\doiref{10.1016/j.nuclphysb.2011.10.014}{Nucl.~Phys.~B855,~333~(2012)}},
\nbbsteprint{\arxivref{1103.3008}{arxiv:1103.3008}}.

\bibitem{Bern:2006vw}
\nbbstauthor{Z.~Bern, M.~Czakon, D.~Kosower, R.~Roiban and V.~Smirnov},
\nbbsttitle{``{Two-loop iteration of five-point $\mathcal{N}$ = 4
  super-Yang-Mills amplitudes}''},
\nbbstjournal{\doiref{10.1103/PhysRevLett.97.181601}{Phys.~Rev.~Lett.~97,~181601~(2006)}},
\nbbsteprint{\arxivref{hep-th/0604074}{hep-th/0604074}}.

\bibitem{arXiv:0803.1465}
\nbbstauthor{Z.~Bern, L.~Dixon, D.~Kosower, R.~Roiban, M.~Spradlin, C.~Vergu
  and A.~Volovich},
\nbbsttitle{``{The Two-Loop Six-Gluon MHV Amplitude in Maximally Supersymmetric
  Yang-Mills Theory}''},
\nbbstjournal{\doiref{10.1103/PhysRevD.78.045007}{Phys.~Rev.~D78,~045007~(2008)}},
\nbbsteprint{\arxivref{0803.1465}{arxiv:0803.1465}}.

\bibitem{Kosower:2010yk}
\nbbstauthor{D.~Kosower, R.~Roiban and C.~Vergu},
\nbbsttitle{``{The Six-Point NMHV amplitude in Maximally Supersymmetric
  Yang-Mills Theory}''},
\nbbstjournal{\doiref{10.1103/PhysRevD.83.065018}{Phys.~Rev.~D83,~065018~(2011)}},
\nbbsteprint{\arxivref{1009.1376}{arxiv:1009.1376}}.

\bibitem{arXiv:0807.3196}
\nbbstauthor{N.~Berkovits and J.~Maldacena},
\nbbsttitle{``{Fermionic T-Duality, Dual Superconformal Symmetry, and the
  Amplitude/Wilson Loop Connection}''},
\nbbstjournal{\doiref{10.1088/1126-6708/2008/09/062}{JHEP~0809,~062~(2008)}},
\nbbsteprint{\arxivref{0807.3196}{arxiv:0807.3196}}.

\bibitem{Anastasiou:2003kj}
\nbbstauthor{C.~Anastasiou, Z.~Bern, L.~J.~Dixon and D.~Kosower},
\nbbsttitle{``{Planar amplitudes in maximally supersymmetric Yang-Mills
  theory}''},
\nbbstjournal{\doiref{10.1103/PhysRevLett.91.251602}{Phys.~Rev.~Lett.~91,~251602~(2003)}},
\nbbsteprint{\arxivref{hep-th/0309040}{hep-th/0309040}}.

\bibitem{Bern:2005iz}
\nbbstauthor{Z.~Bern, L.~J.~Dixon and V.~A.~Smirnov},
\nbbsttitle{``{Iteration of planar amplitudes in maximally supersymmetric
  Yang-Mills theory at three loops and beyond}''},
\nbbstjournal{\doiref{10.1103/PhysRevD.72.085001}{Phys.~Rev.~D72,~085001~(2005)}},
\nbbsteprint{\arxivref{hep-th/0505205}{hep-th/0505205}}.

\bibitem{arXiv:0707.1153}
\nbbstauthor{A.~Brandhuber, P.~Heslop and G.~Travaglini},
\nbbsttitle{``{MHV amplitudes in $\mathcal{N}$ = 4 super Yang-Mills and Wilson
  loops}''},
\nbbstjournal{\doiref{10.1016/j.nuclphysb.2007.11.002}{Nucl.~Phys.~B794,~231~(2008)}},
\nbbsteprint{\arxivref{0707.1153}{arxiv:0707.1153}}.

\bibitem{arXiv:0709.2368}
\nbbstauthor{J.~Drummond, J.~Henn, G.~Korchemsky and E.~Sokatchev},
\nbbsttitle{``{On planar gluon amplitudes/Wilson loops duality}''},
\nbbstjournal{\doiref{10.1016/j.nuclphysb.2007.11.007}{Nucl.~Phys.~B795,~52~(2008)}},
\nbbsteprint{\arxivref{0709.2368}{arxiv:0709.2368}}.

\bibitem{arXiv:0712.1223}
\nbbstauthor{J.~Drummond, J.~Henn, G.~Korchemsky and E.~Sokatchev},
\nbbsttitle{``{Conformal Ward identities for Wilson loops and a test of the
  duality with gluon amplitudes}''},
\nbbstjournal{\doiref{10.1016/j.nuclphysb.2009.10.013}{Nucl.~Phys.~B826,~337~(2010)}},
\nbbsteprint{\arxivref{0712.1223}{arxiv:0712.1223}}.

\bibitem{arXiv:0712.4138}
\nbbstauthor{J.~Drummond, J.~Henn, G.~Korchemsky and E.~Sokatchev},
\nbbsttitle{``{The hexagon Wilson loop and the BDS ansatz for the six-gluon
  amplitude}''},
\nbbstjournal{\doiref{10.1016/j.physletb.2008.03.032}{Phys.~Lett.~B662,~456~(2008)}},
\nbbsteprint{\arxivref{0712.4138}{arxiv:0712.4138}}.

\bibitem{Anastasiou:2009kna}
\nbbstauthor{C.~Anastasiou, A.~Brandhuber, P.~Heslop, V.~V.~Khoze, B.~Spence
  et~al.},
\nbbsttitle{``{Two-Loop Polygon Wilson Loops in $\mathcal{N}$ = 4 SYM}''},
\nbbstjournal{\doiref{10.1088/1126-6708/2009/05/115}{JHEP~0905,~115~(2009)}},
\nbbsteprint{\arxivref{0902.2245}{arxiv:0902.2245}}.

\bibitem{ArkaniHamed:2010kv}
\nbbstauthor{N.~Arkani-Hamed, J.~L.~Bourjaily, F.~Cachazo, S.~Caron-Huot and
  J.~Trnka},
\nbbsttitle{``{The All-Loop Integrand For Scattering Amplitudes in Planar
  $\mathcal{N}$ = 4 SYM}''},
\nbbstjournal{\doiref{10.1007/JHEP01(2011)041}{JHEP~1101,~041~(2011)}},
\nbbsteprint{\arxivref{1008.2958}{arxiv:1008.2958}}.

\bibitem{ArkaniHamed:2010gh}
\nbbstauthor{N.~Arkani-Hamed, J.~L.~Bourjaily, F.~Cachazo and J.~Trnka},
\nbbsttitle{``{Local Integrals for Planar Scattering Amplitudes}''},
\nbbsteprint{\arxivref{1012.6032}{arxiv:1012.6032}}.

\bibitem{Mason:2010yk}
\nbbstauthor{L.~Mason and D.~Skinner},
\nbbsttitle{``{The Complete Planar S-matrix of $\mathcal{N}$ = 4 SYM as a
  Wilson Loop in Twistor Space}''},
\nbbstjournal{\doiref{10.1007/JHEP12(2010)018}{JHEP~1012,~018~(2010)}},
\nbbsteprint{\arxivref{1009.2225}{arxiv:1009.2225}}.

\bibitem{arXiv:1010.1167}
\nbbstauthor{S.~Caron-Huot},
\nbbsttitle{``{Notes on the scattering amplitude / Wilson loop duality}''},
\nbbstjournal{\doiref{10.1007/JHEP07(2011)058}{JHEP~1107,~058~(2011)}},
\nbbsteprint{\arxivref{1010.1167}{arxiv:1010.1167}}.

\bibitem{CaronHuot:2011ky}
\nbbstauthor{S.~Caron-Huot},
\nbbsttitle{``{Superconformal symmetry and two-loop amplitudes in planar
  $\mathcal{N}$ = 4 super Yang-Mills}''},
\nbbstjournal{\doiref{10.1007/JHEP12(2011)066}{JHEP~1112,~066~(2011)}},
\nbbsteprint{\arxivref{1105.5606}{arxiv:1105.5606}}.

\bibitem{Bullimore:2011kg}
\nbbstauthor{M.~Bullimore and D.~Skinner},
\nbbsttitle{``{Descent equations for superamplitudes}''},
\nbbsteprint{\arxivref{1112.1056}{arxiv:1112.1056}}.

\bibitem{CaronHuot:2011kk}
\nbbstauthor{S.~Caron-Huot and S.~He},
\nbbsttitle{``{Jumpstarting the all-loop S-matrix of planar $\mathcal{N}$ = 4
  super Yang-Mills}''},
\nbbsteprint{\arxivref{1112.1060}{arxiv:1112.1060}}.

\bibitem{arXiv:0807.1095}
\nbbstauthor{J.~Drummond, J.~Henn, G.~Korchemsky and E.~Sokatchev},
\nbbsttitle{``{Dual superconformal symmetry of scattering amplitudes in
  $\mathcal{N}$ = 4 super-Yang-Mills theory}''},
\nbbstjournal{\doiref{10.1016/j.nuclphysb.2009.11.022}{Nucl.~Phys.~B828,~317~(2010)}},
\nbbsteprint{\arxivref{0807.1095}{arxiv:0807.1095}}.

\bibitem{arXiv:0902.2987}
\nbbstauthor{J.~M.~Drummond, J.~M.~Henn and J.~Plefka},
\nbbsttitle{``{Yangian symmetry of scattering amplitudes in $\mathcal{N}$ = 4
  super Yang-Mills theory}''},
\nbbstjournal{\doiref{10.1088/1126-6708/2009/05/046}{JHEP~0905,~046~(2009)}},
\nbbsteprint{\arxivref{0902.2987}{arxiv:0902.2987}}.

\bibitem{arXiv:1001.3348}
\nbbstauthor{J.~Drummond and L.~Ferro},
\nbbsttitle{``{Yangians, Grassmannians and T-duality}''},
\nbbstjournal{\doiref{10.1007/JHEP07(2010)027}{JHEP~1007,~027~(2010)}},
\nbbsteprint{\arxivref{1001.3348}{arxiv:1001.3348}}.

\bibitem{Hodges:2009hk}
\nbbstauthor{A.~Hodges},
\nbbsttitle{``{Eliminating spurious poles from gauge-theoretic amplitudes}''},
\nbbsteprint{\arxivref{0905.1473}{arxiv:0905.1473}}.

\bibitem{Mason:2009qx}
\nbbstauthor{L.~Mason and D.~Skinner},
\nbbsttitle{``{Dual Superconformal Invariance, Momentum Twistors and
  Grassmannians}''},
\nbbstjournal{\doiref{10.1088/1126-6708/2009/11/045}{JHEP~0911,~045~(2009)}},
\nbbsteprint{\arxivref{0909.0250}{arxiv:0909.0250}}.

\bibitem{Alday:2010ku}
\nbbstauthor{L.~F.~Alday, D.~Gaiotto, J.~Maldacena, A.~Sever and P.~Vieira},
\nbbsttitle{``{An Operator Product Expansion for Polygonal null Wilson
  Loops}''},
\nbbstjournal{\doiref{10.1007/JHEP04(2011)088}{JHEP~1104,~088~(2011)}},
\nbbsteprint{\arxivref{1006.2788}{arxiv:1006.2788}}.

\bibitem{Sohnius:1978wk}
\nbbstauthor{M.~F.~Sohnius},
\nbbsttitle{``{Bianchi Identities for Supersymmetric Gauge Theories}''},
\nbbstjournal{\doiref{10.1016/0550-3213(78)90270-5}{Nucl.~Phys.~B136,~461~(1978)}}.

\bibitem{Galperin:2001uw}
\nbbstauthor{A.~Galperin, E.~Ivanov, V.~Ogievetsky and E.~Sokatchev},
\nbbsttitle{``{Harmonic superspace}''},
Cambridge University Press (2001),
Cambridge, U.K..

\bibitem{Siegel:2011sy}
\nbbstauthor{W.~Siegel},
\nbbsttitle{``{New superspaces/algebras for superparticles/strings}''},
\nbbsteprint{\arxivref{1106.1585}{arxiv:1106.1585}}.

\bibitem{Mandelstam:1982cb}
\nbbstauthor{S.~Mandelstam},
\nbbsttitle{``{Light Cone Superspace and the Ultraviolet Finiteness of the
  $\mathcal{N}$ = 4 Model}''},
\nbbstjournal{\doiref{10.1016/0550-3213(83)90179-7}{Nucl.~Phys.~B213,~149~(1983)}}.

\bibitem{Brink:1982pd}
\nbbstauthor{L.~Brink, O.~Lindgren and B.~E.~Nilsson},
\nbbsttitle{``{$\mathcal{N}$ = 4 Yang-Mills Theory on the Light Cone}''},
\nbbstjournal{\doiref{10.1016/0550-3213(83)90678-8}{Nucl.~Phys.~B212,~401~(1983)}}.

\bibitem{Parkes:1992rz}
\nbbstauthor{A.~Parkes},
\nbbsttitle{``{A Cubic action for selfdual Yang-Mills}''},
\nbbstjournal{\doiref{10.1016/0370-2693(92)91773-3}{Phys.~Lett.~B286,~265~(1992)}},
\nbbsteprint{\arxivref{hep-th/9203074}{hep-th/9203074}}.

\bibitem{Siegel:1992za}
\nbbstauthor{W.~Siegel},
\nbbsttitle{``{The $\mathcal{N}$ = 2 (4) string is selfdual $\mathcal{N}$ = 4
  Yang-Mills}''},
\nbbstjournal{\doiref{10.1103/PhysRevD.46.R3235}{Phys.~Rev.~D46,~R3235~(1992)}},
\nbbsteprint{\arxivref{hep-th/9205075}{hep-th/9205075}}.

\bibitem{Beisert:2012}
\nbbstauthor{N.~Beisert and C.~Vergu},
\nbbsttitle{``{On the Geometry of Null Polygons in Full $\mathcal{N}$ = 4
  Superspace}''},
\nbbsteprint{\arxivref{1203.0525}{arxiv:1203.0525}}.

\bibitem{Witten:1978xx}
\nbbstauthor{E.~Witten},
\nbbsttitle{``{An Interpretation of Classical Yang-Mills Theory}''},
\nbbstjournal{\doiref{10.1016/0370-2693(78)90585-3}{Phys.~Lett.~B77,~394~(1978)}}.

\bibitem{Witten:1985nt}
\nbbstauthor{E.~Witten},
\nbbsttitle{``{Twistor-Like Transform in Ten Dimensions}''},
\nbbstjournal{\doiref{10.1016/0550-3213(86)90090-8}{Nucl.~Phys.~B266,~245~(1986)}}.

\bibitem{Bullimore:2010pj}
\nbbstauthor{M.~Bullimore, L.~Mason and D.~Skinner},
\nbbsttitle{``{MHV Diagrams in Momentum Twistor Space}''},
\nbbstjournal{\doiref{10.1007/JHEP12(2010)032}{JHEP~1012,~032~(2010)}},
\nbbsteprint{\arxivref{1009.1854}{arxiv:1009.1854}}.

\bibitem{Mason:2005kn}
\nbbstauthor{L.~Mason and D.~Skinner},
\nbbsttitle{``{An Ambitwistor Yang-Mills Lagrangian}''},
\nbbstjournal{\doiref{10.1016/j.physletb.2006.02.061}{Phys.~Lett.~B636,~60~(2006)}},
\nbbsteprint{\arxivref{hep-th/0510262}{hep-th/0510262}}.

\bibitem{Goncharov:2010jf}
\nbbstauthor{A.~B.~Goncharov, M.~Spradlin, C.~Vergu and A.~Volovich},
\nbbsttitle{``{Classical Polylogarithms for Amplitudes and Wilson Loops}''},
\nbbstjournal{\doiref{10.1103/PhysRevLett.105.151605}{Phys.~Rev.~Lett.~105,~151605~(2010)}},
\nbbsteprint{\arxivref{1006.5703}{arxiv:1006.5703}}.

\bibitem{Bern:1994zx}
\nbbstauthor{Z.~Bern, L.~J.~Dixon, D.~C.~Dunbar and D.~A.~Kosower},
\nbbsttitle{``{One loop n point gauge theory amplitudes, unitarity and
  collinear limits}''},
\nbbstjournal{\doiref{10.1016/0550-3213(94)90179-1}{Nucl.~Phys.~B425,~217~(1994)}},
\nbbsteprint{\arxivref{hep-ph/9403226}{hep-ph/9403226}}.

\bibitem{Gaiotto:2011dt}
\nbbstauthor{D.~Gaiotto, J.~Maldacena, A.~Sever and P.~Vieira},
\nbbsttitle{``{Pulling the straps of polygons}''},
\nbbstjournal{\doiref{10.1007/JHEP12(2011)011}{JHEP~1112,~011~(2011)}},
\nbbsteprint{\arxivref{1102.0062}{arxiv:1102.0062}}.

\bibitem{Beisert:2011pn}
\nbbstauthor{N.~Beisert and B.~U.~W.~Schwab},
\nbbsttitle{``{Bonus Yangian Symmetry for the Planar S-Matrix of $\mathcal{N}$
  = 4 Super Yang-Mills}''},
\nbbstjournal{\doiref{10.1103/PhysRevLett.106.231602}{Phys.~Rev.~Lett.~106,~231602~(2011)}},
\nbbsteprint{\arxivref{1103.0646}{arxiv:1103.0646}}.

\bibitem{Belitsky:2012nu}
\nbbstauthor{A.~Belitsky},
\nbbsttitle{``{Conformal anomaly of super Wilson loop}''},
\nbbsteprint{\arxivref{1201.6073}{arxiv:1201.6073}}.

\bibitem{Beisert:2009cs}
\nbbstauthor{N.~Beisert},
\nbbsttitle{``{T-Duality, Dual Conformal Symmetry and Integrability for Strings
  on AdS$_5$ $\times$ S$^5$}''},
\nbbstjournal{\doiref{10.1002/prop.200900060}{Fortsch.~Phys.~57,~329~(2009)}},
\nbbsteprint{\arxivref{0903.0609}{arxiv:0903.0609}}.

\bibitem{arXiv:0907.5418}
\nbbstauthor{N.~Arkani-Hamed, F.~Cachazo, C.~Cheung and J.~Kaplan},
\nbbsttitle{``{A Duality For The S Matrix}''},
\nbbstjournal{\doiref{10.1007/JHEP03(2010)020}{JHEP~1003,~020~(2010)}},
\nbbsteprint{\arxivref{0907.5418}{arxiv:0907.5418}}.

\bibitem{arXiv:1002.4625}
\nbbstauthor{G.~Korchemsky and E.~Sokatchev},
\nbbsttitle{``{Superconformal invariants for scattering amplitudes in
  $\mathcal{N}$ = 4 SYM theory}''},
\nbbstjournal{\doiref{10.1016/j.nuclphysb.2010.05.022}{Nucl.~Phys.~B839,~377~(2010)}},
\nbbsteprint{\arxivref{1002.4625}{arxiv:1002.4625}}.

\end{thebibliography}
\bibliographystyle{nb}

\end{document}